%% file: ms.tex
\documentclass{aastex61}
\usepackage{lineno}
\usepackage{xspace}
\usepackage{soul}
\usepackage{color}
\usepackage{acronym}
\usepackage{csquotes}
\usepackage{appendix}
\usepackage{longtable}
\usepackage{comment}

\newcommand\aastex{AAS\TeX}

\definecolor{burntorange}{rgb}{0.8, 0.33, 0.0}

\shorttitle{\aastex\ GTN SAG Report}

\begin{document}
\AuthorCallLimit=999
\title{Gamma-ray Transient Network Science Analysis Group Report}

\input{authors.tex}

\section*{Executive Summary} \label{sec:executive}  \pagestyle{empty}
The Interplanetary Network (IPN) is a detection, localization and alert system that utilizes the arrival time of transient signals in gamma-ray detectors on spacecraft separated by planetary baselines to geometrically locate the origin of these transients. Due to the changing astrophysical landscape and the new emphasis on time domain and multi-messenger astrophysics (TDAMM) from the \textit{Pathways to Discovery in Astronomy and Astrophysics for the 2020s}, this Gamma-ray Transient Network Science Analysis Group was tasked to understand the role of the IPN and high-energy monitors in this new era. The charge includes describing the science made possible with these facilities, tracing the corresponding requirements and capabilities, and highlighting where improved operations of existing instruments and the IPN would enhance TDAMM science. While this study considers the full multiwavelength and multimessenger context, the findings are specific to space-based high-energy monitors. These facilities are important both for full characterization of these transients as well as facilitating follow-up observations through discovery and localization. The full document reports the history of this field in Section\,\ref{sec:intro}, followed by our detailed analyses and findings in some 68\,pages, providing a holistic overview of the role of the IPN and high-energy monitors in the coming decades.

We here summarize our core conclusions. Scientific breakthroughs which require capable high-energy monitors are fully detailed in Section\,\ref{sec:sources}. Specific questions of interest include: 
\begin{itemize}
    \item What are the sources of gamma-ray bursts, fast radio bursts, gravitational waves, neutrinos, magnetars, relativistic supernovae, and exotic transients?
    \item What sub-classes exist within these sources, and what causes their differences?
    \item What produces a gamma-ray burst?
    \item What powers fast radio bursts? 
    \item How were the heaviest elements produced?
    \item What governs the cosmological evolution of the universe? Is the universe flat? Is dark energy a cosmological constant? 
    \item What processes underlie ultra-relativistic jets?
    \item What is the equation of state of dense matter?
    \item Is General Relativity the correct theory of gravity?
    \item What caused the matter excess over antimatter near the beginning of time? 
\end{itemize}


In Section\,\ref{sec:requirements} we explore the requisite observational capabilities necessary to answer these questions. We find that the key instrument capability drivers are 
\begin{itemize}
    \item Precise localizations of gamma-ray transients -- prompt precision of at least $\sim$arcminute scale, and prompt or successful follow-up precision to at least $\sim$1\,arcsecond
    \item Rapid alerts of triggers -- ideally within 10\,s of trigger time
    \item Total coverage of the transient sky -- approaching instantaneous 4$\pi$\,sr with $\sim$weeks long contiguous observing intervals
    \item Precise timing -- absolute timing precision to $\sim$1\,ms accuracy and relative timing precision to $\lesssim$100\,$\mu$s
    \item Unbiased, complete, and sensitive observations of sources -- kilosecond scale background stability, temporal coverage with no observing gaps, and fairly uniform sensitivity down to $\sim$1\,photon\,s$^{-1}$\,cm$^{-2}$, though improved sensitivity would greatly enhance scientific return
\end{itemize}

Without continuous all-sky coverage of gamma-ray transients, we will be blind to key diagnostic information. This would result in a total loss of scientific discovery that cannot be recovered through any other means. The IPN is capable of fulfilling the majority of these requirements and thus has a foundational role to play in this new era of TDAMM. We envision an enhancement to the current operations of the IPN to produce products needed by the wider TDAMM community. This is described in Section\,\ref{sec:network}. Key community requests include:
\begin{itemize}
  \setlength\itemsep{0.1em}
    \item \textbf{The automatic collation of gamma-ray transient data and alert dissemination:} A critical IPN product is the localizations of gamma-ray transients. The automation of these localizations and inclusion of auxiliary information is key to enhancing the return of the current network. The automatic association of signals and calculation of localization information would allow for dedicated GCN alert streams that utilize all available information for classifying gamma-ray bursts, solar flares, magnetar flares, and other transients detected by these instruments. This further facilitates joint searches such as gamma-ray bursts in concert with gravitational waves and neutrinos or magnetar X-ray flares with fast radio bursts. Such a system will enable dozens of prompt and afterglow detections of gamma-ray bursts each year with observations from Rubin, Neo Surveyor, and other wide-field telescopes in their normal survey operations.
    \item \textbf{Signal-based gamma-ray transient catalogs:} Most current gamma-ray burst and magnetar flare catalogs are instrument-specific, requiring a great deal of effort by external groups to check observations against all gamma-ray transients of interest. The IPN should take on the responsibility of building signal-based catalogs, which could automatically update via the above described alert streams.
    \item \textbf{A shared TDAMM database and archive:} A necessary step to enable the previous two products is construction of a shared database and archive to gather the relevant data in one place. Given the IPN use of both private and public data this requires dedicated infrastructure.
    \item \textbf{Development of multi-mission coherent analysis:} The shared database/archive would enable coherent analysis of all active gamma-ray burst monitors, facilitating vastly improved sub-threshold searches and deeper upper limits than is possible with individual instruments. This work would require careful inter-calibration studies. Development of new searches for long-duration transients and coherent temporal monitoring of point sources would extend the usefulness of these instruments to new sources.
    \item \textbf{Preservation of historic data:} Ensuring preservation of archival data, enabling ease of access, and applying modern analysis to old observations are in the scientific zeitgeist. A concerted effort to preserve historic gamma-ray burst monitor data including from Vela, Pioneer Venus Orbit, and the Venera series of satellites should be undertaken as soon as possible, while the necessary knowledge is still in living memory.
    \item \textbf{Additional improvements to the IPN:} A variety of other products will rely upon critical IPN work such as the characterization of systematic errors these instruments. These products include information on prioritization  of prompt detections for follow-up, advanced statistical multi-instrument analysis techniques, and shared open-source analysis software. Development of modern corner plots reporting from prompt gamma-ray burst analysis would follow the advancements in related fields. Such results require enhanced data sharing within the IPN and generation of instrument responses of planetary instruments for gamma-ray burst analysis.
\end{itemize}

These products are necessary to meet community needs with current and forthcoming missions and are well matched to the priorities of three Decadal Surveys. There are major roadblocks that have prevented support within the US for development of these products. The IPN is necessarily community-led, which precludes consideration by any sufficient funding mechanism. Additionally, the use of assets from multiple NASA Science Divisions is also a necessity, but the separation prevents the IPN from accomplishing many goals. These findings on enhancing the IPN echo those in the Multimessenger Astrophysics Science Analysis Group final report three years ago. \textbf{We find that elevation of the IPN to a strategic NASA Astrophysics TDAMM role is necessary to ensure the requisite sustained support for the necessary IPN enhancements and improvement operations.}

The benefits are not only scientific. The proposed IPN products will enable ease of access, facilitating use by a growing community. The IPN, as a strategic initiative and with funding to support education and public outreach, can help these transients and TDAMM enter the zeitgeist, perhaps becoming as well known as supernovae. Within the astrophysics community, the IPN could foster collaboration between individuals on instrument teams and in follow-up groups, producing more valuable scientific results and visibility for early-career scientists, analogous to the standard in the Exoplanet community. The United States has played a leading role in astrophysics for more than a century, sometimes alone, and sometimes through international collaboration such as the IPN partnership with Russia. The release of international scientific results is a key form of soft power. \textbf{TDAMM discovery facilities play a special role in American soft power as they enable positive press worldwide, including the Global South and, perhaps uniquely, non-aligned countries.} Within the US, the IPN utilizes public and private data together for public alert, a key theme of \textit{Pathways to Discovery in Astronomy and Astrophysics for the 2020s}, and is thus a unique example of NASA's Transform to Open Science initiative. Adding the numerous gamma-ray smallsats to the collaboration would allow for greater inclusion of the international community, ensure maximum gain of their data, act to facilitate inclusion of private data towards productive TDAMM results, and allow the United States to support the developing space capabilities in other countries.

We identify a number of specific actions that could be taken by NASA to increase the TDAMM return of current, forthcoming, and future facilities, discussed in Section\,\ref{sec:actions}. These include:
\begin{itemize}
  \setlength\itemsep{0.1em}
    \item \textbf{Enhanced space-based communications:} Providing additional communication contacts for gamma-ray burst monitors will lead to faster retrieval of data and data of higher quality (e.g. high resolution time-tagged event data). The immediate allocation of additional contacts for the Neil Gehrels Swift Observatory is important to discovery in the on-going gravitational-wave observing run (the capability was requested in the latest Senior Review round but not mentioned in the report). Similar improvements can be made to forthcoming Astrophysics missions like StarBurst and the Compton Spectrometer and Imager (COSI). This finding also applies to assets in NASA Planetary and Heliophysics, as well as missions led by other space agencies, where additional or more frequent contacts would boost the scientific return of the IPN and other TDAMM facilities. This provides an avenue for a TDAMM program built with international partners.
    \item \textbf{A return to the intentional launch of distant, dedicated gamma-ray burst monitors:} The United States has not launched a dedicated gamma-ray burst monitor on a distant spacecraft since Ulysses in 1990. The IPN has operated by adapting gamma-ray spectrometers of use in mapping element distributions on distant planetary bodies and in studies of solar flares, but the science supported by the IPN could be vastly enhanced if the gamma-ray transient capabilities were intentional rather than a happenstance patchwork. \textit{Origins, Worlds, and Life: A Decadal Strategy for Planetary Science and Astrobiology 2023-2032} identify the forthcoming capabilities of megarockets, such as the SpaceX Starship, and optical communications, which will soon be demonstrated on Psyche, as capability leaps in Planetary research. These advancements would also improve the operation of the IPN, enabling arcsecond precision localizations, if the IPN is sufficiently supported within NASA. The IPN would also benefit from the multi-decade development of high-precision optical clocks in space for use in fundamental physics studies prioritized in the \textit{Thriving in Space: Ensuring the Future of Biological and Physical Sciences Research: A Decadal Survey for 2023-2032}.
    \item \textbf{The Decadal-recommended TDAMM High-Energy Monitor and Program:} For the IPN localization technique to achieve the necessary precision, interplanetary baselines are required. The finite speed of light is inherent to this technique but precludes immediate dissemination of precise localizations. Thus, while this report focuses on the IPN, we consider what would be necessary to meet the needs of the full high-energy transient TDAMM community. We note that despite the enormous advancements in observing gravitational waves and neutrinos, and development of survey instruments across the electromagnetic spectrum, the current 15\,year gap is the longest span of time without a new, dedicated NASA gamma-ray burst mission, at the explorer scale or larger, since these events were first discovered 56\,years ago. This report provides quantitative evidence in support of the \$500-\$800M recommended scale investment by NASA in a strategic TDAMM program for a worthy successor to Swift and Fermi, a fundamental necessity for the multiwavelength and multimessenger community. 
    \item \textbf{Facilitating greater support for the IPN, both across and within NASA science divisions:} The IPN necessarily relies on data from multiple NASA Science Divisions. The stovepipes, while necessary, act against successful cross-divisional work. General enhancement examples include contributions of on-board GRB trigger algorithms, support for improved absolute timing capability, prioritization of GRB data at the start of scheduled downlinks, additional downlinks, and development of instrument responses for the study of gamma-ray transients. We find that linkages across divisions at the level of instrument teams, between program officers, and support for interdisciplinary work in the reviews of missions are important to maximizing the return of the IPN. Dedicated support at headquarters is required; this is especially important with the need to continue the IPN beyond the life of Fermi and Swift. One option could be assignment of the IPN to a program officer within the NASA Astrophysics Division. Specific cross-division opportunities include dedicated GRB monitors on Gateway and the recommended Uranus Orbiter and Probe mission, where minimal investment could enable sub-arcsecond timing annuli for more than a decade. An example of an Astrophysics Division-specific opportunity would be launching StarBurst on the dedicated COSI launch. Lastly, ensuring long-term support for the IPN via dedicated funding mechanisms is critical to successful enhancement of the IPN products.
\end{itemize}

\newpage
\tableofcontents
\newpage

\setcounter{page}{1}
\section{Introduction}\label{sec:intro} \pagestyle{plain}

\begin{figure}[htp]
\centering
\includegraphics[width=0.5\textwidth]{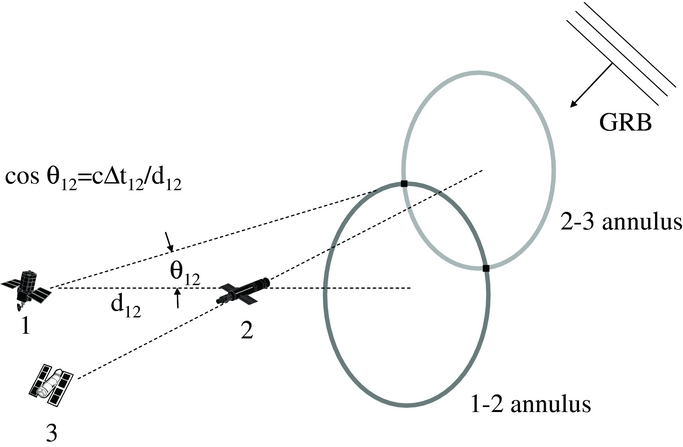}
\caption{Representation of the IPN localization technique. The finite speed of light ($c$) and differential arrival time ($\delta t_{12}$ of a GRB at a pair of spacecraft (spacecraft 1 and 2) separated by a distance ($d_{12}$) allow for geometric determination of an annulus with radius $\theta_{12}$. Analogous variables are shown for the pair comprised of spacecrafts 2 and 3. Multiple pairs allow for multiple annuli. Additional information can be utilized to discriminate between the two regions that remain from the intersection of two annuli. Figure borrowed from \citet{hurley2013interplanetary}.}
\label{fig:triangulation}
\end{figure}

The first Gamma-Ray Burst (GRB) was discovered by the Vela series of satellites in 1967, launched to monitor Earth for atmospheric detonations of nuclear bombs after the signing of the Partial Nuclear Test Ban Treaty. These satellite pairs achieved total global coverage with $\sim10^5$\,km orbits on opposing sides of Earth. The separation distance gives a light travel time difference of up to $\sim1$\,s. As the temporal profiles of GRBs can be measured to $\sim$tens of ms, this allowed for localization by drawing a timing annulus for a pair of detectors, or when the network of satellites was expanded, the intersections of the annuli generated by each pair of spacecraft. These multiple annuli localized the origin of GRBs as outside our Solar System \citep{klebesadel1973observations}. This process of localizing transients is formally called pseudo-range multilateration but is commonly referred to as triangulation, a colloquialism we will use here. A representation of the approach is shown in in Figure\,\ref{fig:triangulation}.

Helios-B, the first mission with an instrument intentionally designed for the study of cosmic GRBs, was launched in 1976 and reached distances of up to 2\,AU, which when paired with the Vela satellites formed the beginning of the InterPlanetary Network (IPN). They were quickly joined by the Pioneer Venus Orbiter, the International Cometary Explorer (originally named the International Sun-Earth Explorer-3) and the Konus and SIGNE-2 series of instruments. Beginning with the discovery of the 1979 magnetar giant flare and subsequent repeating bursts, the IPN began observing flares from what became known as soft gamma-ray repeaters (SGRs). Eventually the IPN triangulations and temporal and spectral studies separated GRBs and SGR short bursts as distinct phenomena \citep{1988ApL&C..27..229E,hurley1989cosmic}.

The early IPN led to evidence of the isotropic distribution of GRBs \citep{mazets1981recent,hartmann1989angular}, pointing to either a particularly local origin within the Milky Way or a cosmological origin of GRBs. The latter was unambiguously proven with the Burst And Transient Source Experiment (BATSE) on-board the Compton Gamma-Ray Observatory, NASA's gamma-ray Great Observatory, launched in 1991, through the combination of isotropy and inhomogeneity inferred via the observed cumulative fluence distribution deviation from a $-$3/2 power-law due to the expansion of the universe \citep{meegan1992spatial}. BATSE remains the most sensitive GRB monitor. 

The failure of the Compton onboard flight recorder led to continuous data downlink via the Tracking and Data Relay Satellite system in order to preserve data and save the mission. This allowed for the automated dissemination of GRB triggers to the community through a system known as the \textbf{BA}TSE \textbf{CO}ordinates \textbf{DI}stribution \textbf{NE}twork (BACODINE) beginning in 1993. The first conception of a GRB jet emitting synchrotron radiation powered by the relativistic, jetted outflow developing up a shock with the surrounding material came out the same year \citep{paczynski1993radio}. The next major change in the field was the launch of the Italian-Dutch BeppoSAX mission, which was built to find GRB counterparts at other wavelengths through localization of the prompt gamma-ray emission at the $3-5'$ level, using a coded aperture mask, reported on the timescale of hours in International Astronomical Union Circulars \citep{frontera2019key}. This combination, more precise than BATSE and more rapid than IPN, led to the recovery of the first GRB afterglow following GRB\,970228 \citep{costa1997discovery} as well as the first optical spectrum which directly proved a cosmological origin for GRB\,970508 \citep{metzger1997spectral}. 

Around this time, the BACODINE system was renamed as the Gamma-ray burst Coordinates Network (GCN), which allowed for automated Notices from multiple missions and human-written Circulars. This system was utilized by the community for several recoveries including the recovery of the first supernova following the BeppoSAX observation of GRB\,980425, proving the collapsar origin of long GRBs \citep{galama1998unusual}. The Robotic Optical Transient Search Experiment performed automated wide-field follow-up of the BATSE localization of GRB\,990123 reported through GCN to find an optical signal in the prompt phase of the burst, giving insight into the prompt emission mechanism of GRBs \citep{akerlof1999observation}. The IPN began disseminating alerts to GCN including marking the active state of SGR\,1900+14 and providing a precise localization \citep{hurley1998reactivation} shortly before its 1998 giant flare. This was shortly after the first proof of magnetars \citep{kouveliotou1998x}, being neutron stars with extreme magnetic fields \citep{duncan1992formation}, made by X-ray observations of the precise IPN position of SGR\,1806-20 \citep{1999ApJ...523L..37H}. Continuing progress was made with the first recovery of afterglow from a short GRB and the IPN discovery of the SGR\,1806-20 giant flare. 

In 2004, the launch of the  Neil Gehrels Swift Observatory (Swift) combined a large coded aperture mask, the Burst Alert Telescope (BAT) with autonomous repointing of the X-ray Telescope (XRT) and Ultraviolet and Optical Telescope (UVOT). This enabled rapid and precise localizations which were disseminated through GCN every few days, on average. This and the corresponding upgrades of ground-based telescopes led to the modern era of multiwavelength characterization of explosive transients across the electromagnetic (EM) spectrum. The launch of the Fermi Gamma-ray Space Telescope in 2008 provided a sensitive low Earth orbit broadband spectrometer for the prompt phase in the Gamma-ray Burst Monitor (GBM) and broader coverage of the prompt and afterglow phase by the Large Area Telescope (LAT). Discoveries in this time period, enabled by these missions and GCN, include the isolation of short GRBs as predominantly arising from neutron star (NS) mergers \citep{fong2015decade}, the first convincing claim of a kilonova following GRB\,130603B \citep{tanvir2013kilonova}, the identification of a low-energy excess in the prompt phase whose origin is still debated \citep[e.g.][]{guiriec2010time,ravasio2019evidence}, and the compilation of a large populations of well-studied events.

In 2015, Laser Interferometer Gravitational-Wave Observatory (LIGO) began distributing GCN notices for gravitational wave (GW) triggers. GW Notices are now distributed by the International Gravitational-wave Network (IGWN), currently comprised of LIGO, Virgo, and KAGRA. Also in 2015, the SuperNova Early Warning System (SNEWS) set up to alert the community to the detection and localization of a Galactic supernova through MeV neutrinos, joined by the dedicated Super-Kamiokande stream in 2020. IceCube began delivering rapid alerts and localizations of high energy neutrinos through GCN in 2016. Thus, GCN is now the backbone for multimessenger transient astronomy. In light of these changes, the GCN was recently renamed to the General Coordinates Network. GCN was key to the announcement of GRB\,170817A, the identification and association of GW170817, and the follow-up observations which recovered the off-axis afterglow and the kilonova signature, marked as the beginning of the GW era of multimessenger astronomy. Other discoveries in this time include several recoveries of GRB afterglows with very-high energy (VHE) telescopes enabling study of inverse Compton components in GRBs for the first time \citep[e.g.][]{magic2019teraelectronvolt} and the discovery of a long GRB with a prompt phase powered by fast cooling synchrotron emission followed by a subsequent kilonova \citep{rastinejad2022kilonova,gompertz2023case}. The identification of GRB\,221009A as the brightest of all time (a.k.a the BOAT) was enabled also by the collaboration tool {\it Slack}, which allowed various instrument teams to compare results more rapidly than is possible through GCN alone.

\begin{figure}[htp]
\centering
\includegraphics[width=\textwidth]{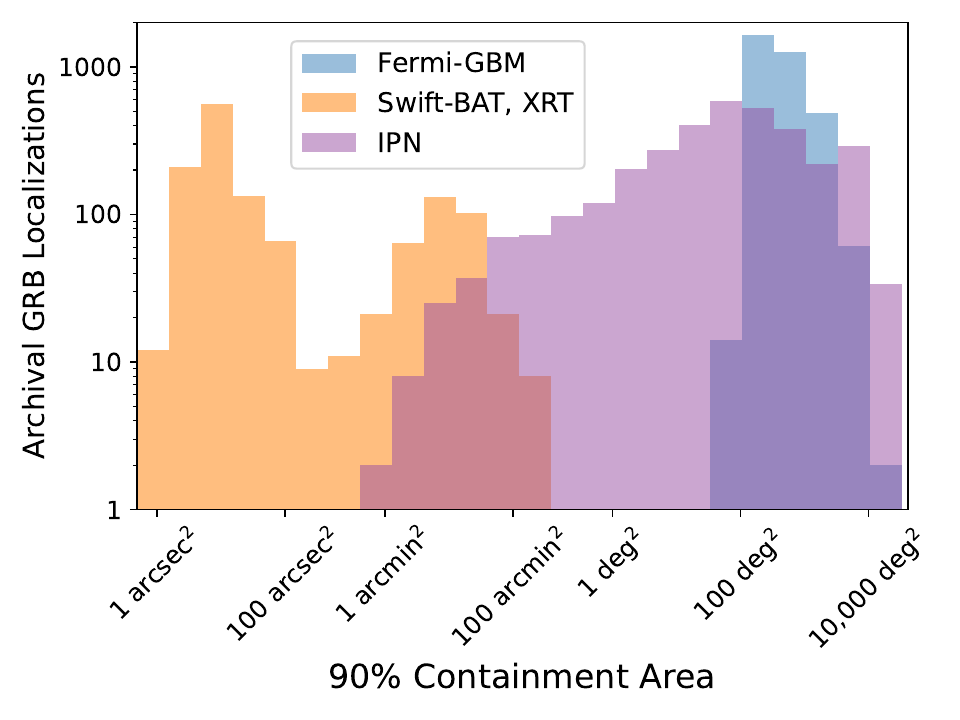}
\caption{The GRB localization precision samples of Fermi-GBM, Swift-BAT and Swift-XRT, and the IPN. On an annualized basis the IPN improves localizations for roughly half of GBM GRBs, which is comparable to the rate of Swift-localized events.}
\label{fig:localizations}
\end{figure}

Swift became the dominant source of prompt, precise localizations of GRBs and the dominant discoverer of magnetars following its launch, resulting in a general perception that the IPN was no longer necessary. As will be described in the following sections, among the most important figures of merit for GRB monitors is detection rate and localization precision. To motivate the continued need for the IPN we point to Figure\,\ref{fig:localizations}. The IPN provide systematic improvement over Fermi-GBM localizations for a large number of events, including all bright events, with localization precision approaching that of Swift-BAT. As bright and nearby \enquote{Rosetta Stone} TDAMM astronomy events provide unique advances in understanding their broader populations, the IPN is indispensible. 

For direct scientific examples, since the launch of Swift the IPN has been instrumental in proving that some short GRBs arise from extragalatic magnetar giant flares \citep{frederiks2007possibility,mazets2008giant,svinkin2021bright,burns2021identification}. Key to those observations is Konus onboard the Wind spacecraft at Earth-Sun L1, which has provided near continuous all-sky coverage of the high-energy sky since 1994. This, coupled with the background stability in the Wind orbit, makes Konus the dominant identifier of the bursts which belong to the ultra-long class of GRBs as well as key to searches for high-energy emission around optically-identified relativistic transients \citep[e.g.][]{margutti2019embedded,andreoni2021fast,ho2022cosmological}. The development of sensitive wide-field observatories has allowed tiling of large localization regions, where higher latency IPN localizations of the prompt phase can be used to unambiguously associate optical counterparts, as was done to identify the collapsar with the shortest prompt phase identified \citep{ahumada2021discovery}. The flagging of active magnetar states by Swift and the IPN was key to radio and X-ray observations of SGR\,1935+2154, leading to the first joint detection of an SGR short burst with a contemporaneous fast radio burst (FRB), proving magnetars as an origin of FRBs \citep{bochenek2020fast,2020ApJ...898L..29M}. Long-term monitoring of the SGR\,1935+2154 flares proved this particular X-ray flare as unique \citep{ridnaia2021peculiar}. Most recently, the rapid IPN localizations of GRB\,230307A enabled most direct measure of heavy element (r-process) creation \citep{levan2023grb}. 

\begin{figure}[htp]
\centering
\includegraphics[width=\textwidth]{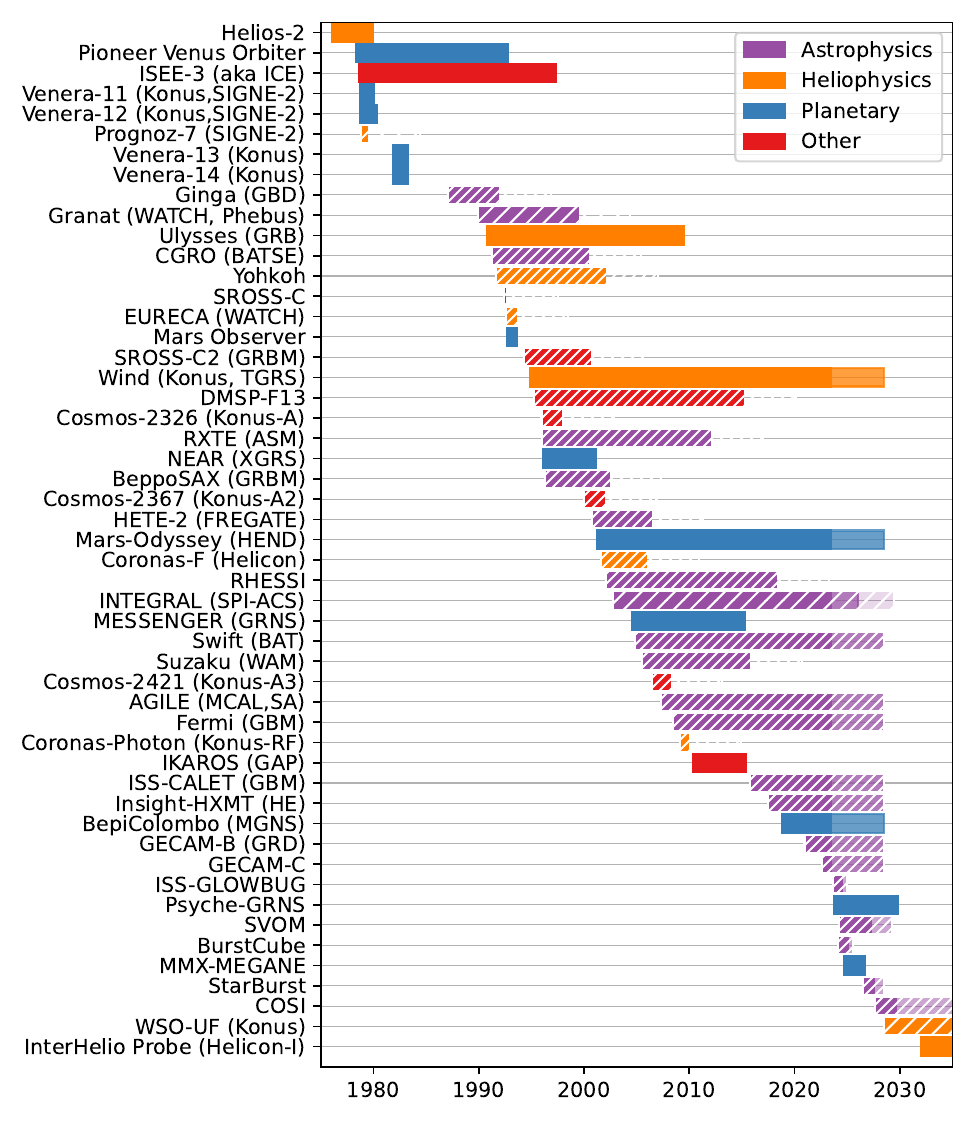}
\caption{The 51 past, present, and future missions of the IPN. Color corresponds to primary spacecraft discipline. Low Earth orbit satellites have many hashes, decreasing to solid colors for distant spacecraft. 
Faded regions correspond to nominal extensions which are assumed to be 5\,years for active missions, unless precise dates are known. INTEGRAL shows both the planned shut off date and the Earth reentry date.}
\label{fig:ipn_instruments}
\end{figure}

The brief historical overview of the study of short-duration gamma-ray transients serves to highlight a number of themes which will be evident throughout this report. For 50\,years there have been a steady supply of major breakthroughs in high-energy astrophysics. While these discoveries were initially made by the high-energy monitors alone, for 30\,years it has been the joint advancement in the capabilities of the high-energy monitors, ground-based facilities, and growth in the adoption of GCN that has allowed for continued advancement. The decision of the high-energy monitors to release open alerts fostered collaboration within the community and has led to grander characterization of these explosions. The importance of rapid, precise, and accurate localizations are evident. Throughout this time the IPN has contributed unique capabilities and complemented the larger-scale, dedicated NASA GRB missions. The numerous list of IPN missions and instruments is shown in Figure\,\ref{fig:ipn_instruments}. With implemented upgrades to other wavelength and messenger capabilities, NASA investment in a major upgrade to GCN, and as we are in the longest gap since field inception for a dedicated NASA GRB monitor at Explorer scale or larger, it is time to plan for the next phase of gamma-ray transient monitors. 


The emphasis on time-domain and multimessenger astronomy by \textit{Pathways to Discovery in Astronomy and Astrophysics for the 2020s}, the Astro2020 Decadal report, led to this Science Analysis Group tasked with quantifying what could and should be done with the IPN. Section\,\ref{sec:sources} contains subsections focused on the study of magnetars, compact mergers, and collapsars by high energy monitors. Each discusses the various signals, scientific results, instrument requirements, and high energy monitor products of relevance for these sources and affiliated communities.
In Section\,\ref{sec:requirements} we explore the science unlocked with fiducial values of various capability requirements for high-energy monitors. These two sections take a holistic view of the field and conveys the science that can be done with the IPN.
Section\,\ref{sec:actions} provides a specific set of actionable items that NASA can undertake to maximize scientific return with current, forthcoming, and potential future gamma-ray instruments.
Section\,\ref{sec:network} outlines the requested data products from the IPN, identifies where specific operational and programmatic improvements to the IPN can be made, and discusses the additional benefits which may arise from sustained investment in the IPN.
We close with conclusions in Section\,\ref{sec:conclusions}. The sections contain some duplication of information but are separated to allow for 
self-contained information for the various use cases of this document, including members of the IPN, instrument teams, the community, and NASA.

\section{Sources}\label{sec:sources}
GRB monitors recover short-duration transients in the keV to MeV regime. Below are subsections focused on 
magnetars, compact mergers, collapsars, and other transients. Each of these subsections discusses the various signals expected from these sources, the scientific results that can be made through their study, the instrument requirements and capabilities that enable this progress, and a discussion on the products from the high-energy teams that would be of great value for the broader community which studies those objects.

The most critical requirements, which have never been met with a single observatory, are localization precision, alert latency to follow-up facilities, uninterrupted all-sky coverage, and sensitivity. Localization precision and alert latency are often deeply intertwined. Uninterrupted viewing requires long-term (ideally, continuous) contiguous viewing intervals as well as a high total observing fraction which is set by the instantaneous field of view (often more than half the sky) and livetime (generally in excess of 80\%). The sensitivities quoted are characteristic order-of-magnitude scales and are baselined against the typical observing range for gamma-ray transients. Temporal requirements include absolute timing precision (offset from a standardized reference time), relative timing precision (offset between events observed on-board), and temporal resolution of the data. Energy range is discussed only when the requirements exceed the fiducial 50-300\,keV detection range. We additionally discuss background stability, maximum photon rates, spectral resolution, and polarization, when necessary.

\subsection{Magnetars}\label{sec:magnetars}
A supernova explosion, or more rarely low-mass mergers or accretion-induced collapse of a white dwarf, can leave behind a NS remnant, a dense stellar object supported by quantum degeneracy pressure rather than fusion. Only $\sim25$\,kilometers across, NSs are so dense they exceed the mass of the Sun; these objects sit near the density limit of matter. A significant fraction of these new NSs are magnetars \citep{beniamini2019formation}, hosting the most powerful magnetic fields in the cosmos with flux densities of order $10^{13}-10^{15}$\,Gauss. 

Magnetars are classified as soft gamma-ray repeaters (SGRs) or anomalous X-ray pulsars (AXPs). SGRs are the subclass of interest for high-energy monitors as they are observed to release X-ray and soft gamma-ray short bursts, magnetar giant flares (MGFs), and FRBs. Short bursts are brief high-energy flashes with a duration on the order of $\sim$1-1000\,ms, luminosities observed from $\sim10^{36}-10^{44}$\,erg/s, and emission generally between 1 and 150\,keV. As the spectral hardness tends to increase with luminosity, the faintest events are observed with narrow-field, pointed X-ray telescopes, and the more energetic events by wide-field GRB monitors. SGR flares are distributed with energy and waiting time distributions similar to earthquakes \citep{cheng1996earthquake} tying the emission trigger to the NS crust. These objects can power SGR storms where numerous flares are released in rapid succession \citep[e.g.][]{younes2020nicer}. 

The most powerful SGR flares are MGFs. The first signal seen from a magnetar was an MGF originating in the Large Magellanic Cloud in 1979 \citep[e.g.,][]{cline1981precise}, though its origin was unclear at the time. Two more have since been observed in the Milky Way. These flares have prompt spikes less than 10\,milliseconds long with luminosities in the range $\sim10^{45}-10^{48}$\,erg/s which are followed by an energetically weaker, fainter, periodic tail that decays exponentially over time. It was these signatures that led to the theoretical conception of magnetars where some of the plasma ejected in the initial energy release that powers the spike is entrained to the surface of the NS by the extreme magnetic fields with the rotation giving rise to the periodic behavior \citep{duncan1992formation,thompson1995soft}. 

SGR flares are softer than GRBs, repeat, and always have short durations. These unique identifiers led to the separation of these transient classes by the IPN, and the precise triangulation of SGRs \citep{hurley1994network,hurley1998reactivation} led to successful X-ray follow-up and proof of the existence of magnetars \citep{kouveliotou1998x,kouveliotou1998discovery}.

\subsubsection{An Origin of Short Gamma-ray Bursts}\label{sec:eMGF}
MGFs are sufficiently luminous to be detected from nearby galaxies outside the Milky Way \citep{hurley2005exceptionally,svinkin2021bright}. The short spike is visible to tens of Mpc, but the periodic tail emission falls below the sensitivity of existing instruments. The detection then looks very similar to a typical short GRB from a NS merger. The method to separate MGFs from cosmological short GRBs is association to nearby host galaxies. There are 5 extragalactic MGF candidates so far which appear as short GRBs, proving the third known progenitor of these events \citep[][Trigg et al. in prep]{burns2021identification}. 

These objects are of interest for several reasons. To start, they constitute $\sim$0.3\% of all detected GRBs (2\% of short GRBs). They are additional giant flares from magnetars that can be used to probe the science described in the following subsections. The initial prompt spike of galactic MGFs saturate all viewing GRB monitors; extragalactic events can thus be studied in more detail (though these too can saturate large, sensitive detectors). Studies of past events include identifying multiple pulses in giant flares, measurement of a $\sim$-5/3 power-law index of the intrinsic energetics distribution, and confirmation of a peak energy correlation with total energy output \citep{roberts2021rapid,burns2021identification,chand2021magnetar}, behaviors which may connect with known properties of the more common short bursts and suggest the origin as a relativistic wind. If sensitivity and time resolution are sufficient, detection of quasi-periodic oscillations (QPOs) across populations of extragalactic magnetars could inform on physics of their interiors. 

Larger samples of extragalactic MGFs with multiwavelength information of associated host galaxies (galaxy type, age, star formation rate, metallicity, potential offsets, etc), would enable deeper understanding of formation channels of magnetars, discussed further in Section\,\ref{sec:magnetar_formation}. This would result in direct comparison with similar measures on other transients such as core-collapse supernova (CCSN) explosions, superluminous supernovae, collapsars, NS mergers, and FRBs and inform on their potential relation to magnetars or magnetar formation. Finally, constraints on the lensing of MGFs (which requires broadband spectra with high temporal and spectral resolution) may constrain primordial black holes as dark matter \citep{2018JCAP...12..005K}.

\begin{figure}[htp]
\centering
\includegraphics[width=\textwidth]{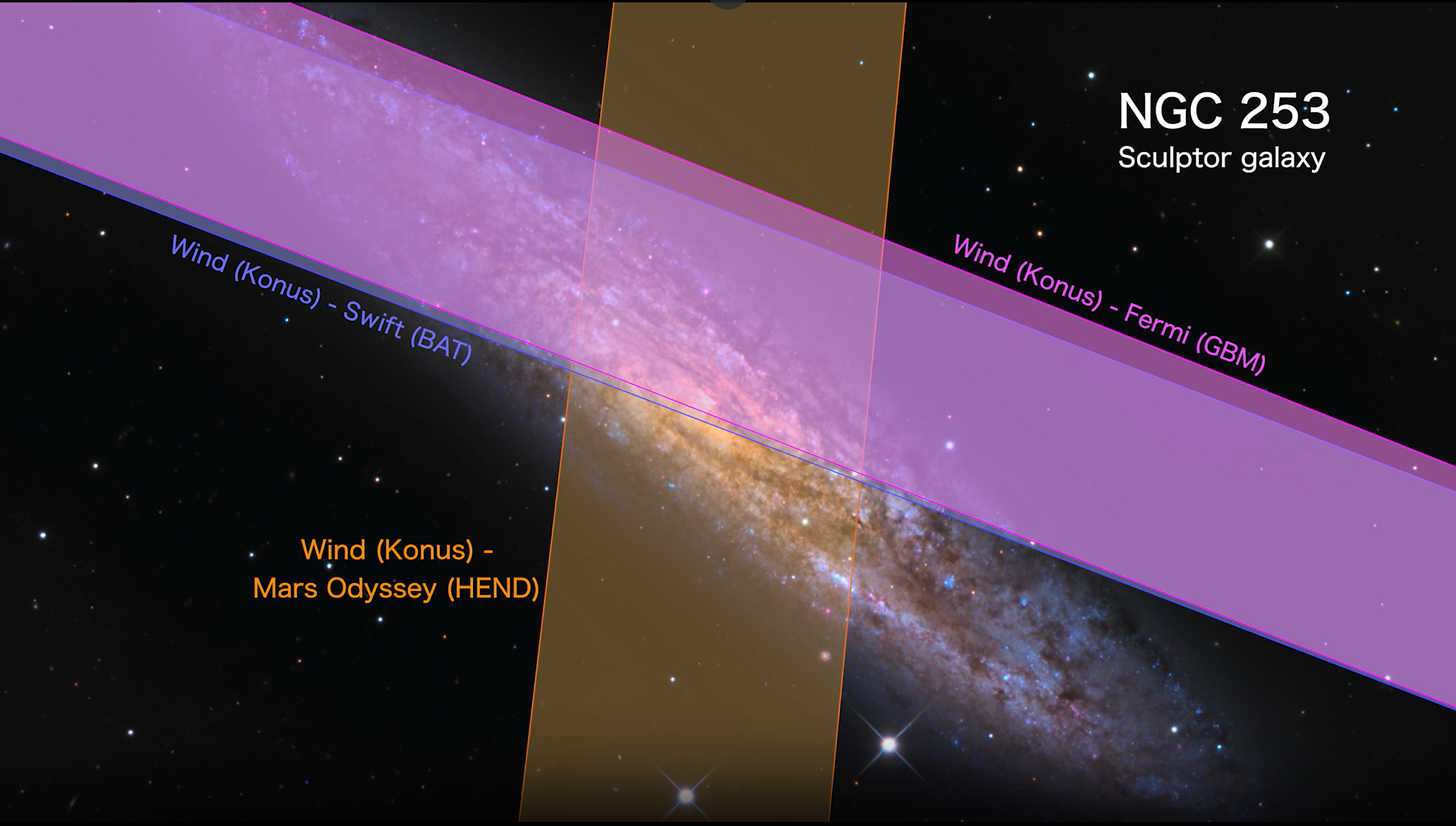}
\caption{IPN localization of GRB\,200415A to NGC\,253, aka the Sculptor Galaxy. Figure credit NASA's Goddard Space Flight Center, DSS, SDSS, and Adam Block/Mount Lemmon SkyCenter/University of Arizona utilizing data from \citet{svinkin2021bright}.}
\label{fig:200415a_localization}
\end{figure}

With reasonable assumptions \citep{burns2021identification} four extragalactic MGFs have been associated to nearby star-forming galaxies with IPN 90\% contour region sizes  between $\sim10-150$\,arcminute$^2$ (typical Swift-Burst Alert Telescope (BAT) localizations at the same confidence are $\sim$30\,arcminute$^2$). An example localizations is shown in Figure\,\ref{fig:200415a_localization}. When utilizing known properties of MGFs (sharp rise times, short pulse durations) to reduce the considered sample, it is possible to associate flares from the likeliest host galaxies even with $\sim$10\,deg$^2$ localizations (A. Trigg, in prep). As the detection horizon of these events increases, the necessary spatial precision requirement becomes stringent, with uncertainties on the few arcsecond scale necessary beyond tens of Mpc. Short GRBs with these properties could be identified as candidate MGFs with larger localization regions, allowing for observations of the putative host. The known counterpart is the decaying X-ray tail. Swift-XRT can recover a typical tail at 3.5\,Mpc if it points within $\sim$400\,s. Earlier observations or a more sensitive X-ray telescope which would relax this stringent timescale would give information on the rotation period of the star, of relevance for understanding its origin. 

We emphasize the local overdensity of star formation rate is particularly helpful for our study of extragalactic MGFs. These events are best observed by all-sky coverage over long timescales. At current sensitivities, $\sim5\times10^{-6}$\,erg\,s$^{-1}$\,cm$^{-2}$, all-sky detection rates are roughly 1 per year within 10\,Mpc though most are not separated from the cosmological short GRB sample. An order of magnitude increase in sensitivity would capture 3 events per year within the same distance and additional further events, but this requires more precise localizations given the larger number of galaxies considered. Study of these events benefit from sub-millisecond temporal resolution and energy range up to several MeV to constrain spectral curvature and thus intrinsic energetics.

\subsubsection{The Neutron Star Equation of State}\label{sec:mgf_ns_eos}
The tails of the giant flares, and perhaps also the lesser flares, exhibit QPOs, where the rotational frequency is modulated by higher frequencies. These frequencies have been attributed to oscillations of the stellar crust or interior, i.e. global NS eigenmodes \citep{watts2007neutron,huppenkothen2014quasi}. A leading model is that these disturbances are either cause or effect of quakes associated with the NS crust. These would ring up mechanical modes which are observed in the X-ray QPO data. It may also be possible to detect potential NS oscillation mode frequencies in extragalatic MGFs \citep{roberts2021rapid,castro2021very}.

Observations of these QPOs may give unique insight into the equation of state of dense matter, the crust-core interface, and the physical origin of the SGR flares \citep{watts2007neutron}; however, the current Galactic sample is too limited for reliable NS core/crust inferences because of confounding factors such as the influence of the magnetic field on the eigenmodes. Such studies would clearly benefit from much larger collections of QPO detections enabled by extragalactic studies. The natural modes of NSs run up to a few kHz. This sets a limitation on the relative timing precision necessary to perform these searches, being a factor of a few better than the inverse of these frequencies, i.e. $\sim$100\,$\mu$s resolution. These searches would also be aided by greater sensitivity than existing instruments.

\subsubsection{A Potential Source of Gravitational Waves}\label{sec:sgr_gw}
One specific class of NS natural modes, the $f$-modes, are a plausible source of GWs. Because of the lack of observational constraints, models for the GW energy emitted by such events varies widely. LIGO and Virgo have searched for GW emission associated with magnetar X-ray flares during the observing runs in which such activity was present. No GWs correlated to magnetar flares have been detected through the 3rd GW observing run \citep{O3LVKmagnetar}, known as O3. Giant flares are of particular interest as they may be energetic enough to activate NS modes \citep{duncan1998global}. The last recorded giant flare from a Galactic magnetar (SGR 1806--20) occurred in 2004 \citep{palmer2005giant}, where LIGO reported upper limits \citep{abbott2007search} but the sensitivity of the nascent GW detector network was $\sim 100$x worse than the current network. We may expect a similar $\sim 100$x gain in the IGWN over the next 20\,years as planning has begun for the third generation (3G) GW interferometers. Even before the 3G network is active, we should enter an era where GW observations of extragalactic MGFs are more sensitive than the current limits on the Galactic SGR\,1806--20 event.

SGR storms may provide yet another particularly promising target. While each burst may impart only minor energy into the modes, the integration of this effect from hundreds or thousands of bursts may enter the interesting regime. Again, the only current observations are from the previous LIGO generation \citep{abbott2009stacked}. Stacking of individual bursts during a given observing run may also prove fruitful. 

The ever-increasing sensitivity of the IGWN network, coupled with the possibility of a more energetic burst, nearby magnetar, or SGR storms, inspires optimism. Even improved upper limits can provide astrophysically interesting constraints on these mysterious and observationally elusive objects. We may perhaps expect other GW signals from magnetars. As one example, deformation of the magnetar following an MGF would produce GWs at twice the rotational frequency, though this is generally below the frequency range accessible to current ground-based interferometers. 

Complete cataloguing of the X-ray bursts will be crucial to the success of this program. This requires continuous monitoring of the high-energy sky during the observing run, and identification of SGR flares. These bursts are preferably associated to specific known magnetars which requires at least $\sim$1\,deg scale localizations. While discovery of new magnetars in crowded fields might demand arcminute-scale or better capability, this precision can be met by stacking numerous localizations of individual flares \citep[e.g.][]{hurley1998reactivation,1999ApJ...523L..37H}. These targeted GW searches would be greatly aided if QPOs could be identified in the high-energy bursts, providing specific frequencies to target. SGR flares identified and studied in real time, particularly giant flares, may allow for rapid targeted GW follow-up. If a GW signal is identified then additional follow-up resources can be marshalled for greater characterization of the magnetar behavior.

\subsubsection{The Origin and Physical Mechanism of Fast Radio Bursts}\label{sec:magnetar_frb}
FRBs are $\sim$\,ms long radio flashes whose dispersion measures place their origin at cosmological distances, with the first bursts detected decades ago \citep{1980ApJ...236L.109L,lorimer2007bright}. Yet, only in the last decade have FRBs garnered community interest as their genuine astrophysical origin is now firmly established \citep{2019A&ARv..27....4P,2019ARA&A..57..417C,2022A&ARv..30....2P,2022arXiv221203972Z}. In many ways, the quest to identify the sources of FRBs followed the same path as GRBs: searches for repetition, counterparts in other wavelengths and messengers, and identification and characterization of host galaxies. The searches for repetition have led to the phenomenological classes of non-repeating and repeating FRBs, the latter of which show similarities and are thought to be related to SGR short bursts \citep{2016ApJ...833..177S,2019ApJ...879....4W,2020ApJ...891...82W,2020A&A...635A..61O,2021ApJ...920..153W,2021ApJ...920L..23Z,2023ApJ...949L..33W,2023ApJ...947L..16L}. However, radio observations of the dispersion measure, polarization, and rotation provide additional and unique diagnostic information on the origin of these sources \citep{petroff2019fast}. 

In April 2020, the first Galactic FRB was observed, and was associated with a frequently flaring magnetar, SGR 1935+2154 \citep{bochenek2020fast,2020Natur.587...54C}. This provided new insight into FRB progenitors, in particular for repeating FRBs, and provides additional impetus to the need for excellent X-ray monitoring and localization. Of crucial importance to models is the time delay between radio and X-rays after de-dispersion (i.e. infinite frequency-equivalent arrival time of the radio). For some models, the FRB counterpart can never arrive before the magnetar short burst. For the SGR 1935+2154, the FRB arrived several seconds after its corresponding high-energy short burst \citep{2020ATel13686....1T,2020ATel13687....1Z}. However, when corrected for dispersion the FRB lagged the high-energy spikes by several milliseconds \citep{2020ApJ...898L..29M,2023arXiv230200176G}, favoring a magnetospheric origin for the FRB. Therefore accurate absolute timing precision for the X-ray flares of $<1$\,ms is an important requirement. The FRB--X-ray time delays, and potential rise time of X-rays, can also inform on how spatially co-located respective the emission regions are and constrain plasma conditions or radiative processes processes responsible for the X-ray counterpart.

A crucial question is why there is not a one-to-one correspondence with SGR X-ray short bursts and FRBs. Hundreds of X-ray bursts were observed during the recent SGR\,1935+2154 active state, but the one seen with the FRB is spectrally harder than the others \citep{ridnaia2021peculiar,2021NatAs...5..408Y}. Additionally, other bright FRBs were seen from SGR 1935+2154 without corresponding X-ray flares \citep{2020ATel14074....1G,kirsten2021detection}, though the X-ray upper limits are insufficient to be interesting. In 2022 another FRB from SGR\,1935+2154 \citep{dong2022chime} was seen in coincidence with an X-ray short burst \citep{wang2022gecam,frederiks2022konus}. Continued monitoring of the high-energy sky is necessary to determine active and quiescent times of the Galactic magnetars and to provide contemporaneous observations of FRBs. A key goal is to determine if other Galactic magnetars also produce FRB emission, or if something specific about SGRB\,1935+2154 makes it unique (e.g., the environment). Construction of complete (in time and with a well-defined, stable, and uniform flux threshold) and detailed  SGR burst catalogs with spectral properties is likely key to understanding the origin of SGR activity and FRBs. 

Localization precision is necessary to associate SGR flares to specific magnetars, though $\sim$deg scale localizations are likely sufficient for flares from magnetars known to be in active states. Improved sensitivities of the high energy monitors and lowering the low energy threshold to $\sim$1\,keV would result in more complete coverage of the flares and bursts.

We emphasize that the association of FRBs to the SGR short burst X-ray emission is the only identified counterpart for FRBs. This is despite enormous observations across the EM, GW, and neutrino spectra. This is, so far, the only method of studying FRBs with detections with multiple diagnostics.

\subsubsection{Discovery of New Magnetars}\label{sec:new_magnetar}
Continued monitoring of the high-energy sky is also necessary to identify the rest of the Galactic SGRs which is generally achieved by observing their X-ray flaring and burst activity. The first magnetars discovered were SGRs identified and localized by the IPN, while Swift now discovers them at a rate of approximately once per year. As SGR flares can be dormant for decades and faint bursts can be undetected, sensitive all-sky X-ray monitoring is key to complete this sample. 

Given the decades-long quiescence seen from many SGRs, it is critical to maintain continuous coverage of the high-energy sky to complete the Galactic census of magnetars. This matters for understanding stellar formation and evolution. As one example, the current constraint that $>10$\% of NS produced in CCSN are magnetars is hard to explain against other observations, e.g. the rates of superluminous supernovae. It may also be possible to identify SGR flares from previously unidentified magnetars (and also past flares from known magnetars) in historic data, if a dedicated search were performed. Lastly, identification of extragalactic MGFs allows the study of magnetars in other galaxies which enables studies on their origins and formation channels.

\subsubsection{Magnetar Formation Channels}\label{sec:magnetar_formation}
Observations generally support that magnetars are produced from a subset of CCSN, as these cataclysmic events produce the majority of NSs. However, several formation channels exist beyond the base CCSN model, including superluminous supernovae \citep{nicholl2017magnetar}, low-mass mergers \citep{price2006producing,levan2006short}, and some white dwarfs which undergo accretion-induced collapse \citep{dessart2007magnetically}. Resolving this question can uncover the fundamental question of how the most powerful magnetic fields are created, probing whether they arise from the compression of the progenitor material or are the result of emergence in dynamos of the proto-NS. It is also unclear what role, if any, stellar binarity plays in CCSN formation of magnetars, as no magnetars are known to exist in binaries \citep[with a possible exception,][]{2023Natur.614...45R}. However, magnetars may have been born and kicked from their originating binary system as most massive stars that form NS are formed in binaries and triples and interact before the supernova occurs \citep{sana2012binary}. 

Specific outcomes of these studies can provide insight into the origin of superluminous supernovae and GRBs, even if magnetars are simply a rare byproduct of basic CCSN. Modeling the broad behavior of known Galactic magnetars (next section) favors the fraction of CCSN that produce magnetars to be $\gtrsim$10\% \citep{beniamini2019formation} which is supported by the high intrinsic rate of MGFs \citep{burns2021identification}. If proven, then CCSN modelers must consider extreme magnetic fields, and formation of NSs over black holes, in a non-negligible fraction of these events. Studies of MGFs thus offer a method of studying populations of relatively young NSs beyond our galaxy.

We can determine the relative importance of magnetar formation channels through several means. Greater characterization of the Galactic magnetar population and new magnetars as they are identified is critical \citep{beniamini2019formation}. Identifying more MGFs and further constraining their rate to a high value would require the formation channels to also have high rates, likely limiting the options to common CCSN or accretion-induced collapse of white dwarfs. Completion of this picture will involve association of magnetars to specific hosts. In the case of Galactic sources, this means classifying the origin of the remnant they are associated to, either to CCSNe or Type Ia SNe, which may be done with COSI through the study of nuclear lines from their supernova remnants. For extragalactic MGFs or repeating FRBs (presuming the magnetar origin) the host galaxy and location within the host galaxy is key, again repeating the approaches of the GRB field. Events from star-forming regions favor massive star origins while association to older regions or large offsets from host galaxies favor the merger origin.  

Accomplishing this science thus requires advancements across several aspects of the field. The requirements specific to the GRB monitors are those listed in other sections here, i.e., the identification of new magnetars, identification of more SGR flares and MGFs, providing those locations to the community, and determining the origin of FRBs.

\subsubsection{Magnetar Properties}\label{sec:magentar_properties}
To understand the magnetar formation channels, it is important to understand the behavior of magnetars themselves. A more complete, well-characterized sample will permit additional insight into the intrinsic physics of magnetars, such as understanding the lifetime and activity rate of magnetars and connections to the physical mechanism of their magnetic field evolution and decay \citep{beniamini2019formation}. These are basic questions on the most powerful magnets. 

One additional question is whether each magnetar can produce at most one MGF. It is possible that a single giant flare is triggered by a nuclear or magnetohydrodynamic phase transition that prevents another from occurring from the same magnetar \citep{mallick2014phase}. This involves the behavior of ultra-dense matter with extreme magnetic fields, particularly stability conditions associated with the cores of NSs. A direct answer to this question would be identifying two giant flares from the same source with low false alarm probability. This requires monitoring the three magnetars in the Milky Way and Large Magellanic Cloud which have undergone MGFs since observations began in the 1970s, requiring continued all-sky coverage. An alternative approach is precise localizations of extragalactic MGFs from nearby star-forming galaxies, which may have more active magnetars ($\sim1'$ at 3.5\,Mpc corresponds to a physical scale of 1\,kpc). A future IPN could achieve $1''$ precision, capable of monitoring magnetars in many local galaxies for repetition, and could provide a decisive answer. Otherwise, an indirect statistical method, requiring improved identification of extragalactic MGFs, can be used to infer a giant flare rate that would only be possible with repetition \citep{burns2021identification}.

\subsubsection{Physics of Magnetar Bursts}\label{sec:magnetar_burst_physics}
The exact mechanism or trigger that produces magnetar bursts (in both SGRs and AXPs) remains unknown, although we know it ought to involve the crust of the NS. Diagnostics including time offset from FRB emission, QPOs, GWs, rotational phase dependence of bursts, and broadband time-resolved spectroscopic information can all provide useful information to answer this question. X-ray studies with GRB instruments have shown most magnetar bursts can generally be described by two blackbodies of similar luminosities \citep{2008ApJ...685.1114I,2012ApJ...749..122V,2014ApJ...785...52Y}, one of low area and high temperature ($kT \sim 10-20$ keV and the other high area and lower temperature ($kT \sim 3-5$ keV). This is consistent with a confined Comptonized fireball trapped within a flux tube of the magnetar magnetosphere. Above $\sim 50$ keV, single photon splitting (a third-order, untested, but standard quantumelectrodynamic process, $\gamma \rightarrow \gamma\gamma$) is expected to operate and be relevant for magnetar bursts by imprinting spectral and polarimetric modifications onto some bursts, depending on the viewing geometry \citep{2019MNRAS.486.3327H}. Long term monitoring of SGRs and AXPs is insightful into understanding what separates this class, especially with the rare SGR-like flares seen from some AXPs \citep{savchenko2010sgr}. Continued all-sky monitoring to determine the waiting time distributions and any identifiable behavior on the active versus quiescent period durations would also be constructive.

We additionally emphasize that polarimetry of bright SGR flares, or the stacking of many bursts, may probe the physical emission region, the geometry of the magnetic field, and the extrinsic properties of inclination with respect to magnetic and rotation axes. Similar inferences can be made in the polarization observations of MGF tails \citep{yang2015polarization}. Success here from high-energy monitors requires polarization sensitivity down to tens of keV. 

New observations of Galactic MGF tails with more sensitive instruments which have higher temporal resolution could reveal the presence of a deviation from a single peak phase-gram, giving information on the geometry of the emitting region on the surface, which may not be compatible with a simple fireball model.

Magnetar burst spectra have been found to have a predominately Comptonized or thermal spectrum. Recent observations of unique events from the outburst of prolific magnetar SGR 1935+2154 strongly suggest that QPOs sometimes occur over the peak energy of $\nu F \nu$ spectra \citep{Robert2023}. These quasi-periodic spectral oscillations, can be explained by magnetospheric density and pressure perturbations where burst emitting plasma consisting of purely e$^{+}$e$^{-}$ pairs propagate along a highly magnetized flux tube of several tens to a hundred NS radii. Some of these events occur within days of each other which provides an interesting constraint on the activation timescales over which the environment of the magnetosphere of SGR J1935+2154 (and perhaps, magnetars in general), change. Therefore, detailed time-resolved analyses of other magnetar bursts are encouraged to evaluate the rarity of these events.

\subsubsection{Summary of Capability Requirements}\label{sec:req_magnetar}
The most stringent localization requirement for high-energy monitors observing magnetar flares is driven by extragalactic MGFs. For particularly nearby events ($\lesssim5$\,Mpc) from bright, star forming galaxies, robust association of individual events to 3$\sigma$ requires localizations at 90\% confidence of order $10$\,deg$^2$. For events to $\sim$20\,Mpc, or those from less likely host galaxies, this requires uncertainty at the level of a few arcminutes. At even greater distances arcsecond scale localizations are required for robust association. For science from the high-energy observations alone there is no alert latency requirement. 

To recover the known fading tail counterparts with Swift-XRT requires observations beginning in minutes. Rapid and precise localizations could enable this. For rapid localizations up to $\sim$100\,deg$^2$, MGF candidates and likely hosts could be identified with fiducial cuts on the lightcurve, allowing for galaxy-targeted follow-up. Higher latency data could confirm or reject classification.

For the less energetic short bursts localizations of $\sim1$\,deg$^2$ are sufficient to associate bursts to catalogued magnetars. As flares often repeat in active periods an alert latency on the order of a day is sufficient to flag that magnetar as active for X-ray, radio, or other follow-up efforts. The discovery of Galactic magnetars requires localizations (either individual or stacked) to $\sim0.3$\,deg, i.e. the typical field of view of X-ray telescopes. 

Monitoring the Galaxy for potential SGR-FRB joint detections, or sensitive X-ray upper limits of Galactic FRBs, requires continuous all-sky monitoring as the timescales are so short the only method of counterpart detection is to have made contemporaneous observations. Discovery of new magnetars or extragalactic MGFs can be made with smaller fields of view but benefit from 100\% sky coverage. 

A substantial increase (factor of several) in sensitivity over the historic average of the IPN would increase the rate of extragalactic MGF identification from every several\,years to $\sim$1 per year, sufficient for major advances in understanding. MGFs drive the high-energy threshold requirement to $\sim$5\,MeV, necessary to constrain spectral curvature. The detection of many normal SGR short bursts requires sensitivity down to $\sim$5\,keV and a high-energy threshold of $\sim$150\,keV. Lowering this threshold further would recover more, softer events allowing for a more complete sample. Pointed X-ray telescopes can detect fainter events than high-energy monitors, but they may not be observing contemporaneously with interesting events such as Galactic FRBs. Substantial polarization sensitivity down to $\sim$tens of keV increases is required for polarimetric study of these events given their very short durations.

Magnetar flares drive timing requirements for high energy monitors. Absolute timing of millisecond precision or better allows for physical insight into the SGR and FRB emitting regions and processes. Relative timing precision and temporal resolution to 50\,$\mu$s allows for study of the shortest features in giant flares.

Galactic MGFs provide the most stringent requirement for maximum photon rates. The SGR\,1806--20 giant flare was the most energetic with $\sim1.5\times10^7$ photons/s/cm$^2$ \citep{frederiks2007giant}. A closer or more energetic event could hit $10^8$ photons/s/cm$^2$, though the other two Galactic events were a factor of 10 less energetic. Extragalactic MGFs can exceed 1000\,photons/s/cm$^2$ \citep{svinkin2021bright,roberts2021rapid}, suggesting an instrumental capability threshold of a few thousand ph/s/cm$^2$.

There are only weak requirements on contiguous viewing with individual instruments and on background stability given the short emission intervals. These objects are not known to have sharp spectral features so they do not drive energy resolution requirements.

Complete SGR studies require extremely long baselines to map the long end of their waiting time distribution, to study the durations of active and quiescent periods, to search for SGR flares from AXPs, and to discover new magnetars.

\subsubsection{Requested Deliverables from GRB Monitors}
The science possible from high-energy monitor observations of magnetars would benefit from several new or improved community products. A key product is an SGR flare naming convention to be adopted by the GRB monitor teams. This would be important to implement for a high-latency SGR flare catalog. This catalog collates information from multiple legacy and current missions and includes entries on each signal, its arrival time, and any duration and spectral information. Given the particularly short durations and finite light travel time, association of flares observed by distant instruments is non-trivial, especially when the specific magnetar origin is unknown.

A real-time collated Galactic SGR flare alert stream would also be beneficial. This would provide a central access point for multiwavelength and multimessenger services, allowing for identification of joint detections fast enough for prioritized follow-up that would help characterize the event and its effect on the source magnetar. Such an alert stream would also allow automatic flagging of active and quiescent state informative for pointed observations at other wavelengths, e.g. radio. Of particular interest would be an automated flagging of Galactic giant flares for incredibly bright bursts in this stream, which would enable automated repointing of various telescopes like Swift-XRT. 

Lastly, automated searches to promptly flag short GRBs as MGF candidates may be key to recovering X-ray tails from extragalactic MGFs. This requires collation and calculation of localization information as rapidly as possible, within minutes, and convolution with local galaxy catalogs.

\subsection{Compact Mergers}\label{sec:mergers}

Binary neutron star (BNS) mergers and neutron star-black hole (NSBH) mergers, both cases herein defined generally as NS mergers, are canonical multimessenger sources, producing high frequency GWs (up to a few kHz), light across the EM spectrum, and must also produce neutrinos. These messengers arise from a variety of emitting regions and processes. Successful understanding of the extreme physics occurring in these events and the corresponding signals is difficult, but is critical to achieve: the detection and characterization of BNS mergers yield a cornucopia of science cases across astrophysics, including (i) the nucleosynthesis of heavy elements, i.e. those heavier than iron \citep{JuBa2015,WuFe2016,RoLi2017,AbEA2017f}, (ii) the nature of matter in the densest environments known \citep{BaBa2013,AbEA2017b,RaPe2018,BaJu2017}, (iii) what drives the expansion rate of the Universe and other cosmological parameters \citep{2017Natur.551...85A,HoNa2018,CoDi2019c,du2019future}, (iv) studies on ultrarelativistic particle acceleration \citep{abbott2017gravitational,burns2020neutron,salafia2021accretion}, and (v) unique tests of fundamental physics \citep{abbott2017gravitational,creminelli2017dark,burns2020neutron}.

The NS merging event is preceded by a long, slow inspiral caused by loss of energy to GWs. The inspiral accelerates and emission strengthens until the objects merge producing a crescendo of GW emission at $\sim$kHz frequencies that exceeds the EM luminosity of GRBs. At least some NS mergers produce bipolar ultrarelativistic jets which release the prompt GRB signal. As these jets propagate outwards they build up an external shock with the circumburst material, releasing synchrotron radiation across the EM spectrum. 

EM-bright mergers shed material quasi-isotropically through various processes. This expelled material has high density and is neutron rich, and thus combines into heavy, radioactive elements. As the initial cloud is opaque, this creates a quasi-thermal signature referred to as a kilonova \citep{li1998transient,metzger2010electromagnetic,metzger2020kilonovae}. These events emit predominantly in the ultraviolet (UV) to infrared (IR) wavelengths on timescales from hours to months.

\begin{figure*}[htp]
	\centering
	\includegraphics[width=\textwidth]{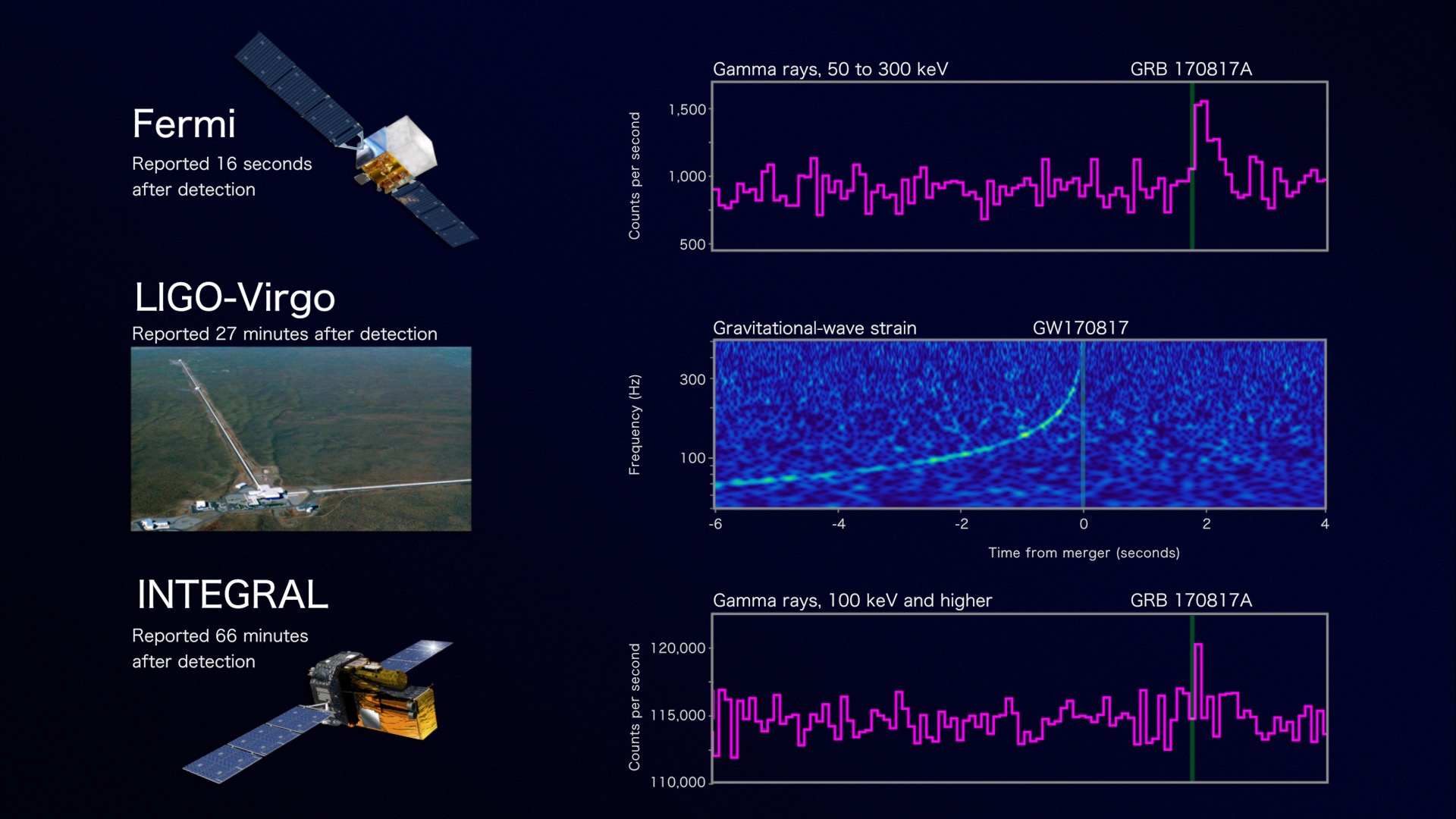}
    \caption{The first multimessenger detection involving a GW source. LIGO and Virgo observed GW170817 and Fermi-GBM and INTEGRAL SPI-ACS observed GRB\,170817A only $\sim$2\,s later. Image credit NASA's Goddard Space Flight Center, Caltech/MIT/LIGO Lab and ESA. }
    \label{fig:gw-grb}
\end{figure*}

More than 1000 short GRBs have been observed in the prompt phase, and $\sim$10\% of them have detections from follow-up observations at other wavelengths. Follow-up detections lead to arcsecond-scale localizations which help determine the host galaxy and the source offset from the host galaxy. Distances are generally determined from a redshift measurement of the associated galaxy. Broadband observations of the afterglow over multiple epochs gives insight into the jet structure, the circumburst density, and total kinetic energies. $\sim$\,10 short GRBs have evidence of an associated kilonova, though few are well characterized. There are still only a few GW detections of BNS and NSBH mergers. The first GW detection of one of these events was the BNS merger GW170817, associated to GRB\,170817A. These multimessenger joint detection is shown in Figure\,\ref{fig:gw-grb}.

There are a number of additional signatures known from observations or predicted from theory, as discussed and referenced in the following subsubsections. Observations show some short GRBs are followed by plateau emission with durations $\sim10-10^5$\,s that can be seen in the GRB monitors or in rapid follow-up observations in X--rays. There are also claims of non-thermal precursor activity to the high--energy emission. The ultraviolet (UV) emission of AT2017gfo, the name of the optical counterpart to GW170817, was unexpectedly at its brightest when first observed at 12\,hours. The origin of this signal remains debated with options including cocoon / shock-breakout of the jet through the previously ejected kilonova material, the decay of free neutrons, energy ejection from a magnetar central engine, or base kilonova emission. There are several theoretical predictions for which observations remain ambiguous. Examples include the kilonova afterglow, radio precursors, X-rays from late-time fallback, and late-time radio excess from a magnetar central engine. Those which require early observations are described below, while the others can likely be recovered in follow-up of GW-detected NS mergers that do not have prompt GRB counterparts. The numerous counterparts cause confusion in interpreting observations, but they provide an opportunity to more deeply understand the physics occurring in these events.

Specific to the study of GW-detected NS mergers, multimessenger studies are enabled with identification and association of EM counterparts. As the GW interferometers become more sensitive the detection distances grow linearly and thus the rates as a cubic factor (until source rates evolution becomes important, first increasing then decreasing). At any projected era the total detected sample of GW localizations will remain poor for the majority of events \citep{PeSi2021}. Even major follow-up facilities like the Vera Rubin Observatory will begin struggle to identify, isolate, and associate kilonovae at these distances. Thus, beginning perhaps as quickly as O5, GRB monitors will be the dominant discovery mechanism for EM counterparts to GWs, especially at greater distances. Ideally, they provide localizations sufficient for any telescope to utilize for full characterization across the EM spectrum.

\subsubsection{Plateau Emission}\label{sec:plateau}
BATSE discovered the existence of a class of short GRBs which have extended emission visible in the prompt detection \citep{connaughton2002batse,norris2006short}. This signature has subsequently been observed in bursts detected by Fermi-GBM and Swift-BAT \citep{kaneko2015short}. Follow-up of Swift bursts established that short GRBs with and without extended emission have similar offsets from their host galaxies, pointing to a merger origin in both cases \citep{fong2009hubble}. Utilizing the autonomous repointing of Swift's narrow-field telescopes to BAT-triggered events, a large sample of early afterglow observations from X-ray through optical has been constructed. A common signature is an X-ray plateau at early times that sometimes transitions to a steep decay most often between $10^2$ and $10^5$\,s \citep{evans2009methods}. Different arguments have been invoked to explain these plateaus, including magnetar central engines \citep{gompertz2014magnetar}, fallback onto a black hole central engine \citep{gompertz2020search}, as well as high-latitude emission from structured jets observed on-axis \citep{oganesyan2020structured} or also off-axis \citep{ascenzi2020high}. These options will be delineated with GW measures of masses and spins, and perhaps additional GW observables \citep[e.g.][]{corsi2009gamma}. The answer to the physical origin of these plateaus is far reaching, as highlighted below.

Observations of extended emission drive requirements for background stability on the order of a few hundred seconds and sufficient sensitivity  to recover the signal. 
Extended emission is often, but not always, softer than the typical prompt spike, favoring a lower energy threshold of 10\,keV, with increasing benefit to $\sim$1\,keV. Capturing the signal of X-ray plateaus requires observations beginning within 100\,s after the start of the prompt GRB. The end of the plateau phase occurs between $10^2$\, and $10^5$\,s after the GRB, giving a decreasing plateau identification probability as response time increases. With current X-ray facilities, this requires localization accuracy on the order of $\sim20'$.

\subsubsection{Prompt Radio Signatures}\label{sec:prompt_radio}
Of all the follow-up signatures which may drive temporal reporting latency requirements, radio observations may be the most stringent. This is because the merging systems may produce signals prior to full merger \citep{1996A&A...312..937L,2001MNRAS.322..695H,2012ApJ...757L...3L,2012ApJ...755...80P,2019MNRAS.483.2766L,2020ApJ...890L..24Z,2022arXiv220808808P,2022MNRAS.515.2710M,2023MNRAS.519.3923C,2023PhRvE.107b5205L}, which can be detected in the radio via follow-up observations due to dispersive delays by free electrons in the interstellar and intergalactic media. That is, the precursor signal arrives in the solar system {\it{after}} the GW or conventional higher-frequency EM counterpart.
Looking ahead to the near future, accurate and prompt GRB localizations will enable detections or limits on pre-merger pulsed FRB-like coherent radio emission from NS mergers \citep{1996A&A...312..937L,2019MNRAS.483.2766L,2023MNRAS.519.3923C} by rapidly improving radio facilities (whose capabilities such as FOV increase directly with improving computational resources). This radio emission would be an independent test of the chirping orbital dynamics prior to merger. The expected detection horizon to such radio emission for advanced facilities such as DSA-2000 and SKA-mid is generally larger than that of the current IGWN to BNS mergers, out to several Gpc. Free electrons along the line of sight lead to dispersive delays of radio emission of order 
\begin{equation}
t_{\rm delay}^{\rm radio} \approx 40 \,\, \rm \left(\frac{DM}{1000 \rm \, pc \, cm^{-3}} \right) \left(\frac{\nu}{\rm 300 \, MHz}\right)^{-2} \quad \rm seconds
\end{equation}
where DM is the radio dispersion measure that can vary based on sky location and host galaxy properties, and $\nu$ is the characteristic observing frequency. This delay is manifestly several tens of seconds to minutes at CHIME, LOFAR, CHORD, SKA-mid, MWA and SKA1-Low frequencies. Facilities such as LOFAR can form many independent software beams on the sky to tile localizations and wait for the signal after the GRB alert. Beamformed LOFAR, and SKA1-Low attain ${\cal O}(10^\circ)$ fields of view \citep[see Table\,2 and refs in][]{2023MNRAS.519.3923C}, yet their horizon for detection (and thus expected event rate) is somewhat limited. On the other hand, more sensitive and higher frequency facilities such as CHORD, DSA-2000 and SKA-AAmid whose horizon extends to several Gpc would require sub-minute timescale alerts to repoint.

\subsubsection{Early Ultraviolet Emission}\label{sec:early_uv}
Another driver for latency requirements is understanding the early ultraviolet emission seen in AT2017gfo \citep{arcavi2018first,metzger2020kilonovae}. There are multiple possible options and much additional science relies upon understanding which causes of early UV emission are important in what events. If the ejecta achieves sufficient velocity, it may be possible to see an early UV signal powered by the heating from the decay of free neutrons, which have a half life of $\sim$10\,minutes elongated by the relativistic Doppler factor. If this occurs, it may give the most direct insight into the velocity gradient / distribution of kilonova ejecta, necessary to properly utilize observations to understand the underlying event (e.g., the nuclear yield). Alternatively, if the jet is launched after the initial ejection of material into the polar regions, it may be possible that an early UV signal from shock break-out / cocoon heating is produced. This would have monotonic decay and cooling over time; distinguishing between these two options requires observations beginning no later than $\sim$30\,minutes, with 10\,minutes preferable. 

For the cocoon and the free neutron cases, the prompt monitors play a role both in localizing the event for follow-up and in providing information on inclination, as these may be strongest for face-on events. Thus, full understanding requires early UV observations of a population of events discovered both through GW-only, GRB, and GW-GRB detections of NS mergers. UV-identified events will generally lack inclination information; optical and IR identified events will likely lack early UV observations. 

Bright and early UV emission may also occur from a magnetar remnant powered by the neutrino-heated, magnetically accelerated wind \citep{metzger2018magnetar}. This could be disentangled with UV observations in the first few hours. Lastly, advances in modeling since 2017 have shown that highly ionized ejecta can limit opacity issues at early times, perhaps powering early UV emission from base radioactive contributions \citep{banerjee2020simulations}. 

The key requirements for GRB monitors are localization precision and alert latency. There are a limited number of UV telescopes, and currently only one that can repoint within 30\,minutes of an alert. Swift-UVOT has a field of view of $17'$x$17'$ requiring strict alert latency and precise localizations. ULTRASAT will have a $\sim$200\,deg$^2$ field of view in a geostationary orbit, permitting the ability to immediately repoint. The GRB monitor requirements are thus an alert within $<10$\,minutes and a localization precision of $\sim$200\,deg$^2$. However, we note the extended annuli that may come from rapid IPN localizations may require tiling. Given the limited depth to which ULTRASAT can recover these events ($\sim200$\,Mpc), an all-sky field of view would be most beneficial. Future facilities will likely have fields of view between UVOT and ULTRASAT, but may be significantly more sensitive. For the cocoon case, the UV observations can be matched to the immediate prompt GRB signal at earlier times, if the observing monitor achieves a sufficiently low energy threshold of 10\,keV as a requirement with additional gains to lower values.

\subsubsection{Binary Neutron Star Merger Classes}\label{sec:bns_classes}
BNS mergers can form four different objects after merger, largely determined by the masses of the progenitor objects. If the merging NSs are particularly light, they may be able to form an indefinitely stable NS, supported against gravitational collapse by degeneracy pressure and nuclear forces. The maximum mass of a non-rotating NS is referred to as M$_{TOV}$ \citep{tolman1939static,oppenheimer1939massive}, beyond which a black hole is formed. However, metastable NSs can exist past this threshold if supported against collapse by adding rotation to aid the balance against gravity. If uniform rotation is sufficient, then the NS is referred to as ``supramassive'' and if internal differential rotation is also required then it is ``hypermassive.'' Heavier BNS mergers can, in principle, form these objects. The heaviest BNS mergers will promptly collapse to a black hole. 

Determining when these different cases occur as a function of the masses of the merging NSs must be a key goal in astronomy this decade, given the broad scientific results that will follow. Direct determination can be made with GW interferometers with sufficient sensitivity at a few kHz for particularly nearby events, or perhaps out to a few hundred Mpc when 3G interferometers exist \citep[likely 2035+][]{radice2017probing}. The EM signatures provide a viable path to these results \citep{margalit2019multi,burns2020neutron}. 

A key question to be answered is whether magnetar central engines can power GRBs. Either we observe GRBs from low-mass mergers, possibly with a significant delay due to an engine cleaning time \citep{zhang2019delay}, or we observe GRBs from short-lived hypermassive and longer-lived remnant cases. The non-thermal plateau emission signatures have been modeled as originating from magnetar-driven winds. Longer-lived remnants are thought to create brighter and bluer kilonova, and may be the case capable of releasing free neutrons. 

Mapping out the various remnant cases requires calibration on a few particularly well-studied events with information from across the EM spectrum, with follow-up observations needing to begin within $\sim10-30$\,minutes. The key wavelengths for rapid follow-up are X-ray and UV, where the smallest field of view drives the localization precision required, on the order of $\sim$0.3\,deg. This work also benefits from a sample of NS mergers with total, sensitive prompt GRB coverage, driving both sensitivity and an all-sky field of view. The prompt GRB monitors need absolute timing precision of $\sim$0.1\,s to properly determine the GW-GRB timing offset. 

\subsubsection{Neutron Star-Black Hole Classes}\label{sec:nsbh_classes}
We can similarly separate NSBH mergers into those that are EM-bright and those that are EM-dark. Larger black holes will swallow a merging NS whole, while small ones will disrupt the NS outside the innermost stable circular orbit \citep{foucart2014neutron}, releasing nuclear material which will power a kilonova and perhaps GRB. Obtaining early panchromatic observations of EM-bright events is necessary to understand the origin of the various observed signatures and provide a complete picture of NSBH and BNS mergers. Total, sensitive GRB coverage of NSBH mergers is necessary to fully exclude bright prompt emission from these sources. The requirements are similar to those necessary to understand BNS merger cases.

Establishing the existence and rate of EM-bright NSBH mergers, and characterization of their multimessenger emission is required for much of the broader compact merger science. These include unique and precise tests on the NS equation of state, insight into the spins of BHs and BH fallback, the formation of ultrarelativistic jets, an origin of short GRBs, and may contribute a sizable fraction of actinides given the potential for substantially larger elemental yields with lower electron fractions \citep[e.g.][]{metzger2020kilonovae}. EM observations with GW detections will help determine the maximum mass of NSs, the minimum mass of BHs, and probe whether the these two values are identical or whether some overlap or a mass gap exists. These are key questions to understand dense matter, black holes, supernova explosions, and stellar evolution.

\subsubsection{Progenitors of Short Gamma-Ray Bursts}\label{sec:sgrb_progenitors}
NS mergers have long been discussed as progenitors of short GRBs \citep{eichler1989nucleosynthesis} and have been the only convincing candidate for the dominant progenitor for nearly a decade \citep{fong2015decade}. The association of GW170817 to GRB\,170817A directly proved BNS mergers as an origin of short GRBs \citep{abbott2017gravitational}. It is thought that some NSBH mergers produce short GRBs, but direct proof remains elusive. With IGWN now detecting compact binary coalescences with masses in the ranges expected for NSBH mergers, we should either observe joint GW-GRB detections of these events or never observe a GRB associated with these mergers. Given that the elongated morphology of GW localizations are common, and are likely to continue to common in the future, sensitive all-sky prompt GRB coverage is necessary.

\subsubsection{Progenitors of Long Gamma-Ray Bursts}\label{sec:long_mergers}
Shortly after the launch of Swift, a few long GRBs were identified that had no associated supernova to constraining intrinsic limits \citep[e.g.][]{gehrels2006new,DellaValle2006,Gal-Yam2006}. One explanation for this is direct core collapse preventing a strong supernovae but still allowing for accretion and successful jet launch. However, as the theory and observations of kilonovae has matured, there are now several long GRBs with apparent observed kilonova signatures \citep[e.g.][]{yang2015possible}. Of particular note are GRB\,211211A \citep{rastinejad2022kilonova} and GRB\,230307A. GRB\,211211A also showed a late-time GeV signature \citep{mei2022gigaelectronvolt,zhang2022fermi}. A possible explanation is external inverse Compton in the jet with seed photons originating from a kilonova. If correct, it provides another diagnostic of both the kilonova and the jet and only deepens the debate on GRB classification. GRB\,230307A localized by the IPN and resulting in the first late-time NIR spectrum of a kilonova, taken by JWST, being the most direct observation of r-process nucleosynthesis \citep{levan2023grb}. Both of these events show evidence for the prompt signal arising from fast-cooling synchrotron, providing clear insight into the prompt emission mechanism \citep{gompertz2023case,levan2023grb}.

GRBs 211211A and 230307A are unusually nearby with inferred distances based on the redshift of their putative host galaxies of 350\,Mpc (z $\sim$ 0.08) and 300\,Mpc (z $\sim$ 0.065), respectively. Such events are likely detectable through joint GW-GRB searches if observed by IGWN if they have a BNS origin, and are almost certainly detectable if they have an NSBH origin. 

The identification of a kilonova signature, without an accompanying supernova, points to a merger origin for some long GRBs. This motivates continued broadband follow-up observations of prompt GRB detections, particularly in concert with GW observations for incontrovertible proof. The unusual brightness of GRB\,230307A led to rapid dissemination of IPN localizations and tiling of the region, demonstrating that the community can successfully follow-up events localized in this manner. However, the lack of early broadband coverage is affecting on-going analysis. Beyond the localization and alert latency requirements the current understanding of the prompt emission of the events favors broad spectral coverage to probe the apparent synchrotron origin and $\sim$ms scale relative timing precision as a potential method to identify these events in real-time \citep{veres2023extreme}.

\subsubsection{Origin and Classification of Gravitational Waves}\label{sec:gw_classification_mergers}
Multimessenger observations are also key to identifying and classifying sources of GWs. When IGWN detects a compact binary coalescence, they provide estimates on whether the event is from a BNS, NSBH, or binary black hole merger based on predetermined mass thresholds. The detection of an associated prompt GRB signal immediately confirms the event as EM-bright, and thus either a BNS or NSBH merger. This may be of particular interest for higher mass compact binary coalescences where an NSBH or binary black hole merger origin is unambiguous, giving guidance on when to execute expensive follow-up observations and which telescopes to use. In the future, it may be possible to identify the presence of at least one NS from tidal effects in the high frequency range of the waveform, but this is well beyond the capability of current instruments.

Again, the elongated morphology of many of the GW localizations drives the need for sensitive all-sky coverage. Determining the sub-classes of BNS or NSBH mergers, discussed above, is enabled with rapid follow-up observations, driving the need for early and accurate localizations, aided by combining GW and GRB localizations. 

\subsubsection{White Dwarf-Neutron Star and White Dwarf-Black Hole Mergers}\label{sec:wd_mergers}
BNS and NSBH mergers have been the dominant merger scenarios considered for GRBs for several years, but alternatives have existed for decades. White Dwarf-Black Hole (WDBH) mergers were invoked as possible sources of long GRBs \citep{fryer1999merging}. The White Dwarf-Neutron Star (WDNS) scenario was considered for the long GRBs from mergers, i.e. GRB\,211211A and GRB\,230307A \citep{zhong2023grb,levan2023grb}. There is on-going advances in theory and simulation in this area \citep[e.g.][]{kaltenborn2022abundances}.

It is difficult to probe such an origin with existing or funded GW detectors because the merger frequency, driven by the size of the largest object (white dwarf), occurs at $\sim$Hz frequencies. Thus, it is the responsibility of prompt discovery and localization of GRBs with appropriate follow-up to determine if these are a viable progenitor of these events. Key information for such a determination includes the host galaxy type, offset from the host galaxy, and broadband characterization to look for any radioactively-powered signatures. The need to acquire this information drives requirements on precise localizations fast enough after merger to reliably model the afterglow to search for later time signals on top of it. The requirement is typically within an hour, but faster follow-up will only increase the probability of a successful detection. We note the identification of a merger GRB through EM follow-up observations and a local redshift that could also be excluded as a BNS or NSBH merger from GWs would strongly point to a WDNS or WDBH merger origin.

\subsubsection{Stellar Formation and Evolution}\label{sec:stellar_formation}
Compact object mergers are a rare outcome of stellar systems. These events track the star formation rate with the inspiral time distribution adding a delay. Most are expected to arise from field binary evolution, with a subset arising from dynamical capture. One of the latter may have been recently identified \citep{levan2023long}. Field binaries require an unusual sequence of evolution to create the compact objects with initial separations small enough to merger in a Hubble time (i.e., to be observed by us). Thus, an understanding of these systems provides insight into the formation and evolution of stellar systems and their binary interactions such as the stability and physics of mass transfer and common-envelope phases, angular momentum transfer, stellar winds, and rotation of massive stars -- phenomena that are challenging to observe or study in large populations due to the scarcity of massive stars. GRBs provide a unique opportunity to study the lives of massive stars, particularly in environments beyond our own Milky Way. 

The requisite observations include the local rates of the different merger classes, their rate evolution over time, host galaxy types, and offset distributions of the mergers from the host galaxy, as well as the mass, spin, and mass ratio distributions of the merging population \citep{broekgaarden2021impact,mandel2022rates}. The mass and spin parameters are only measured from GW observations. The local rates can be directly measured by GW detectors, but measurement of rate evolution of BNS mergers is not possible via GWs until the 3G era. Until then, an understanding of the origin of GRBs over cosmic history is key to understanding their source evolution.  This requires deep sensitivity over a wide enough field of view for a high rate of events, with rapid and precise localizations of sufficient accuracy to associate the sources to their host galaxy and measure their respective redshifts. 

\subsubsection{Origin of the Elements}\label{sec:origin_elements}
A fundamental research area of astrophysics is mapping the origin of the elements. A key remaining goal in this area is determining the production sites of the heaviest elements, i.e. the sites of r-process nucleosynthesis \citep{burbidge1957synthesis,cameron1957nuclear}. The ingredients are neutron-rich material with particularly high densities, pointing to the moments of creation or destruction of NSs. The constraints include the current amount of heavy elements that exist, which corresponds to the time-integrated r-process production, and the abundance pattern of the elements corresponding to the averaged relative yields. As a key example of multimessenger astronomy, observations of the rate of dust deposition of $^{244}$Pu on Earth \citep{wallner2015abundance} and the identification of r-process element in only one of many dwarf galaxies \citep{ji2016r} point to the dominant site of r-process elements as rare, high-yield events. The viable options are then NS mergers or a rare subset of supernovae (e.g. collapsars).

NS mergers are almost certainly contributors \citep{metzger2020kilonovae}. Observations of AT2017gfo matched theoretical expectations for kilonovae which include the formation of heavy elements, though direct evidence of third peak elements was not found. Mapping out the relative contribution of NS mergers to the total r-process produced in the universe requires a determination of the yield of individual events, the rate of local mergers, and the source evolution through cosmic time. This is complicated as it must be done separately for BNS and NSBH mergers, as well as their sub-classes. For example, low-mass BNS mergers may have a higher overall yield, but they may not be able to produce the heaviest elements.

Understanding the various types of NS mergers and their integrated and differential yields is critical. This requires a set of well-sampled local events, driving greater sensitivity, all-sky coverage, and rapid $\sim$arcminute-scale localizations in $\sim$10--30\,minutes. This also requires a set of well-studied, off-axis events recovered in GWs to map the inclination dependence of various signals. Together, and with substantial improvements in kilonova modeling, individual yields can be measured. Late-time spectra by JWST and the upcoming extremely large ground-based optical telescopes will also be key. Then, mapping out the source rate evolution of BNS and NSBH mergers (and their sub-classes) are necessary to determine the time-integrated history. This requires recovery of redshifts of distant NS mergers detected through GRBs, or beginning in the late 2030s, through GWs.

\subsubsection{Dense Matter}\label{sec:dense_matter}
The behavior of matter with densities near nuclear saturation are poorly understood because the approaches to quantum chromodynamics predictions on large scales are invalid or infeasible in this regime. This regime can be probed through more than 15 orders of magnitude in size by studying heavy element collisions on Earth \citep{reed2021implications} or through astrophysical observations of NSs \citep{ozel2009reconstructing}. In the latter case, equations of state can be used to make predictions on observable quantities such as the NS mass, radius, and tidal deformability. Determining the equation of state of dense matter is key to understanding one of the three fundamental forces.

NS mergers, particularly involving \textit{hot} NSs without an enormous star blocking the direct view, are critical tests of dense matter. Perhaps the most important test is the determine M$_{TOV}$, the maximum mass of a non-rotating NS, as this is an asymptotic limit (for NS, though not for strange stars). With 10 confidently classified events and GW measurements the measurement can approach 1\% precision \citep{margalit2019multi}. Although this relies on an assumption of a mass ratio of $\sim$1, even a mildly weaker result is still significantly better than any existing or expected NS equation of state constraint. The tidal disruption radius for NSBH mergers is a strong function of the radius and we may expect some NSBH merger detections to more precisely measure the NS masses, since the higher mass ratio allows for the detection of higher order modes and breaking of correlations between mass ratio and spin in GW observations, allowing for particularly precise mass-radius tests of any NS equation of state. GW observations directly measure tidal deformability and the kilonova yields are intimately tied to the compactness of NSs and thus the NS equation of state \citep{pang2021nuclear}. 

Observations of the plateau emission may constrain the lifetime of supramassive NSs, or observations of the delay time may constrain the lifetime of hypermassive NSs, though only one of these tests is possible \citep{burns2020neutron}. Constraining the hypermassive NS lifetime, either through this method or through 3G interferometers, can help determine whether QCD phase transitions exist inside of NS mergers \citep{bauswein2019identifying,most2019signatures}. No other known proposal to accomplish this test exists.

Determining the lifetime of hypermassive NSs requires that magnetars cannot power merger GRBs and requires the observation of a particularly fast GW-GRB event. Should one occur, the GRB monitor needs $\sim$\,ms scale absolute timing. For the supramassive NS case, the GRB monitor needs to meet the requirements to identify extended emission. To be capable of precise NSBH mass-radius tests, the GRB monitor needs to meet the requirements to determine if NSBH mergers are EM-bright. Otherwise the science is achieved through the requirements for characterizing and classifying the sub-classes of BNS mergers, e.g. total sky coverage.

\subsubsection{Cosmology}\label{sec:cosmology}
A quirk of General Relativity allows the measurement of luminosity distance through GW observations of compact binary coalescences \citep{schutz1986determining}. NS mergers are the compact binary coalescences which can be detected by both IGWN and EM observations, the latter of which enable precise localization and redshift measurement. The GW-EM Hubble diagram is, in principle, preferable to standard candle measures as the intrinsic zero point energy is determined by General Relativity, rather than empirically determined such as the approaches utilizing Type Ia SNe. The current disagreement in the Hubble Constant \citep[e.g.][]{riess2022comprehensive} would greatly benefit from a third, fully independent determination. While this tension may be resolved before the precision of the standard siren measurement is sufficient to provide an answer, standard siren cosmology is certain to become a fundamental tool of cosmology in the next two decades.

The standard siren measurement precision scales roughly as $N^{-1/2}$, where $N$ is the number of NS merger detections in both GW and with EM determination of redshift. With a single event precision of $\sim$15\% \citep{ligo2017gravitational} and a useful precision of $\sim$2\% for contributing to the Hubble Constant disagreement, this requires redshift measurements for tens of GW-detected NS mergers. This is unlikely to occur until at least O5 in the late 2020s. Additionally, EM counterparts have brightness that varies as a function of inclination, introducing a bias in the redshift recovery of these events. Since inclination is correlated to distance in GW measurements, this can set a systematic error limitation in standard siren cosmology. Additionally, varied peculiar velocities of nearby host galaxies deviating from the Hubble flow can also introduce additional systematics. For kilonova counterparts \citep{chen2020systematic} and because of the unknown GRB jet structure, we are unlikely to reach 2\% total precision with our current understanding of the inclination dependence of these signals.

In the current observing run, more kilonovae may be recovered than short GRB counterparts of GW-detected events. However, as the GW network becomes more sensitive, GW-GRBs will be the dominant multimessenger detections of these events \citep{chen2021program}. Their restriction on the range of inclination may also be beneficial, predicated on using close events to gather understanding of the real structure of these jets. This will allow sub-percent precision cosmology throughout the universe, complementing existing investments and resulting in breakthroughs in measuring the equation of state of dark energy, testing both the current models of cosmology and gravity.

Reduction of the systematic floor requires greater understanding of all the various signals from NS mergers. This favors immediate identification and localization of all nearby merger events, promoting detailed study of all signals. The future GW-GRB program is well served by arcminute localizations with a monitor of great sensitivity, to recover NS merger GRBs as deep into the universe as they occur. This will be an absolutely critical piece of the TDAMM ecosystem during the 3G era.

\subsubsection{Ultrarelativistic Jets}\label{sec:merger_jets}
One surprise with GRB\,170817A was that the core of the jet was oriented away from Earth \citep[e.g.][]{nynka2018fading} leading to lively debate on whether the prompt signature was from a successful structured jet viewed off-axis or a ``cocoon'' breakout, resulting from a delayed jet interacting with the surrounding ejecta to create a pressurized cocoon that expands outward at mildly relativistic speeds \citep{NaHo2014,LaLo2017}. Mapping out the jet structure provides insight into the jet launching mechanism, what shapes the jet, as well as what and where the primary emitting region is. This can be probed with complete gamma-ray coverage of GW events where both high-energy detections and sensitive upper limits can compliment GW information on the source distance and orientation \citep{biscoveanu2020constraining}. Precise localizations of events, particularly nearby ones, enables follow-up characterization of the afterglow of off-axis events from radio to X-rays \citep[e.g.][]{hajela2022evidence}, whose evolution is deeply tied to jet structure \citep[e.g.][]{ryan2020gamma}. For the closest events, radio interferometry can observe proper motion of the jets, giving a direct diagnostic on their velocity and potential extension \citep[e.g.][]{ghirlanda2019compact}. 

Observations of these events also provide insight into how jets form and launch. GRB monitors meeting the requirements to determine the behavior of the BNS merger remnant will determine if magnetars can be central engines of merger GRBs, a question currently unsettled between theory and observation. For particularly well characterized events, measurements of the efficiencies can be used to probe how the jet is powered, i.e. Blandford-Znajek or neutrino-antineutrino annihilation \citep{blandford1977electromagnetic,salafia2021accretion}. In the case of a magnetar central engine or Blandford-Znajek, the jets can be Poynting-flux dominated where ordered magnetic fields would create prompt polarization signatures. Measurement of the efficiencies requires measurement of the prompt spectral turnover, and an energy range that extends to several MeV. 

\subsubsection{Fundamental Physics}\label{sec:fun}
One of the major results from the joint detection of GW170817 and GRB 170817A was the measurement of the speed of gravity, which is responsible for roughly a third of all citations to the joint detection publication \citep{abbott2017gravitational}. Note that this result was so important for the field of gravity that it rivals the number of citations of the kilonova discovery papers. This measurement requires sensitive coverage of the GRB sky to detect these events and absolute timing precision to $\sim$0.01\,s accuracy to ensure that the timing precision is not a limiting factor in improved measurements ($\sim$0.001\,s accuracy to ensure it is never a limiting factor).

One additional, unresolved fundamental physics question is whether gravity violates parity \citep{alexander2009chern} which can be explored most precisely with mergers jointly detected in GWs and with GRBs \citep{yunes2010testing}. There is evidence for parity violation on large scales \citep{philcox2022probing,hou2022measurement} which is most naturally explained as arising from gravitational parity violation. Extensions of general relativity which violate parity are generally motivated from high energy completions of gravity, such as string theory. If gravity violates parity, then it may explain the excess of matter over antimatter near the beginning of time \citep{2006PhRvL..96h1301A} and may contribute to the existence of dark matter. This requires a GW and prompt GRB detection of a face-on NS merger, as well as successful follow-up to recover the redshift, which is necessary to seek the disagreement in the luminosity distance as measured by GWs against the value from EM observations. This test is more sensitive for more distant events, being aided by increased prompt GRB sensitivity. Successful follow-up characterization of the afterglow to confirm the on-axis nature would also be beneficial, but may not be required in all cases (i.e. a particularly powerful prompt emission requires an on-axis event).

The tests of fundamental physics possible with NS mergers are numerous \citep{burns2020neutron}. GW-GRB observations can determine the value of the neutrino mass eigenstates through standard siren cosmology. Tests of extra large dimensions require comparison of the GW measurement of luminosity distance with an EM measurement of the same parameter, requiring successful recovery of redshift. These tests generally require successful follow-up observations, favoring rapid dissemination of precise localizations. 

\subsubsection{Summary of Capability Requirements}\label{sec:req_compact_object}
The advent of GW astronomy enables the majority of new scientific results possible with EM observations of merger GRBs. The IGWN is expected to detect NS mergers at a rate of one per month during its next observing run \citep{PeSi2021}. However, the coarse localizations expected ($\sim$\,100--1000\,deg$^2$) will continue to be challenging for wide-field optical facilities such as Pan-STARRS or the Zwicky Transient Facility (ZTF) to follow up, as these searches and automated filters yield dozens of candidates that require photometric and/or spectroscopic follow-up. As the interferometers advance in sensitivity, the rate of events will vastly increase, but many GW localizations will still have poor precision \citep{PeSi2021}. With the cosmological reach of 3G interferometers and the corresponding high discovery rate of NS mergers, follow-up telescopes will struggle to successfully study the population. 

Of all the EM counterparts to NS mergers, prompt GRBs are unique in their propensity for regular, rapid independent localizations of GW-detected events. They are also the only EM counterparts that have been recovered beyond the local universe and the only ones that can always be associated to the GW given the small temporal offset. The GW and GRB localizations can be combined and distributed to the follow-up community, allowing for earlier and more reliable recovery of the kilonova and afterglow signatures, as well as precise localizations enabling host galaxy determination. Additionally, the short time offset between the GW signal and the GRB can act as signal confirmation and background rejection, allowing joint searches to be far more sensitive than those of only one messenger. 

Thus, prompt GRB monitors are necessary to provide diagnostic information from the gamma-ray signatures, to help localize the events for the full community, and to help identify additional GW events. This work is in addition to the historic role that GRB monitors played as the sole discovery instrument for 50\,years before they were joined by GWs. GRB monitors are absolutely critical to the study of these events.

The ideal scenario for a GRB monitor is immediate reporting of arcminute-scale (or better) localizations within 10\,seconds, with true, all-sky, sensitive coverage. With $\sim$0.01\,s timing precision, significant sensitivity beyond existing instruments, long contiguous observing intervals, high temporal resolution data, broadband coverage ($\sim10$\,keV-$\sim$10\,MeV) optimized to fully characterize short GRBs, and a long mission, all of the above science is possible. The localization precision could be relaxed if sufficient wide-field, sensitive follow-up facilities, with massive fields of view and regard, existed at all other wavelengths. While possible, this would be vastly more expensive and would necessarily limit fully characterized events far a far closer horizon, given the trade-off between field of view and sensitivity. 

For current or forthcoming facilities, the first major issue for weakened alert latency and localization precision requirements would be loss of recovery of X-ray plateau emission. Secondly, if less than one minute alert latency cannot be attained, this excludes most targeted ground-based radio searches for delayed pre-merger emission from compact object coalescence. As rapid localization precision drops, this also prohibits most direct follow-up from radio facilities. ULTRASAT should allow recovery of UV signatures for nearby events, but will struggle at greater distances, and cannot provide localization information on UV-faint events. As the capability to recover each known or expected signature is lost, so is much of the science described above. 

The development of wide-field infrared, optical, and UV facilities allows for recovery of follow-up signatures from localizations on the order of $\sim$100\,deg$^2$ within a few hours. Note that this introduces an inherent latency from the need to characterize a potential counterpart as rising or fading. In this case, advancements will need to be made to understand BNS and NSBH merger sub-classes, the origin of short GRBs, long GRBs, and GWs, and much of the broader science mentioned in these sources. It is difficult to project exactly what science is lost given how much remains unknown about these sources. 

We emphasize that having capable GRB monitors, regardless of alert latency, to help follow-up are still necessary for much of this core science. Sensitive, all-sky coverage with broadband energy ranges and stable backgrounds allows identification of extended emission, provides insight into the BNS merger remnant cases, determines which NSBH mergers are bright or dark, confirms the progenitors of GRBs and origin of GWs sources, probes the properties of dense matter, illuminates the physics of ultra-relativistic jets, and provides breakthroughs in  fundamental physics. High-latency localizations enable robust association of GW and GRBs, and will allow confirmation or rejection of follow-up counterpart candidates identified before arrival of GRB data. Joint searches of GW and GRBs, even in high latency, will further broaden this science.



\subsubsection{Requested Deliverables from GRB Monitors}
Real-time reporting of information from GRB monitors is critical to the success of TDAMM science, as demonstrated by Swift and Fermi. Swift-BAT localizations are extremely precise, but BAT will only localize $\sim$1 in 6 joint events. For all other events, automated inclusion of all relevant localization information would vastly improve the current mode of operation. This includes generation and distribution of IPN timing annuli as data arrives, inclusion / exclusion of planetary-occulted regions for detection / sensitive non-detections of events, and combination of all available autonomous localizations (i.e. Glowbug and Fermi-GBM). 

As one example, INTEGRAL SPI-ACS detects the majority of Fermi-GBM short GRBs, and its data will soon be available within $\sim$20\,s. The Fermi-INTEGRAL annuli reduces GBM-only localizations by a factor of a few for some short GRBs. While the latency of data from any individual distant spacecraft of the IPN is high, when multiple missions are active the latency from one of these distant spacecraft may be low. In particular, Mars Odyssey gets several downlinks per day, which enables an annulus with arcminute-level width. These combined localizations and GRB information should be distributed through a collated GRB alert stream.

An additional approach is the automatic association and localization combination of GW and GRB information. The GW-GRB Working Group, comprised of IGWN, Fermi-GBM, and Swift-BAT, are automating this procedure for GBM and BAT triggers for the O4 observing run. Combined localizations will be distributed through the IGWN alert stream in the exact format as the IGWN-only notices. GBM localizations can reduce the GW localizations by a factor of a few to several, but these will still generally be large localizations appropriate for follow-up by wide-field observatories. The creation of a collated GRB alert stream enables joint GW-GRB localizations with all contributing GRB monitors.

The small time offset between the GW signal and the prompt GRB enables increased sensitivity in joint searches, which can directly provide unambiguous proof of additional GW-detected NS mergers. Blind searches are all-sky, all-time, and have enormous effective trials factors. A localization and a detection time to inform a search reduces both the spatial and temporal trials factors. With a fiducial $\sim$25\% increase in sensitivity in GW searches around GRBs \citep{williamson2014improved} we would expect to double the number of joint detections. This is a major effort within the IGWN, which produces population-level searches after each observing run \citep{LVCO3bGRB}. Improved GRB localizations allows for a vast reduction in computation time for these follow-up searches around GRBs with otherwise poor localizations as well as improved GW sensitivity on individual bases. 

Similarly, the GW network can pass temporal and spatial information to the GRB monitors. This has been done during O1 and O2 for Fermi-GBM \citep{hamburg2020joint} and for Fermi-GBM and Swift-BAT during O3 (GW-GRB Working Group, in prep). This work began due to the expectation that nearby GW-detected mergers may produce subluminous GRBs \citep{burns2016fermi}, which was confirmed by GRB\,170817A. The O2 version of the GBM search could detect 2.2$\times$ as many GRBs as the on-board trigger for the purposes of joint detection, while the O3 version could detect 3.7$\times$ as many (Joshua Wood, private communication). Sustained investment in joint searches directly correlates to an increase in the number of confirmed multimessenger events. Recent developments include the construction of true joint searches, combining the data from GWs and GRBs together directly, without the need for distinct searches of the individual datasets \citep{2023arXiv230604373P}. This should lead to future increases in joint sensitivity.

Note that a form of these searches are also run in real-time. Joint detections of multimessenger transients will be reported in low-latency, on the order of hours, providing the joint localization to the community for follow-up. An obvious improvement for these searches would be coherent searches of multiple GRB monitors together, allowing for similar background rejection capabilities inherent to the ground-based interferometer network. 

Additional information could be automatically calculated and provided for real-time analysis that would help guide follow-up observations. For GW-GRBs, one can check if the cocoon closure relations \cite{nakar2012relativistic} are met, which can be used to prioritize X-ray and UV follow-up. The minimum variability timescale could be calculated on a similar timescale, perhaps identifying long GRBs from mergers, prioritizing follow-up observations to characterize these enigmatic events. Continued advancement of distributing useful information as understanding evolves would greatly benefit the community, similar to the improvements made to the IGWN alerts in each observing run.

The community would also benefit from a collated GRB catalog with inclusion of all localization information. This vastly reduces data gathering efforts of various collaborations, can improve sensitivity and reduce computational expense on joint searches, and provides standardized results. A key piece of information that would be useful is a basic assignment of short or long GRB probability, or even more fundamental, the probability that a given GRB was produced by a merger or collapsar.

\subsection{Collapsars}\label{sec:collapsar}
Supernovae can be powered by either nuclear fusion (thermonuclear explosions believed to make Type Ia supernovae) or gravitational potential energy (stellar core-collapse explosions believed to make Type Ib/c and II supernovae).  For CCSN, the energy arises from the collapse of the core of a massive star down to nuclear densities (and the formation of a proto-NS).  GRBs are also believed to be powered by the release of gravitational energy either through compact binary mergers or the collapse of massive stars.

The standard engine behind normal supernovae ($\sim 10^{51} \, {\rm erg}$ explosions) is the convection-enhanced, neutrino-driven supernova~\citep{1994ApJ...435..339H,2007ApJ...659.1438F,2014ApJ...786...83T,2015ApJ...807L..31L,2015ApJ...808L..42M,2018SSRv..214...33B} that taps the energy in this collapsed core which escapes as neutrinos.   This engine can not produce the extreme energies and asymmetries in extreme supernovae, a.k.a. hypernovae or broad-lined (BL) supernovae~\citep{nomotohne98,valentibroadlinesn08}, or GRBs.  Both hypernovae and long-duration GRBs, although also believed to be powered by the release of potential energy, require more extreme conditions than normal supernovae.  For these transients, the black hole accretion disk engine would occur in progenitors that are rotating sufficiently rapidly to form a disk around the collapsed core~\citep{woosley1993gamma}.  Magnetic fields generated in this disk drive a powerful jet that produce both the asymmetries and higher total explosion energy.

One of the leading progenitors behind this black hole accretion disk scenario for long-duration GRBs is the collapsar model~\citep{woosley1993gamma} that invokes the collapse of a rapidly spinning massive star to a black hole.  A number of scenarios have been suggested to spin up massive stars~\citep{fryerprog99}.  In addition, mechanisms have been suggested that merge compact objects with massive stars to produce the same conditions as a collapsar, e.g. the helium-merger model~\citep{fryerhemerger98}.  

The jet disrupts the star, producing a supernova/hypernova associated with the long GRB \citep{Sobacchi2017,Barnes2018}. Radio observations of the Ic-BL SN\,1998bw~\citep{Galama1998}, which revealed a relativistic outflow \cite{Kulkarni1998}, built support for this collapsar or collapsar-like progenitor.  Historical observations of SNe Ic-BL do not always reveal direct evidence of a GRB, but may still show implicit evidence of a GRB central engine \citep[see][]{Berger2002,Soderberg2010,Pignata2011,Milisavljevic2015,corsi2022search}, suggesting a jet may form but fail to reach ultrarelativistic velocities.  Despite its success, a number of unsolved gaps in our understanding of the collapsar model for long GRBs persist.  We will discuss these unsolved problems in the following subsections.


While short GRBs are a focus of the field given the advent of GW multimessenger astronomy, long GRBs have long been the preferred class of GRBs to study as they are far more energetic, are detected more frequently, and generally have higher fidelity data. Study of these events is key to understanding engine-driven supernovae and the collimated, ultrarelativistic jets that power GRBs. 

\subsubsection{Collapsar and Collapsar-Like Progenitors}
\label{sec:collapsar_et_al}

The requirements to make a black hole accretion disk engine work in the case of collapsars or other massive star progenitors are three-fold:
\begin{itemize}
    \item Efficient conversion process: The jet energy is correlated with the disk energy, arguing both for high accretion rates and BH, not NS, central objects~(\citealt{pophambhad98}; although see \citealt{Metzger2011}).
    \item High angular momentum: There must be enough angular momentum to form an accretion disk, which argues for a black hole engine, instead of a NS, because achieving the high angular momenta is easier with black hole progenitors~\citep{woosley1993gamma}
    \item Baryon-free jets: The high Lorentz factors in the jets of long GRBs require baryon-free jets; therefore, the engine must be able to clean out the jet region. To do this, it is assumed that the engine itself lasts longer than the timescale for the jet to propagate through the star and the jet itself will first create this baryon free region.  This limits the stellar size to a compact star~\citep{macfadyenjet99}.
\end{itemize}
These restrictions led to a number of predictions:  the angular momentum requirements should only occur in a small subset of collapsing stars which explains the low rate, long GRBs from these rare and massive progenitors will form primarily in star-formation regions, and the stellar explosions will arise from compact He- or carbon/oxygen stars leading to Type Ib/c supernova associations. Furthermore, a supernova from this system will be produced as pressure waves from the jet unbind the star, producing very peculiar BL emission in the supernova spectrum.

With the identification of SN\,1998bw following GRB\,980425 \citep{galama1998unusual} as well as subsequent detections of unusual supernova in GRB error boxes \citep{cano2017observer}, the collapsar progenitor became the standard model for long GRBs. The basic model explains the rate of long GRBs, their distribution within their host galaxies~\citep{bloomlgrbloc02}, and the fact that the progenitor must be compact arguing for a progenitor stripped of its hydrogen envelope~\citep{macfadyenjet99}.  But it does not explain why the GRB-associated supernovae appear to be stripped of their helium envelopes as well (GRB-associated supernovae are Ic, not Ib supernovae). Progenitor models also struggle to produce the needed high angular momenta~\citep{WoosleyBloom2006,fryerkitp07}. For many of the progenitor scenarios, the model also predicts a large number of Type I and Type II BL supernovae with baryon-loaded jets that do not produce prompt GRB emission, which is in tension with the rates of optically-identified GRB afterglow \citep{ho2022cosmological}. Of related interest is whether the progenitor systems of cosmological collapsars is the same as those of ultra-long GRBs and low-luminosity GRBs, as discussed below.

There is a sub-class of CCSN that are relativistic, identified by radio emission from a faster moving ejecta powered by a central engine \citep{soderberg2010relativistic}, but the inferred rate appears to be on par with the GRB rate rather than the large number predicted by these scenarios \citep{Corsi2022}.
Also, while the exact requirements (e.g., the baryon-loading fraction) to produce a successful GRB jet are still not well understood, we now know GRB\,980425 is part of a class of low-luminosity (or sub-energetic) GRBs \citep{Bromberg2011}.
In fact, many GRBs with associated supernovae have low luminosities \citep{cano2017observer}, as both can only be seen to limited cosmological scales. Some of these low-luminosity GRBs are thought to be emitted by relativistic shock breakout \citep{Campana2006,Waxman2007,nakar2012relativistic} while others require a different, non-thermal emission mechanism, likely the typical internal dissipation mechanism of GRBs \citep{chand2020peculiar}. 
To make things more complex, for normal CCSN, relativistic CCSN, low-luminosity GRBs, and fully successful collapsars, the ejecta velocity and kinetic energies of the slow moving ejecta component are comparable, while the same parameters for the faster moving ejecta span a continuum over orders of magnitude \citep{margutti2014relativistic}. 

One of the biggest unsolved mysteries of collapsar or collapsar-like GRBs is understanding the nature of the progenitor.  Given the low rate of GRBs and BL supernovae, it is clear that jet-driven supernovae are rare; therefore, the process to produce them is uncommon. Spin-down of giant stars could explain why no Type II supernovae are known to have relativistic ejecta (see \citealt{Smith2012} for an argued jet-driven Type II).  But to explain why He-stars also do not produce long GRBs is more difficult. It has been argued that a large fraction of jets are ``choked'' inside the star (or ``failed''; \citealt{Bromberg2012,Lazzati2012}), which could be because they become so baryon-contaminated that the shock is not relativistic; however, there is no decisive observational evidence that any given supernova truly harbors a choked jet.

While the discovery of collapsars and low-luminosity GRBs is the domain of GRB monitors, optical facilities are key to identifying supernovae of interest. These include searching for BL Type Ib and Ic supernovae. While this began nearly 20\,years ago \citep{berger2003radio}, modern optical facilities and the modern TDAMM ecosystem for follow-up to characterize the fast ejecta have vastly increased identification rates of these events \citep[e.g.][]{corsi2017iptf17cw,ho2020broad,Corsi2022}. 

To add another creature to this mix, the discovery of AT2018cow \citep{Prentice2018} and similar events \citep{Coppejans2020,Ho2020koala,Perley2021,Yao2022_AT2020mrf} led to the recognition of ``luminous fast blue optical transients'' (LFBOTs; \citealt{margutti2019embedded,Metzger2022}).
It is generally accepted that these events involve an active compact object or central engine \citep{Ho2019,margutti2019embedded}, with mildly relativistic speeds in some cases \citep{Coppejans2020,Ho2020koala}.
These LFBOTs have been argued to involve shocked jets and may also be key neutrino and VHE sources \citep{fang2019multimessenger,guarini2022neutrino}.
However, no prompt emission has been detected.

The nature of LFBOTs is uncertain. They may arise from massive stars, e.g., failed supernovae \citep{Perley2019fast,margutti2019embedded} or stellar-mass tidal disruption events (TDEs) \citep{kremer2021fast,Metzger2022}.
Alternatively, they may represent intermediate mass black hole TDEs \citep{Kuin2019,Perley2019fast}. 
In order to understand the origin of LFBOTs, continuous and sufficiently sensitive coverage is necessary to recover their shock breakout or low-luminosity GRB emission or to fully rule them out.

Probing the evident continuum from normal CCSN through relativistic CCSN, low-luminosity GRBs, and ultrarelativistic collapsars must be a key goal in TDAMM this decade given the broad range of facilities coming online to identify and characterize these events. Understanding whether and where X-ray flashes, LFBOTs, and other exotic supernovae fit into this picture will increase our understanding of both supernovae and GRBs. The exact questions to be answered are vague due to the lack of knowledge of these sources. Mapping out the relation between supernovae and GRBs explores one of only three supernova explosion mechanisms known. Understanding why the lower energy form of engine-driven supernova appear similar to the traditional CCSN will undoubtedly drive advances in modeling of these events. Observation and characterization of shock breakout will provide the most direct diagnostic and understanding of the structure of massive stars near the end of their life. Understanding when the escaping jets transition from only shock breakout to the traditional GRB emission gives a handle on the necessary conditions for jet formation, propagation, and energy dissipation. 

The high-energy monitors are to be responsible for detection and characterization of the initial prompt signatures from these events. For ultrarelativistic collapsars, the signature is that of a traditional long GRB. The signature for low-luminosity GRBs are generally softer, longer, and less luminous. For relativistic supernova the shock breakout will have peak energies around a few keV and durations of hundreds or thousands of seconds. For normal CCSNe, the shock breakout can peak as low as $\sim$0.1\,keV. Identifying a population of these events is critical to meeting the Decadal recommended TDAMM science. The information carried in these signals is impossible to recover through other means, i.e. observations of the shock breakout in UV (as it cools below the initial $\sim$X-ray temperature) will probe the circumburst material surrounding the star, but only the X-ray signature can probe the stellar structure itself. We do not understand massive stars at the end of their lives, and this is the only known method to directly study them. 

Successful collapsars are served with a range of a few tens of keV to a few MeV. Low-luminosity GRBs often have peak energies between 10 and 100\,keV. X-ray flashes, likely to be part of the continuum towards less successful jets, have been observed between $\sim$\,1 -- 10\,keV. This is the range appropriate to study shock breakout of relativistic supernova. Observing normal supernovae push the requirement down to 0.1\,keV. Ideally these events are reported fast enough to recover the UV signatures of the shock-cooling as well as the potential VHE signatures (discussed later), requiring localizations with tens of deg$^2$. For characterization of more distant events, this precision requirement is more stringent. 

Sensitivity requirements are generally similar to that of studying short GRBs. Existing sensitivity is sufficient, but deeper sensitivity, similar to BATSE from 30\,years ago, would provide far greater detection rates and deeper characterization of detected events. The study of long GRBs requires greater background stability on the order of their duration, with a median value of 30\,s, but a tail extending to several hundred and sometimes thousands of seconds. This timescale is well-matched to that of the durations of the initial shock breakout emission of supernova. 

As mapping out the GRB to supernova continuum is a key goal, the GRB monitors must be designed as partners to the UV, optical, and infrared time-domain surveys. While fortunate events discovered by these facilities may have rise times constrained to tens of minutes, a more typical value is on the order of a day, and a less optimistic scenario is an uncertainty of a few days. This drives the need for all-sky coverage (or at least coverage of the full night sky), sensitivity, and long contiguous viewing windows. The sensitivity must be sufficient to exclude traditional long GRB emission of these events, which is ideally an order of magnitude more sensitive than Konus-Wind, the current most sensitive all-sky instrument. The contiguous viewing windows should be longer than a week, as quantified in Section\,\ref{subsec:fov_lt_cvi}. This additional dimension to the observing fraction requirement is specific to longer-duration discovery windows. 


\subsubsection{Dirty Fireballs and Optically-identified Transients}\label{sec:dirty_fireballs}

Transients resembling the optical afterglows from collapsar GRBs are now routinely detected via their afterglow signatures independently of prompt GRB emission \citep{Cenko2015,Stalder2017,Bhalerao2017,Ho2020_AT2020blt,andreoni2021fast,ho2022cosmological}. \citet{ho2022cosmological} show that the majority of optically-identified GRBs either have associated prompt GRB emission or the GRB upper limits are insufficient to exclude a prompt signature. This implies that, if dirty fireballs have a similar energy per solid angle as clean fireballs, the rate of dirty fireballs is within the rate of GRBs. However, it is possible that less energetic jets may be more likely to become baryon loaded, and that it is more difficult for a mass loaded jet to escape the progenitor star. Therefore, absence of evidence is not evidence of absence: dirty fireballs may exist, but with less luminous optical afterglows to clean fireballs. Ongoing optical searches, and particularly searches for soft X-ray transients (expected from a dirty fireball), will resolve this question.

Matching the discussion for general engine-driven supernova, the start times of these transients are sometimes as fast as tens of minutes but more often are on the timescale of a day, given the nightly observing cadence of many facilities. Confirming an associated prompt GRB or providing constraining upper limits requires contiguous viewing intervals over the full sky covering the entire start time uncertainty. Determining if there is an associated GRB requires checking the data of several facilities with varying amounts of coverage, livetime, and sensitivity. Comparison of optically-identified with gamma-ray-identified collapsar jets is informative on whether they arise from the same underlying population. 

\subsubsection{Origin of Short GRBs}
As noted, recently there have been convincing cases of long GRBs that arose from a merger origin. Conversely, there are short GRBs that arise from collapsars, one of which is GRB\,200826A \citep{ahumada2021discovery}. The burst was discovered in the prompt phase by Fermi-GBM and in the afterglow phase by ZTF. Despite the $\sim$1\,s duration, the expectation was a merger origin, but follow-up identified an excess inconsistent with a kilonova origin but consistent with a supernova. 

The prompt duration is generally shorter than is expected to be possible from collapsars. One explanation is the jet was only partially successful at escaping. It is also feasible that the inferred duration is incorrect due to the tip-of-the-iceberg effect. This drives a need for vastly improved sensitivity over Fermi-GBM. While these events cannot be particularly common, otherwise Swift would have identified more, it is unclear how to prioritize follow-up of these events. However, the focus on characterizing short GRBs will be beneficial in finding more of these events, which provide insight into what makes a successful collapsar jet. 

\subsubsection{Ultra-long GRBs}\label{sec:ultra-long}
On the other end of the collapsar prompt GRB duration distribution, ultra-long GRBs have particularly long durations \citep{levan2013new}. The exact delineation from normal long GRBs is not agreed upon, but is on the order of 1,000-3,600\,s. The most extreme event surpasses 10,000\,s in duration as seen by Konus-Wind \citep{golenetskii2011konus}, with evidence from follow-up observations for a prompt duration of up to 25,000\,s \citep[i.e., 7\,hours;][]{gendre2013ultra}. It is generally believed that normal long GRBs arise from compact Wolf-Rayet progenitors, which should not be able to power the longest duration bursts. It is an open question if ultra-long GRBs arise from different stellar progenitors, including some cases where no supernova is identified to deep limits. Intriguingly, there also appear to be different types of ultra-long GRBs, e.g. a precursor pulse followed by quiescence before the main emission (e.g. GRBs 160625B, 221009A) and others that follow a single pulse with an exponential decay, which may be indicative of external shock from jets that did not have internal dissipation \citep[see discussion in the appendix of][]{kann2018optical}. 

Resolving these questions are key to understanding collapsars. That is, the mechanism that powers their supernova is all the same, but identifying differences in the prompt emission, their shock breakout, or supernova signals will lead to understanding of all the types of progenitor stars capable of ending their lives this way and which stars are incapable of doing so. For example, why has a Type Ib supernova never been found following a long GRB? It is perhaps related to the size of the star, which may suggest these events appear as LFBOTs or other engine-driven supernova. This is another step to understanding the exotic zoo of relativistic transients. 

These events are identified either in cases where they remain bright enough to be recovered by Swift-BAT over multiple orbits \citep{lien2016third} or through the unbiased coverage by Konus-WIND (Svinkin et al., in prep). Having two large GRB monitor with background stability over several hours (requiring a non-LEO orbit) would allow for discovery and confirmation of additional GRBs. This would determine whether an ultra-long sample fully distinct or the extreme of the general long GRB population based on prompt properties (i.e. via duration).

An additional capability that would be a large advancement in the study of these sources would be rapid classification of a given burst as belonging to this class. In a few cases, Swift-BAT has re-triggered on the same burst over multiple orbits, but a dedicated search identifies additional events \citep{lien2016third}. Konus is capable of identifying these events, but the high downlink latency generally prevents follow-up at sufficiently early times.

\subsubsection{High Redshift GRBs}
\begin{figure}[htp]
\centering
\includegraphics[width=\textwidth]{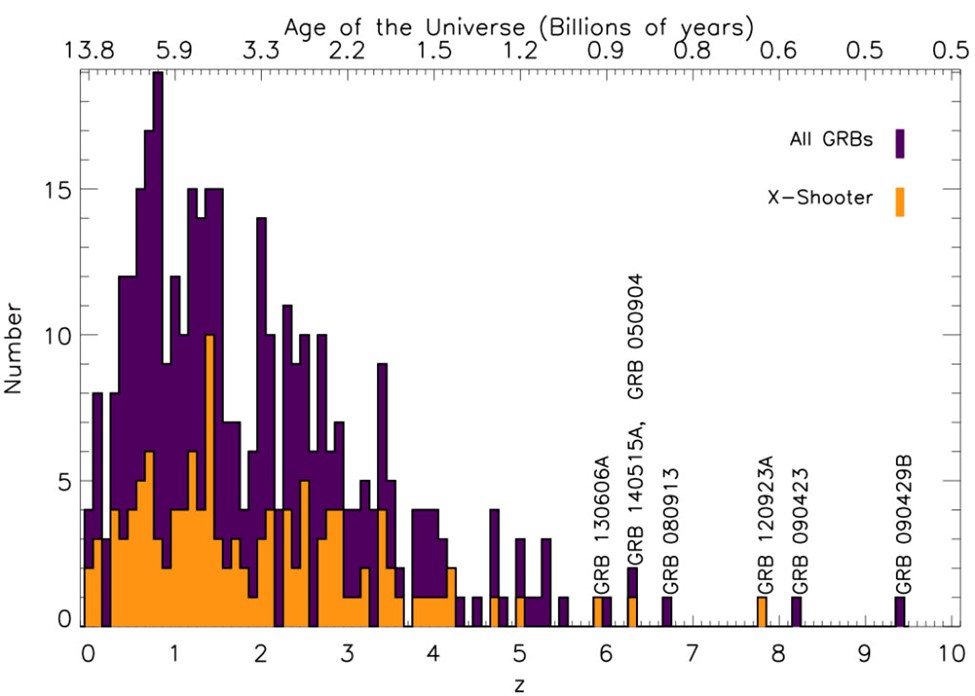}
\caption{The observed redshift distribution of long GRBs \citep[from][updated through 2020]{decia2011thesis}}
\label{fig:highz}
\end{figure}

High-redshift GRBs are one of the most sought after events as they provide a key cosmological probe of the high-redshift universe \citep{lamb2000gamma}. They have been detected out to a redshift of $\sim$9 (\citealt{salvaterra2015high}; Figure\,\ref{fig:highz}), encompassing the era of reionization and beyond. In principle, they can be recovered deeper into the universe than other objects, e.g., the BOAT could be seen with existing facilities to a redshift of $\sim$15-20 \citep{frederiks2023properties}.

The featureless, power-law behavior of GRB afterglows and their extreme brightness make high-z GRBs the best probes to understand the epoch of reionization, which could be mapped with tens of events \citep{tanvir2019fraction}. They also uniquely allow follow-up observations to study small, early galaxies. GRB afterglow measurements also provide a direct measure of the optical depth of the host galaxy to Lyman continuum radiation and thereby constrain the fraction of UV emission that escapes to reionize the intergalactic medium \citep{kawai2006optical}. 

Spectral observations of the afterglow measure absorption lines from elements in the host galaxy and the intervening intergalactic medium, probing the chemical enrichment history of the universe \citep{hartoog2015vlt}. Follow-up observations of the host galaxy after the afterglow has faded allows study on the mass-metallicity relation allowing for tests of the earliest phases of galaxy formation \citep{laskar2011exploring}. A sample of these events also allows inferences on the earliest phases of star formation rate evolution \citep{robertson2011connecting}. Lastly, some of these events may arise from the first stars in the universe.

Several missions have been conceptualized to specifically study and identify high redshift GRBs. These concepts generally achieve a low-energy threshold of a few keV with wide-field monitors to handle the effects of extreme cosmological redshift and design toward extremely sensitive instruments, necessary to identify events at these distances. High redshift GRBs comprise perhaps 5-10\% of the total GRB population. In order to trigger the large near-infrared telescopes like JWST and the future extremely large telescopes these events must be identified rapidly, preferably within 15\,minutes. Thus, these concepts also generally have on-board infrared telescopes to make this measurement and to recover the $\sim$arcsecond scale localization for spectral follow-up. 

\subsubsection{Very-High-Energy Emission}\label{sec:vhe}
The search for Very-High-Energy (VHE) EM emission from GRBs has been a holy grail in the field for decades. The first potential detection occurred in 2000 with Milagrito \citep{atkins2000evidence} but it took nearly 20\,years for unambiguous proof. 
We separate our discussion into the two main types of VHE detectors: Imaging Atmospheric Cherenkov Telescopes (IACTs) and water Cherenkov arrays. We first focus on detections by the IACTs.

IACTs are pointed instruments with fields of view of a few degrees but excellent sensitivity above tens of GeV. IACT arrays react to GRB detections by pointing at or tiling the localization regions disseminated by satellites, resulting in an irreducible delay with respect to the prompt emission.
The first publicly announced VHE GRB was GRB 190114C, discovered by Swift-BAT and Fermi-GBM, which was detected in the TeV energy range by the MAGIC Telescopes during the early afterglow \citep{GRB190114cMagic}. The TeV emission provided the first potential evidence for inverse Compton radiation --- more specifically, synchrotron self-Compton --- which has been long theorized to be an emission channel during both the prompt and afterglow phases of GRBs. This synchrotron self-Compton component would provide a measurement of microphysical parameters of the jet which remain otherwise unknown after decades of study. Significantly, this GRB has also opened a new channel for high-redshift GRB identification. 

Although GRB 190114C was a particularly bright GRB, the VHE detection of the early afterglow phase was only possible due to a rapid autonomous IACT response to the well-constrained localization provided by the \emph{Swift}-BAT through GCN.
However, late-time observations by IACTs can also be fruitful, as GRB 190829A \citep{hess2021revealing} remained detectable in the VHE regime up to 56\,hours after the prompt emission, despite being a low-luminosity burst. Curiously, the VHE emission was more consistent with a synchrotron than a synchrotron self-Compton origin. If true, this would indicate a strong incompatibility with the standard one-zone scenario. VHE GRBs therefore offer a unique opportunity to move beyond the standard simplified scenario and explore GRB jet properties such as the magnetic field structure, as well as relativistic shocks in general. 

The detection of high-energy emission by the LAT following GRB\,211211A, one of the long GRBs that arises from a compact merger origin, suggests an external Inverse Compton (EIC) origin \citep{Mei2022,zhang2022fermi}. EIC is where the electrons are from the jet but the seed photons are from another source, in this case possibly from the kilonova. This detection brings up the feasibility of recovering EIC from nearby GRBs, e.g. low-luminosity GRBs.


The next-generation IACT array, the Cherenkov Telescope Array (CTA), will be roughly 10$\times$ as sensitive as the current IACTs and, in sub-hour timescales, 10,000$\times$ more sensitive in the 75-250\,GeV range than Fermi-LAT \citep{fioretti2019cherenkov}. With $\sim$6 VHE detections of GRBs by the existing IACTs in the past $\sim$5\,years and a handful of detections of $\sim$100\,GeV photons by the Fermi-LAT in 15\,years of observing, CTA could detect tens of GRBs per year, so long as it has fast access to sufficiently accurate localization information. The largest of the telescopes, which have the lowest energy ranges (25-150\,GeV) and are therefore more sensitive to extragalactic sources (as the highest energy photons are preferentially absorbed by the Extragalactic Background Light), will be able to repoint to any position in the sky within 20\,s of reception of an alert\footnote{\url{https://www.eoportal.org/other-space-activities/cta}}. These Large-Sized Telescopes will have fields of view of 4.3$^\circ$ radius, and for poorly localized sources, CTA can cover a larger region with lower depth by staggering the pointings of its telescope array, allowing for an adaptive follow-up program. Thus, CTA can follow-up initial localizations on the order of tens of square degrees and have robust association of low significance VHE signals through comparison with higher latency precise localizations of the prompt signal. For the brightest GRBs, CTA will have the most sensitive view into the GeV energy band; for the larger number of more moderate GRBs, CTA will provide the statistics necessary to better characterize the population of VHE GRBs.

The water Cherenkov telescopes are wide-field survey telescopes, and are the better instrument for VHE observations during the prompt emission. They cover vast regions of the sky (few sr) at a much shallower depth than the IACTs, but targeting a higher energy (above a hundred GeV) and with the advantage of having a much larger duty cycle, as they do not require the dark observing conditions that IACTs do. The premiere telescopes are the High-Altitude Water Cherenkov Observatory and the Large High Altitude Air Shower Observatory (LHAASO), both in the Northern Hemisphere, with a proposed southern installation called the Southern Wide-field Gamma-ray Observatory. Thus far, the only detection of a GRB by a water Cherenkov telescope is of GRB\,221009A by LHAASO \citep{LHAASO_WCDA}, with confirmed photon energies of up to 7~TeV and tentatively as high as 18~TeV \citep{huang2022lhaaso}. As the GRB went off in the LHAASO field of view, this also marked the first VHE detection of a GRB during the prompt emission, though the smooth nature of the VHE emission was more consistent with being due to forward shock emission. While this has been the sole detection of a GRB by a water Cherenkov telescope to date, the enormous number of counts seen by LHAASO suggests other bright events may be recovered if they are fortuitously observable by these arrays. If all three water Cherenkov observatories are in operation, then a large fraction of the sky will be simultaneously observed at any time. These arrays do not require prompt localizations, only delayed localizations of sufficient accuracy for association.

However, a VHE detection on its own --- whether by IACTs or water Cherenkov telescopes --- is not sufficient to constrain GRB properties such as the magnetic field structure or the microphysical parameters of the jet. Rather, this is only achievable when there are longer wavelength observations to provide the crucial context for the VHE measurements. In particular, sensitive instruments operating in the UV to GeV range will be absolutely crucial to characterize the true nature of the transition between the synchrotron and inverse Compton components. Thus, the prompt localizations need to meet the needs of telescopes across the EM spectrum.

\subsubsection{Origin of Neutrinos and Ultra-High Energy Cosmic Rays}\label{sec:neutrino_et_al}
There are strong connections of ultra-high energy cosmic rays (UHECRs) with particle physics and high energy astrophysics. Three messengers, cosmic rays, gamma-rays, and high-energy neutrinos (HEN), are linked, providing complementary information about the same underlying physical phenomena in astrophysical environments. A golden multimessenger era lies ahead of us and with the next-generation cosmic ray experiments it will be crucial to seek out neutral ultra-high-energy particles related to transient events in order to acquire insights into a number of nature's most energetic processes. 

Among the outstanding questions in these interdisciplinary fields is the origin of UHECRs, of which GRBs are prime suspects. Assuming they produce UHECRs, then HEN emission is also expected as the UHECRs will collide with photons \citep{WaxBach}. Unlike UHECRs, HEN will travel in a straight line through magnetic fields, making it possible to pin point their origin. A HEN signal from a GRB would be considered \enquote{smoking gun} evidence for GRBs being hadronic accelerators. 

HEN emission from GRBs has yet to be detected and they have been ruled out as the main contributor to the all-sky HEN flux \citep{ICGRB}. The most optimistic models for HEN emission have also been ruled out, but more realistic models are yet to be excluded \citep{2015ICGRB}. Measurements or limits on the HEN flux of a GRB can be used to constrain the parameter space of model parameters, such as the baryon loading factor \citep{MuraseGRB221009A}. In cases where there is a detection of VHE gamma-rays from a nearby GRB, a limit or measurement of the HEN flux can set a constraint on how much of the VHE gamma-rays are from hadronic versus leptonic origins. 

The IceCube Neutrino Observatory \citep{IceCube} is currently the most sensitive experiment to transient sources of HENs. The most important factors for improving the sensitivity are increasing the number of GRBs with localizations near to the angular resolution of IceCube's through-going tracks of ${\cal O}(1^\circ)$. IceCube is undergoing an upgrade and pursuing support for a second generation instrument. Similarly, in Europe ANTARES is being upgraded to KM3Net. Baikal-GVD continues improvements in their analysis. Future treatment of these telescopes as a single effective instrument may further increase sensitivity. As these non-EM telescopes advance, they require continued capability in the detection and localization of prompt GRBs.

Related to these works is exploration of whether GRBs can be the origin of UHECRs even if they do not substantially contribute to the HEN flux. For example, in the ICMART model of prompt GRB dissipation \citep{zhang2010internal} the internal dissipation radius is substantially larger than in other models, where the lower density will result in fewer proton interactions, allowing for contribution to the UHECR flux without producing a luminous HEN signal. Continued joint observations of the high-energy sky in neutrinos and gamma-rays is key to understanding the prompt emission mechanism of GRBs, both in population \citep{abbasi2022searches} and individual studies \citep{abbasi2023limits}. 

The IceCube observations of GRB\,221009A and upper limits from analyses designed for lower energy studies has also proven informative in new ways \citep{abbasi2023limits}. For example, \citet{murase2022neutrinos} show that the non-detection of GeV neutrinos from IceCube places constraints on the bulk Lorentz factor of the jet at an earlier phase than has been done before or require an even more pure (low baryon content) jet than previously shown. As analysis techniques improve in the existing and forthcoming neutrino telescopes, we will continue to learn more about GRBs, provided sufficient discovery missions are still observing.

While the above discussion focused on HENs from fully successful collapsars, the most promising neutrino source here may be so-called \enquote{choked} grbs. These events would produce significant amounts of neutrinos given the far higher proton interaction likelihood \citep{meszaros2001tev}. Choked GRBs likely sit in the continuum from normal CCSN to ultrarelativistic collapsars, with fully choked GRBs appearing as relativistic supernovae and partially choked GRBs as low-luminosity GRBs. LFBOTs, X-ray flashes, and similar may also be key neutrino sources. If true, optical identification of these events may preclude robust association of related neutrinos due to the large temporal uncertainty on the EM side. The precise timing and spatial information from detection of shock breakout or GRB emission of a choked GRB will enable association of even untracked high energy neutrinos. This should be a key goal of the new multimessenger era.

When speaking strictly about understanding the origin of neutrinos, there is no strong alert latency driver for high energy EM monitors because the HEN telescopes are effectively continuously observing the entire sky. It is sufficient to provide precise localizations, necessary for robust association of any neutrino signature, in high latency. However, should such a joint detection occur, it would be one of a few known multimessenger source types and full temporal and spectral characterization across the full EM spectrum would be critical. The GRB monitor requirements are thus those described above for general follow-up of collapsars. Follow-up would be aided by the neutrino localizations, though tracked IceCube events require tiling from narrow field-of-view facilities. It is reasonable to assume the first neutrino detections of GRBs will occur for nearby events, which supports all-sky coverage even with less sensitivity to GRBs, though low-luminosity events may be of particular interest in this case.

\subsubsection{Gravitational Waves}\label{sec:collapsar_gws}
The relationship between collapsars and long GRBs is in a state similar to that between BNS mergers and short GRBs prior to GRB 170817A. A GW observation associated with a collapsar, identified by the long GRB, would provide a watershed of new insights. As the sensitivity of the GW network continues to improve, such a possibility becomes more realistic. Some specific models for GW emission, for example, the accretion driven instability model \citep{ADI2014}, suggest a long GRB at $\sim 50$ Mpc would already be observable by the current GW networks \citep{LVCO3bGRB}. Only a small sample of long GRBs within this distance have been observed to date; however, the rate of redshift determinations is very low. For example, of the 86 GRBs used in the analysis for \citet{LVCO3bGRB}, none had a measured redshift. Moreover, gamma-ray brightness is a poor indicator of distance, therefore, the GW analyses follow up all GRBs, since any one of them may be within the detectable horizon. While it is expected that detections will not be possible until the 3G era, observations prior to that epoch will provide some of the only direct information on what occurs in the interior of collapsars. An improved sky localization from IPN or other mission could facilitate a follow up redshift determination to bolster the case for a putative GW detection. The requirements on high energy monitors are similar to those of multimessenger searches with high energy neutrinos, as the IGWN also has archival all-sky observations (when operating). Precise, high-latency alerts are sufficient for GW-GRB specific science and the GW-detected events will likely be nearby, favoring all-sky coverage even with lower sensitivity. Similarly, identification of these events rapidly enough to allow successful follow-up and detailed characterization is highly beneficial. We additionally note the effects of failed jet may be detectable in GWs \citep{gottlieb2022jet}, but prompt EM identification likely requires X-ray recovery of the shock breakout.

\subsubsection{Polarization and the Blandford-Znajek Mechanism}\label{sec:polarization}
An additional diagnostic in astronomy is polarization, providing insight into geometry and ordered magnetic fields. Both geometric and intrinsic polarization may be expected in prompt and afterglow emission. Geometric polarization will arise when the viewing region crosses structure from the jet. Intrinsic polarization may arise from the creation of ordered magnetic fields at the jet launch site by the central engine; these will become disordered due to turbulence as the jet propagation outwards.

From time-integrated measurements in the prompt phase alone it is not possible to determine the origin of a significant polarization signal; however, if the majority of GRB jets have highly ordered magnetic field and their prompt emission mechanism is synchrotron, it is possible to identify this population from the rest of GRBs through a population of prompt polarization detections \citep{toma2009statistical}. This requires a GRB polarimeter and determination of the peak energy for a sample of at least tens of GRBs. External determination of spectra is beneficial for more confident determination of polarization. 

To better constrain the prompt emission mechanism and understand jet physics and evolution, time-resolved polarization is necessary. Historically, theory has focused on time-integrated measurements as a substantial number of photons are required for robust spectropolarimetric fits. With the next generation of polarimeters being built and proposed it may be possible to perform time-resolved fits for an number of GRBs. Theoretical and simulation work can explore distinguishable time-resolved spectropolarimetric behaviors, allowing for advanced understanding on the prompt GRB emission mechanism and on microphysical jet parameters \citep[e.g.][]{Deng2016,Gill2021,parsotan2022photospheric}.

Very early observations of GRB afterglow may be able to capture intrinsic polarization and their destruction as the jet propagates, as possibly seen in GRB\,120308A \citep[e.g.][]{steele2017polarimetry}. At later times, afterglow polarization measurements generally capture only geometric polarization, which can be used to understand jet structure in concert with observations of jet breaks.

Bursts with sensitive polarization observations in both the prompt and afterglow phase are of particular scientific interest \citep{negro2023ixpe}. If follow-up observations of the afterglow find low polarization, but the prompt emission has high polarization, then the prompt phase must arise from a jet with ordered magnetic fields. As ordered fields will only be destroyed during propagation they provide insight into the central engine. Jets formed by black hole and neutrino-antineutrino annihilation central engines \citep{eichler1989nucleosynthesis,meszaros1992high} will tend to carry most of their energy in the matter entrained in the jet and will have low prompt polarization. Jets with ordered magnetic fields arise either from magnetar central engines or black holes whose energy is extracted to power the jet via the Blandford-Znajek process \citep{blandford1977electromagnetic}. 

Bursts with magnetar central engines have a maximal energy output set by the rotation energy imparted on the object. Bursts with total energy, being the sum of the energy released as light in the prompt phase and supernova (or kilonova) phases as well as the kinetic energy in the jet and omnidirectional ejecta, can only be powered by black holes. Thus, an energetic burst with prompt polarization and no or little afterglow polarization would be direct proof of the Blandford-Znajek mechanism. While it is known that black holes are voracious consumers of surrounding material such a result would be proof that black holes occasionally return significant power to the universe, i.e. the first proof of a Penrose process \citep{penrose1971extraction}. This is one of the few specific questions listed in the Astro2020 Decadal report.

\subsubsection{Prompt Emission Mechanism}\label{sec:prompt_mechanism}
The prior discussions on understanding the origin of the VHE signals, seeking neutrino counterparts, and applying polarimetry to the study of GRBs in the prompt and afterglow phase are deeply related to understanding the prompt GRB emission mechanism. The mechanism, or mechanisms, has escaped confident determination for decades \citep[see][for a general review relevant for this section]{2018pgrb.book.....Z}. The current state of the field is an on-going debate on whether a low-energy excess is due to an additional thermal component or an additional break in a synchrotron spectrum. Working out if the jets are Poynting flux or matter dominated, how they are launched, whether they have ordered magnetic fields, and understanding their microphysical parameters are all key diagnostics to understand the viable prompt GRB emission mechanisms. A sample of these well-studied events is necessary to understand if more than one mechanism is viable in distinct GRBs or even if multiple contributions contribute to the signal seem in individual bursts. The GRB monitors need to enable rapid, broadband follow-up of the EM spectrum, provide precise localizations (even in high latency) for the non-EM messengers, and cover a broad range of the prompt emission, ideally $\sim$1-10,000\,keV. Coverage up to 20\,GeV would be well-suited to determine the inverse Compton crossover region in partnership with CTA.

Long GRBs are often easier to use than short GRBs for the study of the prompt emission mechanism as they are generally brighter, as previously discussed. Additionally, their durations allow for the possibility to observe the prompt phase across the EM spectrum (noting it is inherently covered by the neutrino monitors and IGWN when they are observing). This can occur for any GRB if it is in the field of view of a given survey instrument, such as the LHAASO observation of GRB\,221009A or the TESS observation of GRB\,230307A. However, it is unlikely that this will occur for multiple survey instruments at the same time. Swift and Fermi alerts can be received and distributed to the community on the order of 10--30\,s. If observations begin within a minute after trigger, 20\% of long GRBs are longer than this. The first success was with a wide-field monitor observing GRB\,990123 \citep{akerlof1999observation}, with one case of early polarimetry observations \citep{troja2017significant}, and in total, order tens of events have been detected with prompt optical observations. Recovery of the prompt signal from optical (or lower) up to TeV energies would be a stringent test of any prompt emission model. This requires immediate alerts. Arcminute localizations are the ideal scenario, and is perhaps the only way to recover the prompt signature in more than one follow-up facility at a time. 

\subsubsection{Lorentz Invariance Violation}\label{sec:liv}
Observations of GRBs provide some of the most sensitive search space for Lorentz Invariance Violation (LIV), motivated by the goal to test General Relativity and the search of a quantized field theory of gravity, and is thus a path towards a grand unified theory \citep{burns2020neutron}. Here we use the language of the Standard Model Extension Framework \citep{kostelecky2011data}, which separates potential LIV into sectors (gravity, neutrino, photon, matter), dispersive and non-dispersive, birefringent and non-birefringent, and directional dependent violations. 

Perhaps the most well known LIV test using GRBs is using GRB\,090510 where the constraints fall into the dispersive, non-birefringent, photon case \citep[e.g.][]{vasileiou2013constraints}. However, the most precise test of LIV with GRBs arises from the detection of polarization, as birefringent LIV would destroy a coherent signal with precision scaling with distance \citep[e.g.][]{stecker2011new}. 

In the new age of multimessenger astronomy, these and other EM limits are key to determination of LIV in the other sectors. For example, the speed of gravity was determined from the known speed of light, and dispersive LIV tests in gravity determined from the far more precisely known limits on the photon sector with GW170817 and GRB 170817A \citep{abbott2017gravitational}. Similar results have been calculated for SN\,1987A \citep{ellis2008probes} and the neutrinos seen in coincidence with a blazar flare from TXS+0506 \citep{ellis2019limits}.

For dispersive tests, coverage of the prompt emission over several orders of magnitude is critical. Improvements require a range broader than that achieved for GRB\,090510 ($\sim10$\,keV to $\sim$10\,GeV) or a shorter timescale ($\sim$1\,s), favoring the broadband recovery of prompt emission described in the past subsections. For more precise birefringent tests sensitive polarimetric observations is required to recover polarization from events deeper into the universe.

\subsubsection{Summary of Capability Requirements}\label{sec:req_collapsar}
As expected, a great deal of the science that can be achieved through collapsar detections requires precise localizations. High-latency, precise localizations allow deep multimessenger searches for neutrinos and GWs, and can confirm or reject candidate counterparts identified in independent surveys or from follow-up observations of earlier GRB localizations (e.g. optical, VHE, or otherwise); if precise enough ($\sim$arcsecond scale), these can enable late-time follow-up observations to identify host galaxies and thus redshift determination. A best-case scenario certainly includes rapid reporting of precise localizations from the high-energy monitor and is scientifically well-motivated. It allows broadband observations of the prompt phase to study LIV and the prompt emission mechanism, full characterization of the spectral energy distribution for VHE detections, observations of the early polarization decay, measurement of redshift directly from the afterglow, and recovery of the full range of diagnostics of these events. 

The field of view requirement depends on the specific science case. When paired with other discovery facilities, true all-sky monitoring is preferred, given the usual limited horizon of other diagnostics (optical, GW, neutrino, etc). Studies of the high-redshift universe favor depth in a given direction. These capabilities need not be satisfied with the same instruments. 

Discovery of transients in UV, infrared, optical, and radio drives $\gtrsim$1\,week contiguous observing intervals. The need to determine if ultra-long GRBs are part of the long GRB population or are distinct motivates background stability on the order of at least thousands of seconds, preferably hours. The timing precision and temporal resolution of collapsar GRBs is generally less stringent than that of short GRBs or magnetar flares.

Polarization in a GRB monitor is key to understanding the jets themselves and the prompt GRB emission mechanism. Polarimetry of the afterglow of these bursts may provide proof of the Blandford-Znajek mechanism in specific cases. These drive instrument and precise localization requirements. 

Fully successful collapsars require the typical historic energy range of GRB monitors. To entirely map out the continuum through low-luminosity GRBs/X-ray flashes, shock breakout from relativistic supernovae, and shock breakout from normal CCSN requires low energy thresholds of approximately 10, 1, and 0.1\,keV, respectively. A mission capable of achieving 0.1\,keV would be a revolutionary facility for both GRBs and supernovae.

\subsubsection{Requested Deliverables from GRB Monitors}
Many communities, including IGWN, IceCube, and wide-field survey instruments, would greatly benefit from collated GRB alert streams and catalogs. This aides the discovery of multiwavelength and multimessenger events and vastly reduces the work necessary for multidiagnostic studies. It would also enable real-time multimessenger searches, e.g., for neutrino-GRB joint detections or comparison with optically-identified transients. Reduced localization area increases joint search sensitivity and reduces computational expense. Having the prompt GRB community deliver this information will result in greater science return. Providing access to the detections of the full GRB network, or coherent upper limits for non-detections, will help prioritize follow-up observations of externally identified relativistic transients.

Similarly, studying data as it arrives with the goal being to determine if a given burst could be consistent with a shock breakout origin may aid the study of low-luminosity GRBs, or shock breakouts from supernovae in the case of an X-ray monitor. The automatic flagging of potential ultra-long GRBs while the afterglow is still detectable would be a boon in the study of these sources. Providing information on whether a given event may have broadband spectral coverage, prompt polarization coverage, or is particularly bright would also help the follow-up community determine which bursts to target. These are similar to the findings for merger GRBs; we refer the reader to that section for additional details. 


\subsection{Miscellany}\label{sec:misc}
In addition to the sources where GRB monitors have compiled large samples, they are also important for rare and unusual events. This is a driver for sustained coverage of the high-energy sky over the timescales of human lifetimes, to ensure detection of once-in-a-lifetime events.

\subsubsection{Jetted Tidal Disruption Events}
There have been four relativistic (\enquote{jetted}) TDEs discovered to date. Three were discovered by Swift-BAT, one by the on-board trigger and two in high-latency analyses by the BAT hard X-ray monitor \citep{levan2015swift}. The fourth was recovered in optical by ZTF \citep{andreoni2022very,pasham2023birth}. Key questions to these sources include i) what is the connection between jetted TDEs and other classes of TDEs found routinely by optical surveys \citep{van2021seventeen,charalampopoulos20222020wey,hammerstein2022final,yao2023tidal}? ii) are optically-identified and high-energy-identified jetted TDEs distinct or of the same class? iii) how do their jets carry their energy? Study of these sources allows for observation of a jet where both the initiation and end of the jet can be observed, similar to GRBs, and provide a link to active galactic nuclei. Study of the three source classes may allow for understanding on how relativistic jets are powered, what distinguishes ultrarelativistic from relativistic, and understanding of the ways these jets carry their energy. 

Discovery of more TDEs by high energy monitors likely requires instruments beyond the basic scintillator design, which are non-imaging and background dominated. They can be discovered at a low cadence with coded aperture masks like Swift-BAT. Lobster-eye telescopes may prove to be more capable but will generally detect more distant events. However, as optical surveys become ever more capable it is evident that they will discover additional jetted TDEs. Determination of a high-energy signature requires either early identification in optical with rapid follow-up of pointed X-ray telescopes or a highly sensitive X-ray monitor with at least a $\sim$daily all-sky cadence.

\subsubsection{Non-Astrophysical Sources}
GRB monitors are often used for studies in heliophysics, helping to complete the multiwavelength picture of solar flares. The discovery of terrestrial gamma-ray flashes associated with lightning by BATSE has led to characterization of these sources in X-rays and higher energy gamma-rays by RHESSI and Fermi-LAT, as well as comparison with lightning detection by ground-based optical networks. Thus, GRB monitors have a role to play in multiwavelength studies of solar and Earth science events. As we utilize instruments on planetary and heliophysics spacecraft, it would appropriate to ensure the deliverables from the prompt GRB monitors are also designed to benefit other divisions and cross-divisional studies, e.g. in the study of solar flares.

\subsubsection{Unexpected Discoveries}
The discovery of GRB\,211211A as a long GRB from a merger origin was unexpected, as was the consistency of the prompt emission arising from a fast-cooling synchrotron emission \citep{gompertz2023case}. The analysis of GRB\,221009A identified a 10\,MeV line in the prompt phase of individual, rare events \citep{ravasio2023bright} which has no previous theoretical expectation. Even after 10,000 prompt GRB detections, new insights into GRBs are made though careful study of the prompt phase. As a further example we emphasize the possibility of GRBs being created by the evaporation of local primordial black holes \citep{ukwatta2016investigation}. Such a possibility is best studied with multiple spacecraft at interplanetary distances. Maintaining various capabilities of the prompt GRB monitors over decades is key to maximizing science from rare events, either through the study of their prompt signatures alone or with multidiagnostic studies common in the current era.

\section{Capability Requirements}\label{sec:requirements}
Each sub-topic for the sources discussed in Section\,\ref{sec:sources} discusses the specific capability requirements necessary to advance scientific understanding on that topic. Each source also has a Summary of Capability Requirements section giving the broad needs specific to high energy monitor observations of that source class, i.e. magnetars in Section\,\ref{sec:req_magnetar}, compact mergers in Section\,\ref{sec:req_compact_object}, and collapsars in Section\,\ref{sec:req_collapsar}. When seeking to understand the set of requirements necessary for a specific question or source class we refer the reader to the content of Section\,\ref{sec:sources}.

This section takes a more holistic view on the observational capabilities themselves, commenting on the general importance of various requirement thresholds. This is useful in considering which technologies to use or develop. We discuss how easy it is to meet the various requirement levels. We emphasize all capabilities do not need to be met with each instrument, i.e. a polarimeter does not also need to provide arcminute-scale localizations alone. 


\subsection{Localization Precision}\label{subsec:loc_prec}
The science possible with a given localization precision is summarized in Table\,\ref{tab:req_spatial}. For cosmological GRBs, i.e. those from mergers or collapsars, a position determination to $\sim1''$ accuracy is necessary to robustly associate the event to a host galaxy and to determine the offset from the host galaxy. This permits inference of the redshift (though this can be measured directly from the afterglow in some cases), enabling measurements of intrinsic energetics and the progenitor environments. This precision is beyond the capability of wide-field transient monitors and is enabled through successful recovery of the afterglow in follow-up instruments.

\begin{table}
\begin{center}
\begin{tabular}{|c | l | c |} 
 \hline
 Localization Accuracy & Corresponding Result & Sections \\ 
 \hline
4$\pi$\,sr & Detection of gamma-ray transients by all-sky monitors & \\
 & Chance joint detection of transients with other wide-field monitors & \\
$\sim$1000\,deg$^2$ & Follow-up tiling of GRBs by the widest field UV and optical telescopes &  \ref{subsec:loc_prec} \\
 & Robust association of GWs and GRBs & \ref{sec:mergers} \\
$\sim$100\,deg$^2$ & Identification of MGF candidates and potential host galaxy & \ref{sec:grb200415a} \\
 & Follow-up tiling of GRBs by wide-field optical telescopes & \ref{sec:early_uv}, \ref{sec:vhe}\\
$\sim$30\,deg$^2$ & Follow-up tiling of GRBs by wide-field radio, VHE, and IR facilities & \ref{subsec:loc_prec} \\
$\sim$10\,deg$^2$ & Associate nearby extragalactic MGFs to ideal host galaxies & \ref{sec:eMGF} \\
 & Robust association of GRBs to neutrinos & \ref{sec:neutrino_et_al} \\ 
 $\sim$1\,deg$^2$ & Associate SGR flares to specific magnetars & \ref{sec:sgr_gw} \\
  & Robust association of UVOIR identified transients to GRB & \\
 $\sim$100\,arcminute$^2$ & Extragalactic MGF host galaxy association & \ref{sec:eMGF} \\
 $\sim$30\,arcminute$^2$ & Follow-up observations by the majority of telescopes & \\
 $\sim$1\,arcminute$^2$ & Follow-up observations by effectively all telescopes & \\
 $\sim100$\,arcsecond$^2$ & Follow-up identification of Galactic magnetars & \ref{sec:new_magnetar}\\
 $\sim10$\,arcsecond$^2$ & Robust associations of cosmological GRBs to host galaxy and measurement of offset & \ref{sec:req_collapsar} \\
 \hline
\end{tabular}
\caption{Brief summary of the follow-up observations and association strengths for various gamma-ray transient localization precision. Relevant sections which detail specific rows are provided in the third column.}\label{tab:req_spatial}
\end{center}
\end{table}

Thus, achieving precise localizations is tied to a mix of alert latency and precise localizations. Localizations on the $\sim10'$ scale is sufficient to be observed with the vast majority of telescopes, though $\sim1'$ scale is necessary for some specialized telescopes like the large optical and infrared spectrometers. Localizations at this scale are possible with coded aperture masks \citep[e.g.][]{barthelmy2005burst,chattopadhyay2018blackcat}. Sub-degree localizations, achievable with Compton telescopes and pair-conversion telescopes, can be followed-up with a small number of tiled observations. A common example of this is Swift-XRT and UVOT tilings of Fermi-LAT detections, which have a typical uncertainty of $\sim0.5^\circ$. 

Localizations on larger scales can be followed up with wide-field instruments. Instruments similar to Fermi-GBM can achieve localizations on the scale of $\sim1-50^\circ$. As the localization area increases, the number of facilities capable of observing a significant portion of the localization decreases rapidly. With the advent of GW astronomy, which can have localizations of comparable areas, more wide-field optical facilities have been built \citep[e.g.][]{groot2019blackgem,bellm2018zwicky,steeghs2022gravitational} that are capable of covering thousands of square degrees per night \citep[e.g.][]{coughlin20192900}. Similarly, UV coverage by ULTRASAT, VHE coverage by CTA and the water Cherenkov telescopes, infrared by WINTER, and several wide-field radio facilities contribute to broad wavelength coverage. While these facilities exist and unlock new, critical capabilities in the TDAMM ecosystem, there is a cost to poor high-energy localizations. Successful identification of the afterglow can take more than a day as the requirement for a fading transient and observing cadence following Earth's day/night cycle is a fundamental limit for ground-based facilities. There is generally a trade-off between sensitivity and field of view, limiting their depth to the nearer universe. Especially in the optical, there must be vetting of enormous numbers of sources. 

The independent localizations from joint detections of prompt signals can be combined to produce an improved localization compared to either independent localization alone. A key example is the case of joint GW-GRB localizations where GBM-like localizations will reduce single GW interferometer localizations by factors of $10-100$x and double interferometer triggers by a median of $\sim$5x due to the different localization morphologies. Joint detections are also possible with IceCube, ULTRASAT, TESS (e.g. GRB\,230307A), the water Cherenkov telescopes, etc. Though, again, a precise localization on the high energy signal is the best-case scenario for the entire TDAMM community.

The other main role of localization precision is in associating transient events. For the association of two signals, the significance scales inversely to the localization area. Localization precision can be relaxed in this role if the temporal precision is sufficient to associate signals in time. The temporal offset between the GW signal and GRB is on the order of seconds, so even large (thousand square degree) localizations are sufficient to achieve unambiguous association. For optically-identified transients with large uncertainties on the time of initial explosion, these large localization regions can be insufficient for associating signals. These often require final localizations on the order $\sim$1\,deg$^2$, with discovery claims requiring greater precision. The same is true for most expected neutrino signals from GRBs and CTA follow-up observations. 

\begin{figure*}[htp]
	\centering
	\includegraphics[width=\textwidth]{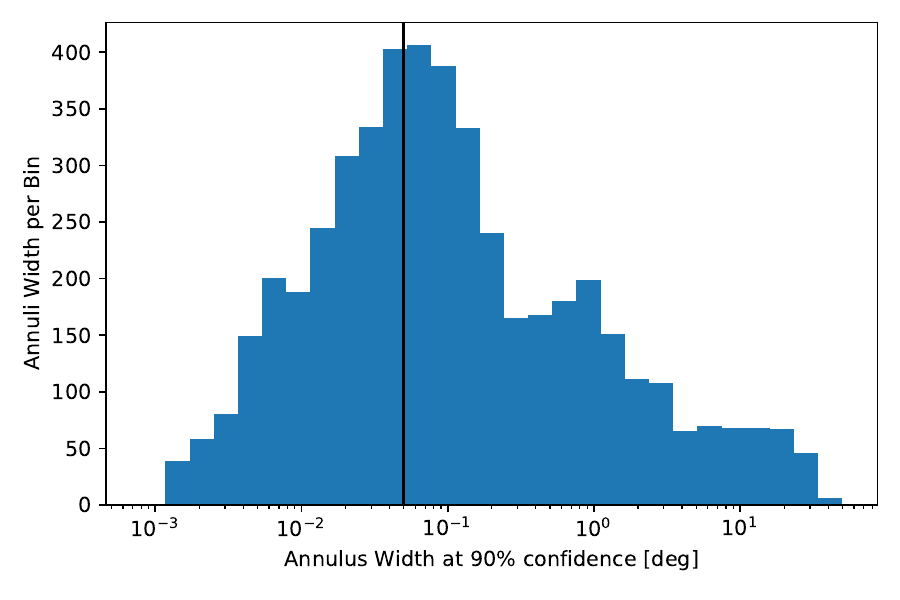}
    \caption{The 1-sigma width distribution of the third IPN timing annuli, through 2020. Overlaid is a black vertical line denoting the typical 90\% confidence error for Swift-BAT. The narrowest 90\% confidence width in the distribution is $4''$ with a median width of $5'$. Additional larger annuli can be generated for satellite pairs in low Earth orbit which is not included in the figure. }
    \label{fig:ipn_annuli_distribution}
\end{figure*}

Timing annuli from the IPN span a wide range of precision, as shown in Figure\,\ref{fig:ipn_annuli_distribution}. Typical annuli widths for IPN bursts recovered by distant spacecraft (which are the bright $\sim$half of GBM bursts) are a few arcminutes wide, though precision down to $4''$ has been achieved. Note that a single annulus with a few arcminute width can span significant total sky area, and at least three spacecraft (two annuli) are required to achieve a localization area comparable to Swift-BAT. IPN localizations have two drawbacks compared to those possible from coded aperture masks. The reporting latency for distant spacecraft is most often measured in hours to days, though fortuitous timing is possible. Second, when the first data from distant spacecraft arrive the initial localization is very elongated, which is a shape that wide-field monitors are not generally suited to tiling. Despite this, successful recovery of interesting events has been demonstrated, most notably with GRB\,230307A as discussed in Section\,\ref{sec:grb230307a}. Initial localizations sufficient for narrow-field follow-up allowed for recovery of a candidate counterpart, which was confirmed in part by the higher latency data providing a particularly precise localization. 

For the study of magnetars, the most stringent localization precision requirement arises for associating extragalactic MGFs to their host galaxies, where local events require precision of $\sim$10-100\,arcminute$^2$ and distant events require precision approaching a few arcseconds. Association of normal SGR short bursts to active magnetars requires order $\sim$1\,deg localizations. New magnetars can be found by stacking localizations of several short bursts, historically achieving accuracy on the order of $100$\,arcsecond$^2$ \citep{1999ApJ...523L..37H}.

\subsection{Alert Latency}\label{subsec:alert_lat}
The ideal worldwide system for transient astronomy would be sensitive all-sky coverage at all wavelengths and messengers. Barring that, the best-case scenario is immediate reporting of a precise localization from the first detectable signal(s) of a given source class. For all types of GRBs, SGR/FRB and magnetar giant flares, and jetted TDEs, this is usually the keV-MeV signal. In the pursuit of multiple diagnostics from the same event, observing the immediate aftermath is absolutely critical to fully understand these sources and unlock multimessenger science. The definition of precise localization in this subsection refers to $<10'$, sufficient for follow-up with a single pointing in most facilities. Achieving this accuracy can be done directly from some high energy monitors, or in concert with high energy monitors and follow-up tiling, as described above. Various telescopes across the EM spectrum can observe with a wider field of view; see Section\,\ref{sec:sources} for specific details. The results possible with a given timescale are summarized in Table\,\ref{tab:req_alert_latency}.

\begin{table}
\begin{center}
\begin{tabular}{|c| l | c |} 
 \hline
 Alert Latency & Corresponding Result & Sections \\ 
 \hline
 $>$10,000,000\,s & Origin of short GRBs, MGF candidate identification & \ref{sec:eMGF} \\
 & Origin and physical mechanisms of FRBs & \ref{sec:magnetar_frb} \\
 & MGF QPOs and NS equation of state & \ref{sec:mgf_ns_eos} \\
 & Sources of GWs & \ref{sec:sgr_gw} \ref{sec:collapsar_gws} \\
 & Discovery of new magnetars & \ref{sec:new_magnetar} \\ 
 & Magnetar formation channels, properties, and burst physics & \ref{sec:magnetar_formation} \ref{sec:magentar_properties} \ref{sec:magnetar_burst_physics} \\
 & Determination of SGRB progenitor fractions & \ref{sec:sgrb_progenitors} \\
 & GRB classification of GW sources & \ref{sec:gw_classification_mergers} \\
 & Speed of gravity measures & \ref{sec:fun} \\
 & Determination of GRB counterpart to orphan afterglow, dirty fireballs & \ref{sec:collapsar_et_al} \ref{sec:dirty_fireballs} \\
 & Origin of neutrinos, ultra-high energy cosmic rays & \ref{sec:neutrino_et_al} \\ 
 1,000,000\,s & Guide fast radio burst searches of active Galactic magnetars & \ref{sec:magnetar_frb} \\
 & Capture rise of supernova & \ref{sec:collapsar} \\
 100,000\,s & Follow-up classification of long GRBs from mergers & \ref{sec:long_mergers} \ref{sec:wd_mergers} \\
  & Latest reliable recovery of afterglow, potential redshift determination, cosmology & \ref{sec:stellar_formation} \ref{sec:origin_elements} \ref{sec:cosmology} \\
  & Guide follow-up of externally-identified transients based on prompt GRB signal & \ref{sec:collapsar_et_al} \ref{sec:dirty_fireballs} \\
  & Capture rise of red kilonova & \ref{sec:bns_classes} \\
 10,000 & Key diagnostic information on relativistic jets & \ref{sec:merger_jets} \\
  & X-ray recovery of plateau emission in afterglow & \ref{sec:plateau} \\
  & Tests of gravity parity violation & \ref{sec:fun} \\
  & Follow-up observations for VHE emission & \ref{sec:vhe} \\
  & Capture rise of blue kilonova & \ref{sec:early_uv} \\
 1000\,s & X-ray observations of fading tail after Galactic MGFs & \ref{sec:magnetars} \\
  & Discrimination of origin of early UV emission in mergers & \ref{sec:early_uv} \\
  & Observation of prompt phase of ultra-long GRBs & \ref{sec:ultra-long} \\
  & Blandford-Znajek test via afterglow polarization observations & \ref{sec:polarization} \\
 100\,s & Multiwavelength characterization of BNS merger classes and associated science & \ref{sec:bns_classes} \\
  & Critical early observations of EM-bright NSBH mergers & \ref{sec:nsbh_classes} \\ 
  & Prioritized follow-up based on GW merger classification & \ref{sec:gw_classification_mergers} \\
 &  X-ray observations of fading tail after extragalactic MGFs & \ref{sec:magnetars} \\
  & X-ray recovery of merger GRB extended emission & \ref{sec:plateau} \\
  & Full tests of dense matter, origin of heavy elements & \ref{sec:dense_matter} \\
  10\,s & Recovery of higher radio frequency (low dispersion measure) precursors & \ref{sec:prompt_radio} \\
 \hline
\end{tabular}
\caption{Summary of the science possible with various alert latency thresholds. Implicit in these timescales are the time to recover the precise position for narrow-field telescopes to use, which may add delays for poorly localized prompt GRB detections. Complete understanding of these transients requires reporting on the timescale of 10-100\,s, though some key science can be done with high latency data access.}\label{tab:req_alert_latency}
\end{center}
\end{table}

The fastest realistic latency is 10\,s, commonly achieved by Swift-BAT. This would allow for the recovery of effectively all signals discussed in Section\,\ref{sec:sources}, given that capable follow-up telescopes are ready. After $\sim$100\,s the radio precursor signature in NS mergers may be lost, as are observations of the early plateau decays (though some plateaus are recoverable for $10^5$\,s). Also after 100\,s the chances to observe prompt emission at other wavelengths begins to rapidly decrease out to $\sim$10,000\,s. After 1000\,s the chance to recover fading, pulsating X-ray tails from extragalactic MGFs with current facilities becomes extremely unlikely. On a similar timescale, the prospects of discriminating some of the potential sources of the early UV emission in NS mergers begin to fade. 

A latency of 10,000\,s results in the loss of the brightest portion of afterglow, making recovery of VHE emission and early polarimetric observations of the afterglow less likely to succeed. Observations at this time are still likely recover follow-up signals, capture the rise of redder kilonova emission, and to observe the pre-break timescales of the jet. After 100,000\,s (i.e. about a day), each of these three are less likely but still possible, but beyond 1,000,000\,s these successful observations become quite unlikely. High latency reporting is still valuable for the purposes of confirming or rejecting candidate counterparts found in other wavelengths and messengers.

Reporting latency is generally less critical for the monitoring of Galactic magnetars. It is necessary to inform the radio community when specific magnetars are active, but high-latency matching of SGR and FRB flares is sufficient, provided other FRB counterparts are not identified. The exception is the detection of magnetar giant flares, as noted, with possible recovery of the fading tail in X-rays if observations begin within minutes of an extragalactic event and within somewhat longer timescales for Galactic events. 

\subsection{Field of View, Livetime, and Contiguous Viewing Intervals}\label{subsec:fov_lt_cvi}
The study of nearby events presents critical diagnostics to provide leaps in the understanding of a given source class. The detection of MeV neutrinos and nuclear gamma-rays from SN\,1987A set our current understanding of the convective engine paradigm in typical CCSN explosions. The detection of SN\,1998bw following GRB\,980425 proved the theoretically predicted collapsar model for GRBs and supernova explosions. The detection of the off-axis GRB\,170817A with GW170817 and AT2017gfo led to breakthroughs in several fields of physics and still sets our standard for the study of NS mergers in astrophysics. The association of an SGR\,1935+2154 X-ray short burst to bright $\sim$ms radio flashes proved that at least some cosmological FRBs originate from magnetars. The limits from the IceCube non-detection of GRB\,221009A are more informative on prompt GRB models than the stacking of several thousand bursts. This is an effect of a multi-diagnostic Malmquist bias, where the nearest maximum detection distance determines the rate of events that can have complete observations. Whenever possible, a key component of the transient and multimessenger ecosystem should approach complete and uniform coverage of the sky. This is approximately met by IceCube and should be met by the IGWN network beginning in $\sim$2027. 

\begin{table}
\begin{center}
\begin{tabular}{|l| c | c | c | c |} 
 \hline
 Sources & Field of View & Livetime & Contiguous Interval & Sections \\
 \hline
 Prompt GRBs & $\sim$100\% & $\sim$100\% & & \ref{sec:magnetars} \ref{sec:mergers} \ref{sec:collapsar} \\
 GWs, FRBs, High-energy neutrinos & $\sim$100\% & $\sim$100\% & & \ref{sec:sgr_gw} \ref{sec:gw_classification_mergers} \ref{sec:collapsar_gws} \ref{sec:magnetar_frb} \ref{sec:neutrino_et_al} \\
 MeV Neutrinos, Shock Breakout & $\sim$100\% & $\sim$100\% & & \ref{sec:collapsar} \ref{sec:collapsar_et_al} \\
 Relativistic supernovae & $\sim$80\% (Night Sky) & $\sim$100\% & $\sim$1\,week & \ref{sec:collapsar_et_al} \\ 
 Orphan Afterglows, LFBOTs & $\sim$80\% (Night Sky) & $\sim$100\% & $\sim$1\,week & \ref{sec:dirty_fireballs} \\ 
 Unknown Unknowns & 100\% & $\sim$100\% & $\sim$1\,week & \ref{sec:misc}\\
 \hline
\end{tabular}
\caption{The Field of View, Livetime, and Contiguous observing intervals necessary for various sources of interest for GRB monitors.}\label{tab:req_fov_et_al}
\end{center}
\end{table}

The use of high-energy monitors in multi-diagnostic astronomy are necessary for the vast majority of future astrophysical advances, and all-sky field of view is often more important than depth for a given effective area of background-dominated instruments, as summarized in Table\,\ref{tab:req_fov_et_al}. This is because nearly all other diagnostics have a more limited detection horizon. NS mergers can only be recovered by current GW interferometers to a few hundred megaparsecs, to an even shorter distance by current UVOIR surveys, and to a smaller volume when observing afterglows than when observing the prompt phase. This effect for GW-GRB joint detections was shown directly in the NASA GW-EM Task Report prior to the Astro2020 Decadal \citep{racusin2019nasa}. For collapsars, low-luminosity GRBs are detectable to only a tiny fraction of the observable volume of typical long GRBs. VHE observations are limited in distance by extinction due to pair production with the extragalactic background light. The first transient neutrino sources will almost certainly be (cosmologically) local. In these cases, all-sky sensitivity for the gamma-ray transients is favored over depth. Exceptions include any hope of recovery of SGR X-ray flares around extragalactic FRBs (perhaps possible with pointed X-ray telescopes) and high-redshift GRBs.

Historically, discovery of relativistic transients was the domain of high-energy monitors. As the survey capabilities of other wavelengths and messengers advanced, they have begun to play a leading role in the discovery of these events. Because the prompt gamma-ray signatures are reasonable contemporaneous or precede other signals, archival coverage of the full event is necessary to make strong statements in the case of sensitive non-detection and to maximize the chance of signal recovery.

For short-duration, well-localized transients, the chance that a GRB monitor will have coverage is determined by its instantaneous field of view multiplied by its livetime. An example of this is observation of a Swift-BAT detected GRB by GBM. For spatially extended localizations, this fraction decreases further when the field of view is under 100\%. For example, the long arcs of two-interferometer IGWN localizations generally precludes total coverage by individual low Earth orbit satellites, driving the need for sensitive all-sky instantaneous field of view. This, and livetime, is driven by the need to provide complete coverage for neutrino events, FRBs, external prompt GRB detections, and other short-duration transients.

For externally identified events with imprecise start times, this fraction decreases even further when the livetime is under 100\%. The lack of sufficiently long contiguous observing intervals prevents direct statements on the gamma-ray emission from individual relativistic transients identified by optical surveys that have start time uncertainties on the order of $\sim$1\,day. This may prevent significant statements on particularly rare events. Additionally, it limits the population comparison of a high-energy-identified and optically-identified transients of the same source class (i.e. collapsar jets, relativistic supernova). With $N$-day fixed contiguous viewing intervals and 1\,day as the representative start time uncertainty, the effective livetime scales as $(N-1)/(N+1)$, i.e. a 3-day window has full coverage of only $\sim$50\% of events and a 7-day window of $\sim$75\%. This an additional dimension to the observing fraction requirement, specific to longer-duration transients. Thus, the development of ground-based wide-field survey telescopes drives the need for long contiguous viewing intervals in GRB monitors.

Similar needs will be needed as the capability to discover transients occurs in new wavelengths and messengers. As one example we point to the millimeter and sub-millimeter wavelengths where orphan afterglows can be found \citep{whitehorn2016millimeter}. The discovery space here is emphasized by the Decadal recommendation for time-domain capability in CMB-S4. 

True 100\% all-sky coverage and livetime is ideal to have in prompt high- energy monitors. This need is generally met by IPN, albeit to a shallower sensitivity than some of the more sensitive low Earth orbit satellites. The best individual instrument, in this regard, is Konus-Wind, which has all-sky coverage, near 100\% livetime, and is more sensitive than the more distant IPN spacecraft. Achieving these properties is effectively impossible to meet with satellites in low Earth orbit, unless there is a fleet. For a mission dedicated to detecting transients that have complete multiwavelength observations, it could be designed to focus on the $\sim$80\% of the sky visible at night on Earth, as most ground-based facilities cannot observe near the Sun.

There is one particularly notable exception to the need for a massive field of view, which is the search for high redshift GRBs. Identifying a sample of these events requires great sensitivity to recover them as far back in cosmic time as they occur. This sensitivity is likely more easily achieved with a narrower field of view.

\subsection{Background Stability}\label{subsec:bkgd_stab}
A related requirement is background stability, referring to the range of time over which background is stable. Stability over $\sim$100\,s is key to identification of extended emission following short GRBs, key to much of the science possible with NS mergers. This requirement can be met with inertially pointed instruments in low Earth orbit as demonstrated by the greater recovery of extended emission in BAT and BATSE and lower recovery rate in GBM. To provide a definitive answer on the overlap of traditional and ultra-long GRBs, and to prompt identify ultra-long GRBs while their emission is still observable, background stability for $\gtrsim$6\,hours is necessary. While Swift-BAT has identified some ultra-long GRBs over multiple orbits, the orbital timescale imprints an effective cut in the timescales. For the study of ultra-long GRBs having multiple instruments with sufficient sensitivity and background stability is key for confirming and localizing these bursts.

\begin{table}
\begin{center}
\begin{tabular}{|c| l | c |} 
 \hline
 Background Stability Interval & Source Type & Sections \\
 \hline
 10\,s & SGR flares and FRB counterparts, short GRBs & \ref{sec:magnetars}, \ref{sec:mergers} \\
 100\,s & Short GRBs with extended emission & \ref{sec:plateau} \\
 1,000\,s & Long GRBs & \ref{sec:collapsar} \\
 10,000\,s & Ultra-long GRBs & \ref{sec:ultra-long} \\
 100,000\,s & TDEs, longest GRBs & \ref{sec:misc}, \ref{sec:ultra-long} \\
 \hline
\end{tabular}
\caption{Brief summary of the background stability necessary for various source classes.}\label{tab:req_background}
\end{center}
\end{table}

\subsection{Timing Capabilities}\label{subsec:time_cap}
Absolute timing refers to the preciseness of the data compared to a given time standard, generally UTC, with accuracy limited by the capability of the onboard clock. For scientific purposes, absolute timing of $\sim$1\,ms is sufficient for nearly all of the discussed outcomes, with the comparison of the emission times of SGR X-ray flares and FRBs imposing the most stringent case requirement. This capability is lost at an absolute timing precision of $\sim$10\,ms. A precision of $\sim$100\,ms is an important limitation on measuring the speed of gravity and LIV. Achieving $\sim$1\,ms timing precision for low Earth orbit satellites is easy, given the capability to update onboard clocks with pings from the Global Positioning System (GPS). It is more difficult for satellites beyond the reach of GPS, but the adoption of atomic clocks and/or tiny X-ray telescopes for pulsar timing and deep space navigation \citep{2014HEAD...1430208R,7500838,2017arXiv171108507R} should prove revolutionary in this regard. Additional approaches to more precise absolute timing would also prove beneficial for these studies. The IPN would benefit from the multi-decade development of high-precision optical clocks in space for use in fundamental physics studies prioritized in the \textit{Thriving in Space: Ensuring the Future of Biological and Physical Sciences Research: A Decadal Survey for 2023-2032}.

\begin{table}
\begin{center}
\begin{tabular}{|c| l | c |} 
 \hline
 Relative Timing & Corresponding Result & Sections \\ 
 \hline
 10\,$\mu$s & Measure shortest duration pulses observed in SGR flares & \ref{sec:magnetar_burst_physics} \\
 100\,$\mu$s & Measure shortest duration pulses observed in cosmological GRBs & \\
  & Search for QPOs in the range of interest for neutron stars & \ref{sec:magnetar_burst_physics} \\
 \hline
 \hline
 Absolute Timing & Corresponding Result & Sections \\ 
 \hline
  100\,ms & Subdominant limitation to most speed of gravity measures & \ref{sec:fun} \\
 1\,ms & Enable minimal (statistical) timing annuli widths & \\
  & Comparison of FRB and SGR arrival time, physical mechanisms & \ref{sec:magnetar_frb} \\
 \hline
\end{tabular}
\caption{Brief summary of the relative timing and absolute timing requirements, with relevant sections.}\label{tab:req_timing}
\end{center}
\end{table}

Relative timing precision sets how accurately the offset between specific events can be measured. There are two main types of temporal data in high energy monitors, time-tagged event (TTE) and binned data. TTE data is an array of a time and energy channel for each event registered, and the timing precision is limited by the instrument and electronics. Binned data are time-series spectral histograms, with predefined temporal resolution, which is nearly always larger than the native timing precision. Most onboard clocks undergo a slow drift away from a reference time, meaning relative timing is often far more precise than absolute timing. 

An effective temporal resolution of 10\,$\mu$s is sufficient for any high-energy monitor. The shortest pulses ever observed have durations of 70\,$\mu$s and 100\,$\mu$s \citep{bhat1992evidence,roberts2021rapid}, which may be lost for instruments with only 100\,$\mu$s timing. This relative timing precision allows for QPO searches up to $\sim$10\,kHz, beyond the range of most NS oscillation eigenmodes. These searches will miss much of the QPO frequency range of interest for these modes at 10\,ms relative timing precision. At 10\,ms the complete searches for SGR flares will be negatively affected. At 100\,ms they will be severely limited, and there will be significant losses in the detection of cosmological short GRBs.

A key example of the importance of temporal resolution is the sub-threshold searches for SGR and short GRBs in Fermi-GBM and Swift-BAT data. With the planned continuous binned data resolutions of these instruments, searches for untriggered SGR flares would be impossible, and these provide a more complete picture on the waiting time and energetics distributions of these events. Many short GRBs would also be lost, preventing associations between GRBs and GW signals. The downlink of continuous TTE data for Fermi-GBM and the saving of Swift-BAT TTE data around externally identified times of interest by GUANO \citep{tohuvavohu2020gamma} has allowed substantially more sensitive searches for, and better localizations of, these types of events. This method of requested TTE is an option that should be explored in all future GRB monitors as the flexibility and resolution of TTE data is a critical asset for TDAMM studies. 

Timing capabilities are intimately tied to the operation of the IPN. The instrument with the lowest absolute timing resolution and relative temporal resolution will dominate the systematic uncertainty. Extragalactic MGFs can, in principle, be timed to sub-millisecond accuracy, setting a floor on this requirement. A level of $\sim$1-10\,ms is likely sufficient for the majority of bright GRBs. There are two hurdles implicit to IPN operation in this regard. The first is the relatively inaccurate absolute timing precision of distant spacecraft. In fact, using the position of known GRBs and inverting the IPN calculation is used to map the timing accuracy of planetary spacecraft. Second, the limited bandwidth through the Deep Space Network (DSN), and similar communication solutions, places a floor on the temporal resolution of the binned data, and has precluded unbinned data. These can be ameliorated in future instruments, as discussed in Section\,\ref{sec:actions}.

TTE data has a critical use in the IPN. Timing annuli are generally derived through a cross-correlation function comparison of two datasets. The ability to analyze data are arbitrarily small timescales ensures minimal timing annuli widths. For example the Konus data has a 2\,ms binning, but the time offset between GBM and Konus for GRB\,200415A is 1.3\,ms. There should always be LEO satellites with TTE data. TTE could be recovered from distant spacecraft with advanced communications solutions, or through limited TTE around intervals of interested identified by an on-board trigger. 

$\sim$1\,ms timing precision is necessary for joint SGR-FRB studies and allows for more accurate triangulation of SGR X-ray flares. This drives both absolute timing precision as well as minimum effective temporal resolution.

\subsection{Energy Range}\label{subsec:energy_range}
Standard GRBs typically have the highest signal-to-noise ratio over the $\sim$50-300\,keV energy range. Broadening this energy range to lower values brings detection of additional relativistic transient classes. Broadening the range in either direction brings greater characterization of detected transients. These are summarized in Table\,\ref{tab:req_energy_range}.

\begin{table}
\begin{center}
\begin{tabular}{|c| l | c |} 
 \hline
 Energy Range & Corresponding Result & Sections \\ 
 \hline
 50--300\,keV & Detection and localization & \\
 \hline
 \hline
 High Energy Threshold &   &  \\ 
 \hline
 1\,MeV & Constraint on peak energy, total energetics for long GRBs & \ref{sec:prompt_mechanism} \\
 10\,MeV & Constraint on peak energy, total energetics for short GRBs and MGFs & \ref{sec:prompt_mechanism} \\
  & Evidence for additional spectral components in prompt GRB emission & \ref{sec:prompt_mechanism} \\
  1\,GeV & New understanding of additional spectral prompt components & \ref{sec:prompt_mechanism} \\
 \hline
 \hline
 Low Energy Threshold &   &  \\ 
 \hline
 10\,keV & Increase in SGR flare detections, FRB counterparts & \\ 
  & Identification of short GRBs with extended emission & \ref{sec:plateau} \\
  & Detection of low-luminosity GRBs & \ref{sec:collapsar_et_al} \\
  & Identification of multiple prompt GRB spectral components & \ref{sec:prompt_mechanism} \\ 
 1\,keV & Insight on prompt GRB emission mechanism & \ref{sec:prompt_mechanism} \\
  & Detection of X-ray plateaus & \ref{sec:plateau} \\
  & Detection Shock breakout of relativistic supernovae & \ref{sec:collapsar_et_al} \\
  & Detection of X-ray flashes & \ref{sec:collapsar_et_al} \\
  0.1\,keV & Detection of shock breakout of most supernovae, full mapping of relativistic transients & \ref{sec:collapsar_et_al} \\
 \hline
\end{tabular}
\caption{Corresponding results possible with specific low and high energy instrument thresholds. Each column implicitly assumes sufficient sensitivity for a given topic.}\label{tab:req_energy_range}
\end{center}
\end{table}

Observing at higher energies improves the spectral characterization. Understanding the spectrum and energetics is vastly improved when spectral curvature can be measured. With a high-energy threshold of 1\,MeV curvature will be measured for most collapsar GRBs. At 10\,MeV it will be measured for the majority of all GRBs and MGFs, although the curvature has been measured above 10\,MeV in some rare cases. Extending the energy range higher is also important to seek potential additional spectral components, in particular an extra power-law component that is likely indicative of afterglow radiation. Sensitivity up to 20\,GeV would match the expected low-energy threshold of CTA and thus provide complete spectral coverage to the VHE regime. This complete spectral coverage would have the ability to determine the spectral turnover from base synchrotron to inverse Compton humps. 

Decreasing the low-energy threshold is also crucial for characterization of GRBs. Sensitivity to $\sim$10\,keV allows recovery of the observed low-energy excess in some bright GRB spectra, the origin of which is still unknown. It also enables recovery of extended emission and mapping of the low-energy quasi-thermal tails seen after GRBs 170817A and 150101B \citep{goldstein2017ordinary,burns2018fermi}. Extension down to 0.1--1.0\,keV would enable routine measurement of three separate regimes in the prompt synchrotron interpretation, increase the recovery rate and insight on extended emission, and increase the recovery rate of the cooling of thermal tails and shock breakout emission.

Additionally, the low-energy threshold increases the capability of the instrument to recover additional transient classes. If the threshold is at 50\,keV, the majority of SGR short bursts are missed, while many are recovered with a threshold of 10\,keV. Given the peak flux-temperature connection for SGRs, pushing to even lower values has limited return. A 10\,keV threshold allows recovery of many low-luminosity GRBs, with near-complete sampling if a 1\,keV threshold can be achieved. This threshold would also allow recovery of relativistic supernova and X-ray flashes, likely helping to understand the exotic zoo of relativistic transients. Pushing down to $\sim0.1$\,keV would also capture shock breakout from normal CCSNe, providing direct insight into most types of massive stars at their end, fulfilling an additional key science goal of the Astro2020 Decadal. These capabilities would also unlock identification of X-ray bright tidal disruption events and undoubtedly uncover new transient classes. 

Fermi GBM and LAT combine to provide unparalleled broadband observations of GRBs. The GBM low-energy threshold of 8\,keV has allowed for the identification of structure in the low energy end of GRB spectra ($\sim$10s of keV) which may be evidence for an additional thermal component an an extra break in the broader curvature. The high energy response from Fermi shows an expected turnover below the LAT energy range in several GRBs. Together they have shown evidence for the onset of a power-law component during the prompt phase, likely attributable to the onset of external shock. Pushing to even lower energies and covering the gap of $\sim$10--50\,MeV promise ever further advances in understanding the prompt GRB components.

\subsection{Sensitivity}\label{subsec:sensitivity}
The sensitivity of high energy monitors drives both the number of events they will observe and the statistics available for detections of a given brightness. Over typical GRB energy ranges, the order of magnitude sensitivity of Fermi-GBM and Swift-BAT triggers is $\sim1\times10^{-7}$\,erg\,s$^{-1}$\,cm$^{-2}$. BATSE triggers and Swift-BAT subthreshold triggers reach to a few$\times10^{-8}$\,erg\,s$^{-1}$\,cm$^{-2}$. An approximate sensitivity for the distant instruments of the IPN (i.e. beyond Konus) is $\sim1\times10^{-6}$\,erg\,s$^{-1}$\,cm$^{-2}$.

Bright bursts are easily recoverable by the IPN instruments, but the IPN sensitivity strongly limits scientific return. The non-detection of GRB\,170817A by any spacecraft more distant than INTEGRAL demonstrates that GW-GRB joint detections require deeper sensitivities. The current all-sky coverage of optically-identified afterglows can exclude most of the viable parameter space for prompt GRB counterparts, but it is not sufficient to fully exclude such a signal, preventing direct inferences on the potential existence of dirty fireballs. 

The deeper sensitivities of GBM and BAT enable far higher detection rates and recovery of events far deeper into the universe. They also allow recovery of far greater numbers of magnetar bursts, though this is also enabled by high temporal resolution data. Neither instrument, even with deep subthreshold searches, is capable of recovering GRB\,170817A beyond 100\,Mpc, while IGWN can already recover BNS mergers out to $\sim$300\,Mpc, a $\sim$30$\times$ larger volume. The deeper BATSE sensitivity resulted in a measurement of the turnover in the cumulative fluence distribution of long GRBs, proving the cosmological origin of GRBs. For temporal and spectral studies, the BATSE data is often far superior than modern instruments (though it has a narrower energy range than GBM). Achieving $\sim1\times10^{-8}$\,erg\,s$^{-1}$\,cm$^{-2}$ with modern flight software detection algorithms and TTE data would enable breakthroughs. 

Increases in sensitivity well beyond BAT and GBM are key to success to observing high redshift objects as well enhanced studies of magnetars in the local universe. However, we emphasize that the characteristic sensitivity values quoted here do not apply to X-ray transients, which generally have lower absolute values as they are integrated over a narrower energy range and we are quoting energy (not photon) fluxes.

\subsection{Maximum Photon Rate}\label{subsec:max_ph_rate}
The most stringent requirement on a maximal photon rate comes from a potential energetic Galactic MGF which, in the most extreme scenario, may exceed $1\times10^8$\,photons\,s$^{-1}$\,cm$^{-2}$, although the SGR\,1806 event had a peak flux of 1.5$\times10^7$\,photons\,s$^{-1}$\,cm$^{-2}$ \citep{frederiks2007giant} and the other two nearby MGFs were more than an order of magnitude less luminous. The highest rates of extragalactic MGFs, normal Galactic SGR short bursts, and the BOAT would be perfectly measured if an instrument could handle $1\times10^4$\,photons\,s$^{-1}$\,cm$^{-2}$; roughly one GRB per year exceeds $\sim1\times10^3$\,photons\,s$^{-1}$\,cm$^{-2}$.

The maximum photon rate a given detector can handle is determined by the instrument size and various aspects of its readout implementation. For example, it is far easier to saturate BATSE than a smallsat with much smaller detectors. It is easier to saturate slower scintillators than fast scintillators. The electronics readout speed is also important. When the rate limit is exceeded there are various instrumental effects that can occur including data losses, data gaps, and spectral distortion. These can sometimes be modeled and corrected, but this process is often imperfect and model dependent. Achieving full, unaffected data of a Galactic giant flare likely requires a tiny dedicated detector built for this purpose. For all other purposes, large instruments can likely make clean observations, but there is a cost to accommodating sufficiently high photon rates; however, exceptionally luminous and energetic events often provide greater physical insights, strongly motivating this additional cost. 

\subsection{Spectral Resolution}\label{subsec:spec_res}
High-energy transients usually have smooth continuum spectra owing to their origin from incoherent radiative processes in relativistic plasmas. Instrument energy resolution is typically 10\%--50\% over $\sim$50--300\,keV, which has proven sufficient for nearly all purposes. For future study of physically motivated models or possible lines in prompt GRB emission \citep{ravasio2023bright} higher-energy resolution may be beneficial. This will be informed by the forthcoming COSI mission. We add that this requires spectral readout into a sufficient number of energy channels in order to maximize sensitivity to transients and enable spectral analysis. If a high-energy monitor low-energy threshold were extended down to $\sim$1\,keV, then spectral resolution capable of resolving lines, e.g. iron lines, would be scientifically well motivated.

High spectral resolution benefits constraints on fundamental physics. For instance, constraining populations of low-mass black holes for dark matter by lensing of fast transients (such as MGFs) potentially requires higher spectral resolution $\Delta E/E \ll 10\%$ \citep{2018JCAP...12..005K}.

\subsection{Polarization}\label{subsec:polarization}
The study of polarization in prompt GRBs is still relatively new. With properly calibrated instruments, scientific advancements can be made with a minimum detectable polarization down to 10\% for bright bursts. Recent works have shown that tests of specific prompt GRB models are best done with time-resolved spectropolarimetric analysis. This requires either studying only the very brightest bursts or vastly increased sensitivity to polarization than prior instruments, e.g. POLAR. Precise requirements will be more easily defined after continued advancements in theory and simulation and with respect to the specific specialized instruments designed for this purpose.

Magnetar bursts at all energies (but particularly above 50\,keV) are expected to be highly polarized. Models of such bursts are at an early stage but forthcoming.

\section{Actionable Items for Missions and Instruments}\label{sec:actions}
Earlier sections detailed the scientific potential achievable through high-energy monitor observations of transient events, as well as the requirements needed to reach this potential. Several actions could be taken by NASA to significantly augment this multi-diagnostic, interdisciplinary science. Some specific actions are summarized below. 

We emphasize that this report has implicitly assumed that space missions will be capable of downlinking data and uplinking new commands with the typical latencies currently in use today. The TDAMM Communications SAG will investigate the space-based communications needs for broader science and instruments than are considered here. We comment directly on how alterations in the use of DSN may improve IPN operations below. 


\subsection{Active Missions}
The Astrophysics Advisory Committee (APAC) recently recommended NASA's Astrophysics Division (APD) to perform a reanalysis of their current portfolio to determine how to maximize TDAMM capabilities. Here we discuss some specific actionable items where NASA APD could enhance return of its own assets as well as assets in other Divisions, similar to its funding a TDAMM enhancement of the Near-Earth Object Surveyor mission. The key enhancements are all effectively different ways to get more data accessed as fast as possible. We note that Swift and Fermi have historic and planned improvements, per the last round of Senior Review, to continue enhancing the TDAMM science return of these workhorses. For example, the plan to lower the on-board triggering threshold of GBM for short GRBs would allow for great scientific returns in partnership with the IGWN.

Perhaps the most immediate example is the over-guide request from Swift for additional ground-station passes in the last round of Senior Review. This would reduce the latency of full data downlink from BAT (both standard and GUANO data), XRT, and UVOT. Identifying counterparts of GWs from any of these instruments $\sim$hours earlier would contribute to understanding of specific NS mergers, and allow the location information to be passed to telescopes worldwide that are ready to characterize these events. The lower latency of data would be generally useful in the study of all Swift time-domain sources. For example, analysis of the Swift UV grism observations of the recent supernova in M101 (the closest in a decade) were key to the decision to repoint Swift for additional observations and as guidance for other UV and  X-ray telescopes, but the existing system resulted in delays. The cost is minimal compared to the scientific gain. The choice of Senior Review to avoid decisions on TDAMM-related over-guides and lack of strategic TDAMM planning since then has overlooked this obvious improvement. Yet, O4 has already begun. This is perhaps the greatest immediate priority action NASA could take to benefit TDAMM.

Additional ground-station (or other downlink) passes could benefit additional NASA missions with respect to gamma-ray transients. This requires discussion with the relevant instrument teams to ensure it is possible and beneficial within the existing mission architecture. Fermi is a specific case where additional downlink passes would accelerate spectral and temporal analysis of on-board triggers and sub-threshold searches for additional events rapidly enough to guide follow-up observations of its core transient classes. 

One of the most important improvements would be additional downlinks for the distant spacecraft of the IPN. Today the IPN includes 11 gamma-ray detectors, of which four are beyond low Earth orbit. Additional types of detectors can be used for particularly bright events (on the order of one per year). The most distant gamma-ray detector alternates between the Mercury Gamma-ray and Neutron Spectrometer (MGNS) on-board BepiColombo Mercury Planet Orbiter, and the High Energy Neutron Detector (HEND) on-board Mars Odyssey. The recent switch of the MGNS data on BepiColombo, en route to Mercury, from 250\,ms to 50\,ms has enabled current IPN localizations to rival Swift-BAT positions. HEND on-board Mars Odyssey has 250\,ms resolution. Konus on-board the US Wind satellite has 2.944\,s background resolution and 2\,ms resolution for on-board triggers. Additional downlinks would support more rapid IPN localization in the multimessenger era and could possibly support higher temporal resolution. However, we emphasize that any change is not trivial and must be done in concert with interest from the instrument and mission teams. Potential issues include aging flight recorders, the necessity to maneuver the spacecraft for downlinks, and potential onboard processor or storage limitation. 

Beyond decreasing the latency for data to arrive on the ground, improvements can be made to data access. For Fermi-GBM, distribution of HEALPix maps that account for systematics of the ground localization would be beneficial for observations beginning under 15\,minutes, as would removal of Earth occulted positions from localization maps. Serializing the roboBA localization into the alert itself rather than waiting for the delay added by HEASARC would allow more rapid use of GBM data. Further, distribution of retraction Notices due to the astrophysical nature or misclassification of events would be beneficial to guide follow-up efforts. Some of these enhancements can be done outside of the instrument team pipelines and could be handled by the IPN (see Section\,\ref{sec:network}), as discussed in Section\,\ref{sec:network}.

Support for improving the robustness of existing IPN data pipelines and enhancing the information shared is key to automating the process and reducing ground-based contributions to alert latencies. Support for creating instrument responses for the purposes of GRB studies is necessary for inclusion in any multi-mission analysis.

\subsection{Forthcoming Missions}
There are numerous forthcoming missions of relevance. NASA APD has selected COSI as an upcoming Small Explorer mission, StarBurst as an upcoming Pioneer, and is funding additional technology demonstration missions through APRA. Space-based multi-band astronomical Variable Objects Monitor (SVOM) is a French-Chinese mission designed with a similar ideology as Swift. China is launching a series of GECAM satellites. Several gamma-ray spectrometers on planetary and heliophysics spacecraft are planned to be built and launched by American and Russian scientists over the next decade.

StarBurst will have several times the effective area of Fermi-GBM, launching as a ride share. StarBurst is using silicon photomultipliers, which will limit the overall instrument lifetime at high inclination orbits due to increased radiation exposure. COSI will have a dedicated mission launch and requires a $<2^\circ$ inclination orbit. Pairing StarBurst to the COSI launch would significantly enhance the scientific return of StarBurst by increasing observing livetime due to fewer passages through high rate particle background regions. It would also increase the operating lifetime of StarBurst by reducing the damage of the silicon photomultipliers. Additionally, increasing the number of ground contacts and bandwidth would enable the downlink of all StarBurst TTE data, instead of TTE limited to on-board triggers.

COSI is slated to launch in 2027, well-timed with the planned beginning of the IGWN O5 observing run. The onboard COSI trigger for GRBs will be initiated by count rate increases in the anticoincidence shield, which will trigger prompt downlink of a limited number of Compton events allowing for a $\sim$deg scale localization within an hour of the trigger. Enhancing the amount of data that can be rapidly downlinked by COSI would improve that initial localization. Enabling the downlink of the full dataset would result in final localizations at a much higher cadence, enhancing the transient science return of COSI. We also support the downlink of all possible data from the COSI shields and single-site events in the main instrument, regardless of likely event class or the likelihood of being recovered as Compton events. COSI will contribute to the IPN with events detected by its shields and will help characterize the IPN systematics through facilitating the recovery of arcsecond localizations. There may also be possible TDAMM enhancements for the COSI student-led BTO instrument which will broaden the energy coverage and can also contribute to IPN.

The Johns Hopkins Applied Physics Laboratory is building the Psyche Gamma-Ray and Neutron Spectrometer (GRNS), the Mars-moon Exploration with GAmma rays and NEutrons (MEGANE) instrument for the JAXA Martian Moons eXploration (MMX) mission, and the Dragonfly Gamma-ray and Neutron Spectrometer (DraGNS). All three spectrometers use germanium which must be cooled. DraGNS will utilize the temperature of the Titan atmosphere for passive cooling, preventing use for the IPN. 

The Psyche GRNS will contribute to the IPN for $\sim$6\,years while MEGANE on MMX should contribute for $\sim$2\,years. These instrument and mission operations have been adapted for use in the IPN. Each has an onboard trigger that will achieve 50\,ms resolution for a fixed time period. At Psyche's greatest distance of 3\,AU this allows a statistical timing annulus width of $\sim10''$. Additional TDAMM support from NASA could ensure the data is accessed as quickly, easily, and reliably as possible, and to support the generation of GRB response matrices.

The Ioffe Institute in Russia is launching gamma-ray spectrometers on three future spacecraft \citep{ulanov2019konus}. Konus-UF will consist of two detector units on the World Space Observatory Ultraviolet, providing all-sky coverage and possibly TTE data from a geosynchronous orbit. If so, it would be the first TTE data from beyond low Earth orbit. The twin InterhelioProbe \citep{kuznetsov2016sun} spacecraft will each contain one detector unit and will be among the few spacecraft to significantly diverge from the ecliptic, which will be valuable for IPN localizations. US contributions to these missions may bolster the IPN and meet the Decadal suggestion of supporting a TDAMM program through international partnership. 

As this report is being delivered, the review of the current Mission of Opportunity call is underway. Both missions, LEAP and MoonBEAM, are dedicated GRB monitors, with different focus areas. We support consideration of additional TDAMM enhancements should one or both instruments be selected.

\subsection{The Decadal-Recommended High Energy Monitor} 
The top priority sustaining activity recommended by the Astro2020 Decadal \textit{Pathways to Discovery in Astronomy and Astrophysics for the 2020s} \citep{national2021pathways} was: \enquote{NASA should establish a time-domain program to realize and sustain the necessary suite of space-based EM capabilities required to study transient and time-variable phenomena, and to follow-up multi-messenger events. This program should support the targeted development and launch of competed Explorer-scale or somewhat larger missions and missions of opportunity.} While much of TDAMM can be done from the ground, the wavelengths only observable from space are critical for the TDAMM ecosystem, as noted by the Decadal: \enquote{The most important of these are wide-field gamma-ray and X-ray monitoring, and rapid and flexible imaging and spectroscopic follow-up in the X-ray, ultraviolet (UV), and far-infrared (far-IR).} We also highlight the following quote to emphasize the introduction of the concept of a workhorse mission to NASA APD: \enquote{NASA’s workhorse hard X-ray and gamma ray transient facilities (Swift and Fermi, respectively) are aging and their longevity is uncertain. Higher 
sensitivity all-sky monitoring of the high-energy sky [...] is a critical part of our vision for the next decade in transient and multi-messenger 
astronomy.} Other NASA science divisions have sustaining capabilities that are ensured across generations of missions; for example, total and complete coverage of the solar flares from the Sun. The message is clear: NASA astrophysics should ensure continuous coverage of the sky with wide-field high-energy monitors. Furthermore, the total NASA TDAMM program was recommended a budget of \$500M-\$800M for use over the coming decade. 

Relativistic transients are a major component of time-domain and multimessenger astrophysics, containing the multi-diagnostic events of supernovae, collapsars, NS mergers, TDEs, FRBs, magnetars, and some signals whose origin is sill unknown. The science and corresponding requirements detailed in this document provide motivation and an outline for a strategic high-energy monitor. In the context of a field where every other wavelength is facing a major upgrade in TDAMM capabilities (e.g., the Decadal recommended next-generation Very Large Array, the Square Kilometer Array, CHORD, Roman, the Vera Rubin Observatory, ULTRASAT, Einstein Probe, and the Cherenkov Telescope Array) including simultaneous upgrades of ground-based GW interferometers and HEN telescopes, a TDAMM-focused high-energy monitor is the missing critical link in the full ecosystem. It is important to bolstering the return of NASA's investment in other missions and infrastructure.

NASA's current largest investment in a high-energy monitor is COSI. COSI will achieve $\sim$degree scale localizations of GRBs that will be reported within 1\,hour, meeting some goals outlined in Section \ref{sec:requirements} and the temporal and polarization requirements listed below. COSI's prime design driver; however, is nuclear astrophysics, and although it will revolutionize this field, COSI will not replace Fermi or Swift for TDAMM studies. As highlighted in the Decadal, another future high-energy monitor is SVOM, a French-Chinese mission being built with similar instrument designs and science goals as Swift, but with worse sensitivity and localization. SVOM is set to launch in 2024 and will provide localizations at $\sim10'$ precision for 50-60\,GRBs per year. Over the nominal 3\,year mission and potential 2\,year extension, SVOM will produce a sample of well-studied events, but its capabilities will not match that of Swift or Fermi. Additionally, as noted in Astro2020, the US has no significant involvement.

A number of nations have built, or are in the process of building, low-cost scintillator-based instruments to search for GRBs from GW sources. A veritable fleet will be operating soon. The largest of these are Glowbug and StarBurst, funded by NASA as a technology demonstration and Pioneer, respectively. These missions will both be more sensitive than Fermi-GBM, a key piece towards improving the number of GW-GRB detections, but have narrower energy ranges, face the same limitations of operating in LEO, and it is impossible to substantially lengthen mission lifetimes, which are $\sim$1\,year each, given the instrument design. Additionally, they are designed for one area of transient gamma-ray science. They will both provide new insight into the universe in partnership with GW detectors, but they are not Fermi or Swift replacements, which is not surprising given the orders of magnitude lower total cost.

Based on the requirements outlined herein, and the currently planned future of this field, there are some obvious missing existing capabilities. A high-energy monitor covering the majority of the sky and capable of $\sim$arcminute scale localizations, provided within $\sim$10\,s, from an instrument that continuously observes at a sufficient sensitivity would fill the missing gap in the transient ecosystem. The operational mode of the IPN precludes this possibility, given light travel time from distant spacecraft. Such a dedicated mission is within the capability of existing technology. If the instrument could achieve 10' scale localizations, a low-energy threshold down to $\sim$1\,keV and transient spectral sensitivity to several MeV, while maintaining background stability, it would additional bring the recovery of the initial shock breakout (necessary to map the properties of the progenitor star) and high-cadence monitoring of the X-ray sky (also highlighted in the Decadal). This seems feasible with developing technologies \citep[e.g.][]{chattopadhyay2018blackcat} and achievable as the TDAMM mission at the scale recommended by the Decadal. Such a mission would prove a critical partner to all other transient facilities including nearly the entire astrophysics fleet of NASA (and the rest of humanity) as well as the low and high energy neutrino facilities, low and high frequency GW interferometers, and the vast numbers of ground-based telescopes used in follow-up observations. 

Such a mission, through collaboration with worldwide partners, would unlock new knowledge on fundamental physics including gravity parity violation, dark matter, antimatter, the speed of gravity, and Lorentz Invariance violation. Some of these tests using astrophysics are orders of magnitude more sensitive than any other proposed method. It would provide the most precise measures on the behavior of dense matter by determining the maximum mass of NSs, the lifetime of meta-stable NSs, the radius of NSs, insight into the proto-NSs formed during core-collapse, and the compactness of NSs. The maximum mass of a NS is an asymptotic value, providing a uniquely constraining test of dense matter, and could be measured to 1\% precision \citep{margalit2019multi}, which is a factor of several better than non-asymptotic tests expected through currently active or selected missions. This mission would probe the origin of GRBs, FRBs, GWs, astrophysical neutrinos, low-luminosity GRBs, ultra-long GRBs, unusual supernovae and LFBOTs, every other relativistic transient class discovered with forthcoming surveys, and rare and unusual individual transients. It will provide insights into how jets form, whether black holes return power to the universe, insights into the physics of magnetars, and the transition regime between dense and condensed matter. It will support multiple diagnostics for identifying different source classes, standardizing NS mergers for precision cosmology throughout the universe, support proper inference on the origin of the elements, standardize CCSN to understand how they explode, and resolve the neutrino mass hierarchy. The all-sky field of view is critical to identify all nearby transients where the diagnostics with the most limited detection distances can still be recovered, a capability that has never existed with prompt and precise localizations. Implementing the requirements will provide significant advances in astrophysics, cosmology, gravity, and fundamental physics, and enable interdisciplinary work in nuclear, particle, plasma, and atomic physics. 

The level of support for space-based TDAMM missions recommended by the Astro2020 Decadal is absolutely necessary to unlock this magnificent breadth of science. While no mature mission concept is capable of these requirements, informed estimation accounting for the cost of fiducial detectors to achieve the requisite sensitivity and including the requirement for achieving at least a high Earth orbit suggest it may not fit within the Small Explorer mission budget cap. The Decadal also highlights the need for \enquote{...rapid and flexible imaging and spectroscopic follow-up in the X-ray, ultraviolet (UV), and far-infrared (far-IR)}, which are wavelengths only accessible by instruments in space. Key to the modern TDAMM system is the decoupling of discovery monitors from the follow-up telescopes onto separate spacecraft. This design principle implies the necessity of dissemination of high-energy alerts to the full community, which can route to low-latency commanding of the follow-up telescopes, as shown by Swift. This allows the high-energy monitors to maintain background stability and vastly reduces the moment of inertia of the narrow-field telescopes. Together the cost is likely to fall into the range of the $500-800M$ recommended in Astro2020.

If the above described monitor is sufficiently sensitive, it will also support high-redshift GRB science; however, this may not be technically feasible, and a dedicated $\sim$1\,sr field of view instrument may be required. It is critical for this instrument to overlap with JWST, allowing for follow-up characterization of the low-mass galaxy hosts in the early universe. In the GW 3G era, it will be critical to recover merger GRBs deep into the universe to build the precision Hubble diagram out to a redshift of a few. Thus, such a mission should be designed to launch before the end of JWST and to overlap with the future GW interferometers. The characterization of explosive transients by the previously described near-universe monitor will prove important to science of this far-universe mission.

\subsection{The Future of the InterPlanetary Network}
The IPN has utilized more than 50 instruments, and outlived 31 of them, providing true all-sky coverage for nearly 50\,years. It is the existing example of a workhorse network in NASA Astrophysics. Regardless of dedicated GRB missions, the use of all-sky monitoring with relatively precise localizations will always be needed. 

Note that precise localizations of GRBs cannot be routinely accomplished by a network of detectors entirely contained to low Earth orbit. A given annulus is described as (RA, Dec, radius, width). The radius is the angle $\theta$, defined relative to the vector joining the two spacecraft, calculated as $\cos(\theta)=c\delta T/D$ with $c$ the speed of light,  $\delta T$ and $D$ the arrival delay time and distance between the two spacecraft, respectively. The width is calculated as $c \sigma(\delta T)/D\sin(\theta)$ where $\sigma(\delta T)$ is the uncertainty on the delay time \citep{hurley1999ulysses}. The interplanetary baselines are of order $10^8$ km. LEO provides baselines  $\lesssim 10^4$\,km. While the temporal resolution of distant IPN spacecraft can limit timing annuli accuracy, GRB lightcurves are not delta functions and have a floor on the arrival time uncertainty determination. 

A representative 10\,ms arrival time uncertainty in LEO gives a best-case 1$\sigma$ annulus width of $\sim$10\,deg. Even for particularly bright bursts where 1\,ms timing is possible, this is still $\sim$1\,deg annulus width. Improved methodologies cannot avoid this fundamental fact. Additionally, small detectors will also generally result in poorer timing precision given the smaller effective area. Careful simulations of a network show that they will achieve degree-scale localizations similar to those provided by Fermi-GBM alone \citep[e.g.][]{thomas2023localisation, hurley2020gamma}. \citet{hurley2020gamma} also raises the issue that increasing the size of the network will decrease the average distance between viewing detectors, thereby limiting the spatial precision gain possible. 

What is required from LEO satellites is a deep sensitivity and TTE data, which is a critical complement to the more distant instruments in the IPN. The more distant instruments will launch predominantly on planetary and heliophysics spacecraft, and perhaps on astrophysics satellites that are more often venturing to the Sun-Earth L2. Often these instruments are enhanced beyond their primary purpose for use in the IPN: planetary missions use gamma-ray spectrometers to map elements on the surface of other bodies in the solar system and heliophysics missions can use them to study the X-ray and gamma-ray emission in solar flares. Enhancements included increasing the temporal resolution of all data and adding onboard GRB triggers (with a pre-trigger buffer) for higher temporal resolution during specific times. Additional enhancements that require very little modification to the existing systems could include prioritization of the GRB data to be downlinked first during scheduled communication passes and contribution of additional DSN time to allow for greater temporal resolution. 

Swift has demonstrated the recovery of TTE data around externally identified times of interest, e.g. GW triggers. An addition to missions that host contributing instruments to IPN would be large onboard storage, which would allow days worth of TTE data to be saved onboard and requested during specific times of interest. 

A more disruptive enhancement may be scheduling DSN downlinks with greater frequency. Given the large distances, only directional antennae are viable, requiring reorientation for communication during coast or acceleration times. This can result in downlink delays of $\sim$1\,week, which are substantial limitations. If DSN time can be scheduled and mission operations can accommodate this change, this would be a vast improvement in IPN operations. 

There are a few potential game changers identified in \textit{Origins, Worlds, and Life: A Decadal Strategy for Planetary Science and Astrobiology 2023-2032} for planetary science that would also enable capability leaps of the IPN. The first is the planned Deep Space Optical Communications experiment on Psyche, seeking the first demonstration of optical communications beyond the Moon. Success would mean potential bandwidth increases by 10-100 fold above the current radio solutions. This would greatly enhance the temporal resolution of distant IPN instruments and may additionally support more frequent contacts. The other development is the potential success of Starship and similar mega-rockets. These heavy launchers would alleviate the greatest limitation of launching distant spacecraft: mass, potentially allowing the launch of multiple planetary spacecraft with one rocket. This would allow for larger gamma-ray spectrometers for the study of element distributions on planetary bodies. Further, it may allow the return of dedicated GRB monitors on distant spacecraft which was routine in the field for decades but has not occurred with U.S. led missions since the launch of Ulysses more than 30\,years ago. Russia plans to contribute dedicated GRB monitors to three spacecraft, an area where the US could explore contribution of downlinks in support of the IPN and the Astro2020 Decadal suggestion of partnering on foreign missions for TDAMM.

The last major change is in the reduction of systematic uncertainty that has previously limited IPN precision. One advancement is the demonstration of precise atomic clocks in space and, in particular, the plan to launch one on the forthcoming VERITAS probe to Venus in 2028. Historically, the onboard clocks of distant IPN spacecraft have been verified or their drift quantified using the GRB instruments \citep[e.g.][]{hurley1994timing}. The precise absolute calibration is critical for IPN operation and key to knowledge of the true spacecraft position. Inclusion of atomic clocks can solve this problem. Alternatively, both timing and position could be enhanced with pulsar timing positioning systems \citep[e.g.][]{2014HEAD...1430208R,7500838,2017arXiv171108507R}. As discussed, the suggested clocks program in \textit{Thriving in Space: Ensuring the Future of Biological and Physical Sciences Research: A Decadal Survey for 2023-2032} would vastly exceed the precision needed for the IPN. The other improvement since Ulysses is the external verification of true source positions, which supports the  verification of IPN precision or modeling of an irreducible systematic uncertainty. The 90\% annulus width of $4''$ was derived from the collaboration between the GRB monitor on Ulysses, a distant heliophysics satellite, and BATSE, but this was limited by temporal resolution of these instruments and the systematics of the IPN at the time.


If optical communications, megarockets, and atomic clocks become pillars of future distant spacecraft, then the future of IPN could be scientifically fruitful, perhaps achieving arcsecond scale localizations. There are $\sim$20 distant spacecraft launched by NASA or partner space agencies. If even half of future distant spacecraft had dedicated GRB monitors, roughly $\sim$10 instruments would be available. Even if each has an average downlink latency of days, then a few will downlink data each day. This could support precise localizations with latency sufficient for successful follow-up of most sources. The addition of a GRB monitor to any of the future Venus probes would be beneficial. 

At a few AU, e.g. the distance achieved by Ulysses and expected for Psyche, TTE data would enable statistical localizations on the order of 0.1$''$ for particularly bright and variable bursts. 1$''$ precision is possible with $\sim$10\,ms temporal resolution. If the atomic clocks and systematics are carefully modeled, the resulting localizations are sufficiently precise that host galaxy association and redshift determination could be made without the need for successful follow-up recovery of the afterglow; only optical spectra follow-up would be desirable. This could produce a complete mapping of intrinsic energetics distributions and source evolution and complete the Hubble diagram alongside 3G GW interferometers. 

This range of enhancements, from non-disruptive to the intentional construction of a dedicated IPN, requires inter-division discussions at NASA. Elevation of IPN to a strategic initiative could be a necessity, as discussed in the next section. This could prioritize taking advantage of new capabilities and options, such as a dedicated monitor onboard Gateway enabling TTE data from $\sim$1\,lightsecond away. 

As an additional example, \textit{Origins, Worlds, and Life: A Decadal Strategy for Planetary Science and Astrobiology 2023-2032} recommend the Uranus Orbiter and Probe mission as the highest priority among new flagship missions. If $\sim$2\% of the total mission mass could be allocated to a dedicated GRB monitor and NASA APD funds the instrument, an on-board trigger, and the technology for precise timing, then sub-arcsecond timing annuli would be available to the IPN for nearly two decades ($\sim$13-15\,year cruise, $\sim$4.5\,year prime mission). This would be a foundation for a new era of the IPN. With an appropriate paired instrument (preferably outside of the ecliptic), host galaxies could be found without requiring an afterglow detection phase. Other opportunities exist including the Venus probes. Additionally opportunities may arise pending the outcome of the Heliophysics Decadal.

\section{Actionable Items for the Gamma-ray Transient Network}\label{sec:network}
The needs of the various communities interested in studying magnetars, compact mergers, and collapsars also lead to a number of requested products from the ideal gamma-ray transient network. We find the best path to creating these materials to be enhancing the operations of the IPN.

\subsection{Support for Community-led Strategic Initiatives}
Of critical importance to NASA's implementation of the TDAMM aspects of the Decadal is long-term, sustained funding in support of community-led strategic initiatives, software development, and formalized renewal of mission-oriented projects that are too small for a Senior Review. There is no general software development call, preventing NASA from tapping expertise in our broad community. These issues have precluded enhancement of the IPN due to limited, sporadic, and fractured support.

Kevin Hurley's operation of the IPN included half of the real-time localization and reporting, archiving of the GRB and SGR flare catalogs, maintenance of the entire system, and providing requested products to various institutions (e.g. catalogs to LIGO). In his later years, this was supported through Fermi and Swift GI proposals, proposed annually, providing a total support on the order of \$100\,k per year. This was both insufficient to support modern development of an improved system and the year-to-year funding nature required time spent on proposals that otherwise could have been used to deliver an obvious community project. Prior roles as participating scientists in planetary missions allowed fostered communication and support from the relevant instrument and mission teams, but the proposed work must be directly for the individual funding missions, resulting in bifurcated community products \citep[e.g.][]{2005NCimC..28..299H,2010ApJS..191..179H,2011ApJS..197...34H,2011ApJS..196....1H,2017ApJS..229...31H}.

Making the IPN, and the products described above, a community-led strategic-initiative supported by NASA with review on a reasonable cadence would enable far greater science with existing and forthcoming missions for a fraction of the cost of a new mission. Such a decision would also allow the IPN to act in an appropriate manner related to its importance in the field. For example, allowing the IPN to request NASA-led press directly.

The advancement of astronomy has necessitated ever greater group sizes. In many cases, the lone astronomer has been replaced by groups that have sometimes evolved into larger collaborations, and a few of those have coalesced into consortia. This change has been particularly prevalent in the area of TDAMM, requiring broad expertise to capture a full story of specific events. Perhaps the most visible is the formalization of IGWN from the LIGO Scientific Collaboration, Virgo Collaboration, and KAGRA Collaboration. This integration has greatly benefited these groups and the astronomical community as a whole. Similarly, the creation of the SNEWS from the global set of MeV neutrino telescopes will be critical to discover and characterize the next Galactic supernova. There are also initial discussions on potential integration of the HEN telescopes.

The gamma-ray transient detectors have not integrated to the same degree. The IPN and the GW-GRB Working Group are the two relevant multinational consortia. The IPN has operated as a university-led project within the US for decades. This setup has allowed for continued operation through the evolution of restrictions related to geopolitics, use of private data that is unlikely to be accessible outside of the IPN, and is a key example of community-led mission-oriented work. The GW-GRB Working Group is now more than a decade old and has been co-led by various individuals at universities, non-profits, and labs.

The integration of these groups is on-going. While leadership could rotate to individuals at NASA institutions, it is preferable to maintain the broad community-led nature of these consortia. There is a sufficient support mechanism to facilitate growth of these projects to meet the needs of the broader TDAMM community outlined in this report. NASA has supported key aspects related to these goals through the Internal Scientist Funding Model, but this method is not accessible for university-led projects. Given that the deliverables described herein would result in maximal TDAMM science return via gamma-ray scientists making otherwise publishable results immediately available to the community, long-term sustained support would act as a proper reward system for these scientists. Otherwise the individual benefit is maximized by restricting public release to ensure authorship on multiwavelength papers.

The purpose of this report is to gather the needs of the wider TDAMM community, and to ensure those needs are met. The specific products to maximize science with these instruments is listed in the next subsection. NASA has facilitated the technical capability for many of these projects through various grants and internal funding; however, to fully implement the community requests, support for a greater integration of the gamma-ray transient community is necessary. The specific implementation likely requires working through the existing consortia, which enable the sharing of otherwise private data.

\subsection{Products}
The analysis of data from multiple missions is generally limited to joint spectral analysis or localization via the IPN. The products listed here would greatly benefit the community, as previously noted in the Multimessenger Astronomy Science Analysis Group \citep{brandt2020physics}.

\subsubsection{Event-based Real-time Alert Stream}
The community has developed signal-based reporting to facilitate coordinated observations and analysis of specific events. This is most evident in the fixed naming conventions of supernovae, GRBs, FRBs, GWs, etc, which are generally distinct from the internal identifiers of each observing instrument. For example, GRB\,170817A is the event-based GRB name for the GBM trigger bn170817529. IGWN has created a single alert stream for GW alerts, allowing the community to receive the best available real-time information, especially spatial information, in a fixed format. SNEWS has similarly created this for temporal and spatial information of MeV detections of neutrinos.

No such stream exists for GRBs. This would be one important benefit the high-energy monitors could deliver to the community: providing a single access point to real-time GRB information to the full follow-up community. This would unify the schema of the data presentation, allowing a single script to be written to listen in to this alert stream rather than requiring work for each individual existing stream. The greatest benefit would be improved spatial constraints to aide follow-up efforts. We emphasize this stream would support equitable access to data in the field. Currently large institutions have an advantage in developing frameworks to handle the vast heterogeneous alert streams. A single access point allows institutions with less technical capability to access the full information from a field with a single script. The Unified Schema from GCN will be a great equalizer in this regard by facilitating ease of access to information from all contributing TDAMM fields.

As currently implemented, the community is losing key early information in rare and interesting events. This is perhaps best exemplified by GRB\,230307A (Section\,\ref{sec:grb230307a}). It was iterative localized by IPN as data became available, but limitations introduced by processing the real-time data and the manual steps required resulted in delaying the alerts. Despite this, the localization was tiled with both Swift, a space-base facility, and ULTRACAM, a large (3.5\,m) ground-based telescope, showing they are willing to tile even with particularly competitive facilities. This enabled the first late-time infrared spectrum of r-process nucleosynthesis, which will be the leading observation point to guide kilonova simulations and theory for a decade. What is missing is key data from the first few hours, which would illuminate the evolution of the separate afterglow and apparent kilonova emission. 


As it stands, for the $\sim85$\% of events not localized by Swift-BAT, the community is missing out on critical science. Prompt identification of the active states of magnetars are missing, and thus they are not observed with radio and X-ray telescopes for potential joint SGR-FRB flares. The prompt follow-up of potential extragalactic MGFs is also missing, which may allow inference on magnetars in other galaxies. Without improvements, the earliest EM signatures of GW-detected NS mergers, will be missed, which is key to advances in understanding dense matter, ultra-relativistic particle acceleration, fundamental physics, and the origin of the elements. As a specific example, recovery of the afterglow before the jet break is important to determining if gravity violates or preserves parity, which translates to tests on potential grand unified theories, dark matter, and the excess of matter over antimatter at the beginning of time. The community will miss opportunities to recover VHE detections of GRBs, which has thus far kept us in the low-count regime. As CTA begins observing, we will need this stream to be constructed and robust. 

Such a stream would maximize success of follow-up, minimize effective trials factors, and facilitate the further development of event-based reporting. A prime example of the latter is the automatic association of GBM and BAT GRB triggers to GW-detected events and the dissemination of joint localizations through the IGWN alert stream in an identical format to the GW-only localizations. This effort provides the full information for follow-up observations without any additional work on the follow-up community. Additional GRB triggers and joint GRB localization information could improve this automated multimessenger search. Other improvements like the construction of a joint GRB-neutrino search or plugging all GRB localizations into new infrastructure such as the use of Treasure Map \citep{treasuremap} would also be substantially easier to implement with a collated GRB stream.

As a critical example we highlight that the Vera C. Rubin Observatory is so powerful it will recover nearly all GRB afterglows (at least, those found following Swift-BAT detections) that occur within its field of regard. Each night will have color information to identify and isolate these candidates, with a 3 night rolling cadence to identify rapid rise and fading events. These will be supplemented by other wide-field optical facilities on the ground and in ultraviolet by ULTRASat in space. Neo Surveyor, comparable to Rubin in the infrared regime, will provide a more complete picture. Thus, the fixed cadence surveys of current and forthcoming facilities will identify GRB afterglows at least as often as weekly. With the improved galaxy catalogs from massively multiplexed spectroscopic surveys, many of these events will immediately be assigned to a host galaxy and known redshift. In partnership with the IPN alert stream this will identify a vast sample of GRBs detected in the prompt and afterglow phase with redshift and thus total energetics information. 

One of the major reasons to implement this stream is to provide smaller localization regions for follow-up or more power in statistical assignment or rejected of counterparts. If Swift-BAT localizes a GRB, then nearly all telescopes can observe the localization. Otherwise, input from multiple instruments would be beneficial. Automatic association and combination of autonomous localizations from scintillator instruments can reduce their separate localizations by a factor of a few to several, (e.g., Fermi-GBM and Glowbug). Value-added steps like removal of positions occulted by planets for all detecting instruments can further refine localizations or generation of HEALPix maps with full systematic errors for the earliest localizations (e.g. Fermi-GBM ground localizations). This can be done in near real-time for prompt reporting of LEO instruments and Earth-occulted positions, and in higher-latency for spacecraft around other planets as their data arrives. With a sufficiently integrated network, it may be possible to refine localizations from non-detections by sufficiently sensitive instruments, such as refinement of GBM localizations by incorporating a non-detection in Swift-BAT; however, this requires careful cross-calibration work.

The inclusion of timing annuli as sufficient data arrives will likely result in the greatest reduction in localization precision. INTEGRAL SPI-ACS data is now available for IPN within 20\,seconds. This can be combined with the real-time Fermi-GBM localizations for the majority of GRBs. Mars Odyssey data sometimes arrives with a latency driven by the light travel time delay, producing narrow elongated regions for nearly half of Fermi-GBM on-board triggers. Perhaps the greatest use of this information would be in the case of joint GW-GRB detections, facilitating the recovery of early signals from these events. Swift has agreed to observe events of interest, GRBs with or without GWs, as identified and localized by the IPN. 

An additional benefit would be integration with a GRB naming server, producing automatic name assignments, which is currently a mostly manual process. Lastly, there is a benefit of sending this stream through GCN. The sky maps can be hosted within GCN directly and/or sent as part of the content of the alerts directly. This bypasses the access delay introduced by HEASARC for hosting files, where Fermi-GBM alerts are distributed with links to HEASARC-hosted sky maps before they are fully available on HEASARC. Lastly, such a stream fosters comparison of detected transients in other survey instruments. It could lead to automatic comparisons with GRBs which are likely to have been occurred within the current TESS field and provide classifying information for potential relativistic transients identified by Rubin, Roman, ULTRASAT, and others.

\subsubsection{Event-based Prompt GRB Catalog}
An event-based GRB catalog would have entries for every prompt GRB detected, and the information would be collated from all available observations. This contrasts with trigger-based catalogs which are instrument-specific and distinct from one another. 

The IPN maintains an archive of GRBs localized by the IPN technique where the primary index is the localization. The underlying database is used to build a catalog of all prompt GRB detections with spatial information, represented in multi-order HEALPix maps, and broadband spectral information when available. Given the current funding mechanisms available, this work is strictly limited to archival events. It would be a benefit to the community if this work could be integrated with the previously described real-time reporting stream and maintained in a manner that all instrument teams can contribute updates as they perform high-cadence catalog analysis. This catalog could be preserved within HEASARC on semi-regular intervals. 

Such a catalog would be an invaluable community service, especially if it includes confirmed but untriggered GRBs. The IGWN and IceCube performs high-latency, sensitive searches around GRB triggers. Currently they collate data from multiple, incomplete, and initial (not final) analyses in order to perform their studies. A dedicated catalog provides ease of use, more unified selections for analysis, and could include additional parameters not otherwise available (e.g. the source classification probability). One result of such a catalog would be the association of additional multimessenger transients. 

Similar use cases have arisen as optical surveys have begun to recover relativistic transients independent of the high-energy monitors. Having a single access point with spatial information presented in a unified manner supports rapid checks of corresponding high-energy triggers. This can aid the classification of an event and enable that information to be used while follow-up observations can still recover events. Such a catalog is therefore also critical to the multiwavelength transient observations.

The localization information is also of use for the gamma-ray transient community itself. Scintillator-based localizations have important localization (and likely spectral) systematics due to imperfect modeling of the instrument, spacecraft, and atmospheric scattering of the signal. Providing the largest sample of well-localized GRBs is key to accurate modeling and minimization of the systematic of these sources. This may be of particular importance for short-lived missions such as StarBurst, Glowbug, or the numerous international smallsats with scintillator-based instruments. 

Additionally, the catalog would be of use for studies of the transient signals themselves. The IPN catalog was a key resource to proving some short GRBs arise from extragalactic MGFs \citep{burns2021identification}. Archival spectral information was important to understanding the true rarity of the brightest GRB ever seen \citep{burns2023grb} and in motivating dedicated tiling of GRB\,230307A, which resulted in likely the first JWST detection of a kilonova \citep{2023GCN.33414....1B}. Those calculations all utilized the first collated GRB archival catalog, which will be released within the year, but the lack of public access to a collated GRB catalog is evidently limiting other works. A collated GRB catalog that is updated in relatively low latency will be a key community resource.

Sustained support would facilitate archiving of extant datasets from historical missions while relevant individuals are still active. A key example is the preservation and modern analysis of the Pioneer Venus Orbiter data. It may also be possible to recover the GRBs seen by Vela during the $\sim$5\,year gap from the end of the 1973 Vela GRB discovery paper and the launch of Pioneer Venus Orbiter in 1978. 

\subsubsection{A Shared Database and Archive}
GraceDB\footnote{https://gracedb.ligo.org/} is described as follows: \enquote{The Gravitational-Wave Candidate Event Database (GraceDB) is a service operated by the LIGO Scientific Collaboration. It provides a centralized location for aggregating and retrieving information about candidate gravitational-wave events. GraceDB provides an API for programmatic access, and a client package is available for interacting with the API.} GraceDB acts as a focal point to collate data from the different GW interferometers into a single access point. The data representation is a list of events, where each event has a page with additional details that are helpful for real-time investigation on whether events are astrophysical, the likely classification of events, which pipelines and instruments contributed, etc. The internal portion of GraceDB contains additional information for trigger vetting, such as additional data streams of interest to identify data glitches.  

It would be ideal to create a similar focal point for prompt gamma-ray transient detections. Note this is referring to all events that trigger GRB monitors, not only GRBs. Multi-mission information can be helpful in immediately classifying events. Events can be immediately classified as non-terrestrial and confirmed as real if viewed in multiple spacecraft. Combined localization information will constrain solar flares towards the sun and SGR flares towards specific magnetars for an assigned origin. Inclusion of external information, for example, solar activity, will help classify transients. The automated comparison of localizations against galaxy catalogs will allow for faster identification of extragalactic MGFs, which can be enhanced with rapid compact merger exclusion distances from IGWN. Additionally, automated comparison of localizations against magnetars will more immediately identify which magnetar produced a Galactic giant flare. Such a system would have more rapidly identified the Fermi and Swift detections of GRB\,221009A as the same event, providing earlier observations of the BOAT.

We here emphasize the database as distinct from both GCN and HEASARC. The distinction for the former can be seen as analogous for how GraceDB operates with GCN, which points to a distinction in the broader TDAMM ecosystem: the same signal being observed by multiple telescopes should be collated into signal-based datasets. Examples of this are GraceDB for GWs, SNEWS for MeV neutrino signals, and IPN for GRBs. Distinct from this is the use of GCN as the central clearing house for event-based searches. This database and repository would act as a focal point for the various instrument teams to collate data, perform classification and verification checks, and allow for additional possibilities discussed later (for example, multi-mission subthreshold searches). The outputs of the real-time system will be distributed to the astronomical community through GCN. 

HEASARC is focused on the preservation of data from NASA high-energy missions. It is not optimized for the use of real-time reporting, for the case of frequent alterations of data, or for common and easy retrieval of data from multiple instruments to perform correlative or joint analysis. The database allows for rapid real-time data sharing and for iterating on analysis until a final result is achieved. Key summary information and data from NASA missions could be archived in HEASARC for posterity; however, associated information from non-NASA instruments and observatories are not able to be stored at HEASARC, therefore another solution must be considered for storing complete multiwavelength and multimessenger datasets.

The database must necessarily be distinct from HEASARC and GCN, and additionally housed external to NASA, for several reasons. With existing restrictions on access to NASA systems, scientists from countries with active GRB monitors would not be allowed access. Several key missions in the IPN do not have private data and are shared through existing agreements. International space agencies may not allow NASA to archive or house their data. Data from NSF-funded facilities are outside the purview and funding of NASA, therefore such a database would need to also be funded through NSF to include relevant multiwavelength and multimessenger data. A community-led database would be necessary to allow for inclusion of all relevant data with the rules set on the use of that data by the various instrument teams and space agencies. This generally supports NASA's goal of Transition to Open Science to include critical private TDAMM data for greater scientific results.

\subsubsection{Additional Automated Analysis and Prioritization of Follow-up Observations}
After 10,000 prompt detections and more than 1,000 afterglow detections, it is necessary to provide prioritization information to the follow-up community. The collation of data into a shared repository allows for additional processing that can guide real-time follow-up. As mentioned, combined localization of flares from magnetars or MGFs can identify these events and their host sources/galaxies in an automated fashion. X-ray telescopes can chase tails of Galactic or extragalactic MGFs. X-ray and radio telescopes will know when specific magnetars are active to search for additional SGR-FRB flares. Likewise for short GRBs arising from BNS mergers, radio facilities may follow up promptly for precursor signals.

Additional analysis will be beneficial for cosmological GRB follow-up as well. Duration and spectral analysis can be automated. Calculations of the likelihood of a given burst as belonging to a merger or collapsar can be provided while follow-up signals are still detectable, with increasing fidelity as more data arrives. The initial classification can be on duration, then duration and hardness, then additional parameters as discrimination power is confirmed. 

Additionally, it may be possible to provide classification probabilities for a specified progenitor channel. Isolation of MGFs through spatial alignment with local star forming galaxies has been discussed. The particularly low minimum variability timescales in GRBs 060614, 211211A, 230307A and their identification as long GRBs from a merger origin suggest automated identification of low minimum variability timescale as a method to prioritize follow-up observations of similar bursts going forward \citep{veres2023extreme}. If machine learning classification schemes are validated, then they too could be utilized at early times \citep{misra2023evidence}. When data arrives from instruments with long-term background stability or imaging capabilities then ultra-long GRBs may be flagged while prompt emission is on-going.

Beyond flagging specific bursts as arising from potentially interesting progenitors, such a system can also flag more traditional bursts where multiple diagnostics have already been captured. For example, flagging BAT bursts that are also detected by GBM, or are likely to be detected by Konus, would select bursts where intrinsic bolometric energetics can be calculated, if a redshift is recovered. With the launch of future polarimeters, one can flag bursts that have precise localizations, broadband coverage, and are also bright enough that they will likely have sensitive polarization information. Flagging joint detections (e.g. SGR flares with FRBs or GRBs with neutrinos, GWs, or VHE detections), would also be beneficial. 

\subsubsection{Event-based SGR Flare Reporting and Catalogs}
The discussion on a collated GRB catalog and a collated GRB stream can also be applied to SGR flares. The IPN maintains a list of SGR flares, updated in high latency, that is of use for GW, neutrino, and FRB searches from magnetars. The list is incomplete, for example it is missing the SGR flares that are detected by GBM but occur during the 10\,minute interval where additional triggers cannot occur. SGR flares are not uniformly tracked by GRB monitors. Additionally, there is not a standard naming convention for SGR flares, which would be beneficial to establish. Back-filling the catalog would allow for identification of past bursts as arising from newly identified magnetars. As mentioned, determining which magnetar powered a given flare in real-time is beneficial for knowing when they are in active states for SGR-FRB joint searches.

\subsubsection{Coherent Multi-mission Searches and Upper Limits}
One of the major advancements since the launch of Fermi and Swift is the development of sub-threshold maximum likelihood searches. Historically, GRBs were detected by on-board or on-ground searches for a significant excess of counts above the background as defined by a signal-to-noise threshold. The threshold to identify events is usually set to minimize false triggers. In contrast to counts-based searches, coherent searches consider information from all detectors of a given instrument in deconvolved flux space, which is more powerful than searching in response-convolved counts. Coherent searches increase the significance of signals that are consistent with having a real origin and typically decrease the significance of signals that are inconsistent with a real signal.  The Fermi-GBM and Swift-BAT searches increase the effective sensitivity of each to short GRBs by factors of a few \citep{goldstein2019updates,delaunay2022harvesting}.

These searches are extensible to searching for signals coherent with an astrophysical origin in data from instruments on separate spacecraft. Such an undertaking would be a significant development effort, handling inter-calibration uncertainties, heterogeneous data, and proper factoring in instrument-specific and network-level systematic errors. Additionally, when distant spacecraft are utilized, searches for a source from a given position on the sky will be observed at different times by the different instruments, adding an additional dimension to the arrays included in the maximum likelihood calculation.

However, the payoff is two-fold: the GRB network can be treated as a single distributed telescope and the inclusion of multiple spacecraft allows for an increase in sensitivity and decrease in false triggers. The latter arises from automated removal of terrestrial and particle sources of signals that may be detected and confused with a real signal in a single instrument as well as suppression of Poisson fluctuations in individual detectors. This benefit results in a multiplicative gain of astrophysical transients given the steep $\log N-\log S$ curves. Proper handling of differential detector sensitivity would provide a further localization refinement compared to the more simplistic combination of separate autonomous localizations. These gains are analogous to multiple GW interferometer networks vs single interferometer observing times. 

A multi-mission coherent search presents an additional new possibility for free: coherent, all-sky flux upper limits. Such an approach will prove invaluable for more stringent constraints from gamma-ray non-detections of externally identified transients, including the vast numbers which will be identified from fixed cadence surveys with current and forthcoming facilities. These include elongated GW localizations where instantaneous all-sky coverage is necessary and optically-identified transients with large temporal uncertainties. These upper limits will prove more stringent than those reported from individual instruments, roughly scaling as the square root of the total network effective area.

One potential hurdle in the pursuit of a complete multi-mission targeted search is the need for an accurate all-sky energy responses for each instrument. Some IPN instruments may not have been fully calibrated for GRB studies and theoretical responses may need to be constructed. For satellites in LEO, improved modeling of atmospheric scattering would be beneficial.

\subsubsection{Intercalibration of High Energy Monitors}
The absolute calibration of these instruments is likely accurate to $\sim$10-30\%. Given the brightness of the sources of interest spans a range covering orders of magnitude, the absolute offset generally does not affect conclusions. However, in order to build the products which rely on coherent analysis of multiple missions in searching for single events or upper limits, it is critical to properly understand the level of calibration agreement between missions. Calibration as a function of energy are both critical.

Proper cross-calibration requires deep involvement of the instrument teams, reasonably large samples of events, and careful study. For cross-correlation of signals with light travel time distances comparable to the duration of the events, the uncertainties in the light travel time must be properly accounted for. When only imprecise localizations exist for cross-calibration, this introduces a limitation in perfect comparison. The IPN can take on the responsibility for network-level calibration of active and past instruments, and updated the calibration as new capabilities come online. This work is tedious and unlikely to be funded through Guest Investigator opportunities, but is necessary to properly perform a network-level analysis of high-energy monitors..

\subsubsection{Systematic Characterization}
Another critical need for the proper use of GRB localizations is characterization of systematic error, which can be substantial for instruments like GBM \citep{connaughton2015localization,goldstein2020evaluation}, BATSE 
\citep{briggs1999error}, and GECAM \citep{zhao2023gecam}. Similarly, systematic errors arise in timing annuli including effects from deadtime, timing errors on individual instruments or satellites, or perhaps unidentified reasons \citep[e.g.][]{pal2013interplanetary,hurley2013interplanetary}. These systematics can be characterized by comparing statistically determined localizations of events with known positions, such as those from catalogued Galactic sources or events with known source positions. These studies empirically determine the systematic error, which can then be folded into the total localization posterior. Without this, use of this data in searching for joint detections or providing information for follow-up observations would be fundamentally flawed.

Estimating systematics is of sufficient importance that it is generally part of the key responsibilities of the instrument teams for autonomous localizations and has been core to the operation of the IPN. These works must continue and be supported, but there are two issues which must be addressed. The first is that Swift-localized GRBs have been critical for this work for nearly 20\,years, but Swift is unlikely to last another 20\,years. The second issue is attempting to characterize systematic error for short-lived missions, such as the numerous ones launching with silicon photomultiplier technology. Both issues can be ameliorated or fully resolved with IPN collated alert streams and GRB catalogs, allowing both precise prompt GRB localizations and successful follow-up to provide the critical comparison sample. This will work continue beyond the end of Swift and provide a factor of a few greater sample for short-lived missions to use to estimate their localization systematics. Generation of full probability-probability calibration plots, showing the correct accounting of systematic error, would be beneficial in convincing the community to utilize these localizations.

\subsubsection{Modernized Results Presentation}
For many instruments, the source of the systematic error on localization propagates as a source of systematic error for spectral, polarization, and sometimes, temporal studies. Inclusion of a systematic error into the spectral fitting and polarimetry algorithms, rather than an ex post facto correction, would allow for a fuller characterization of the the uncertainties regarding spectroscopy and polarimetry. Proper modeling of the true systematic error would allow for the generation of accurate corner plots.

Corner plots show the one and two dimensional posterior probability distributions for parameters considered in a given analysis. Sharing of results in this manner allows for more accurate representation of correlations in a given result and can enable more accurate follow-on inferences. Many groups like IGWN and fields like NS equation of state studies have fully adopted these plots. They are used in the study of GRB afterglow as well as supernova and kilonova recovered in follow-up observations, but are still uncommon in the studies on prompt signals. Adoption of a common analysis framework by the prompt instrument teams would facilitate greater multiwavelength and multimessenger results.

\subsubsection{Shared Open Source Analysis Software}
IGWN has also developed a software stack that contains the de facto analysis tools for GW astronomy. No equivalent tool is yet widely adopted for the study of GRBs. XSPEC is not particularly well suited for time-series spectra. NASA has funded the development of the generalized Gamma-ray Data Tools\footnote{\url{https://github.com/USRA-STI/gdt-core}} to build a core set of functionality for analysis of data from basic scintillators and coded aperture masks, as well as the tools for the implementation of a multi-mission targeted search for scintillators. Additionally, the underlying approach for maximum likelihood in BAT is being explored in other coded aperture masks. A scientific consortium could adopt the best open source tools and continue to improve them. A shared open source software framework will also allow for community contributions, such as physically motivated modeling of GRBs, which are not generally available in existing spectral analysis tools.

\subsection{Examples}\label{sec:examples}
In order to demonstrate the need for the products described above, we highlight a few specific GRBs that demonstrate the limitations of the existing ecosystem.

\subsubsection{GRB\,170817A}
\begin{figure}[htp]
\centering
\begin{minipage}{0.49\textwidth}
\includegraphics[width=\textwidth]{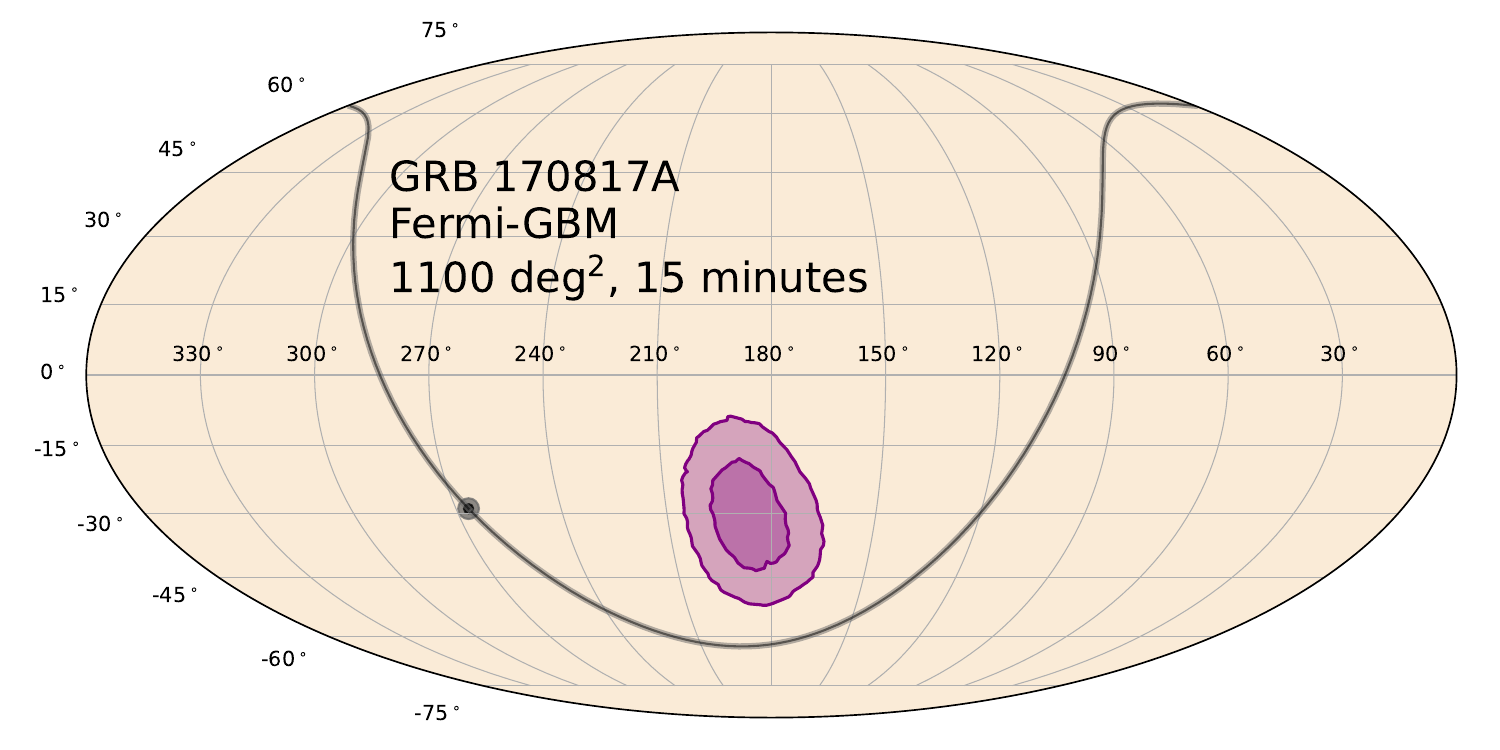}
\end{minipage}\hfill
\begin{minipage}{0.49\textwidth}
\includegraphics[width=\textwidth]{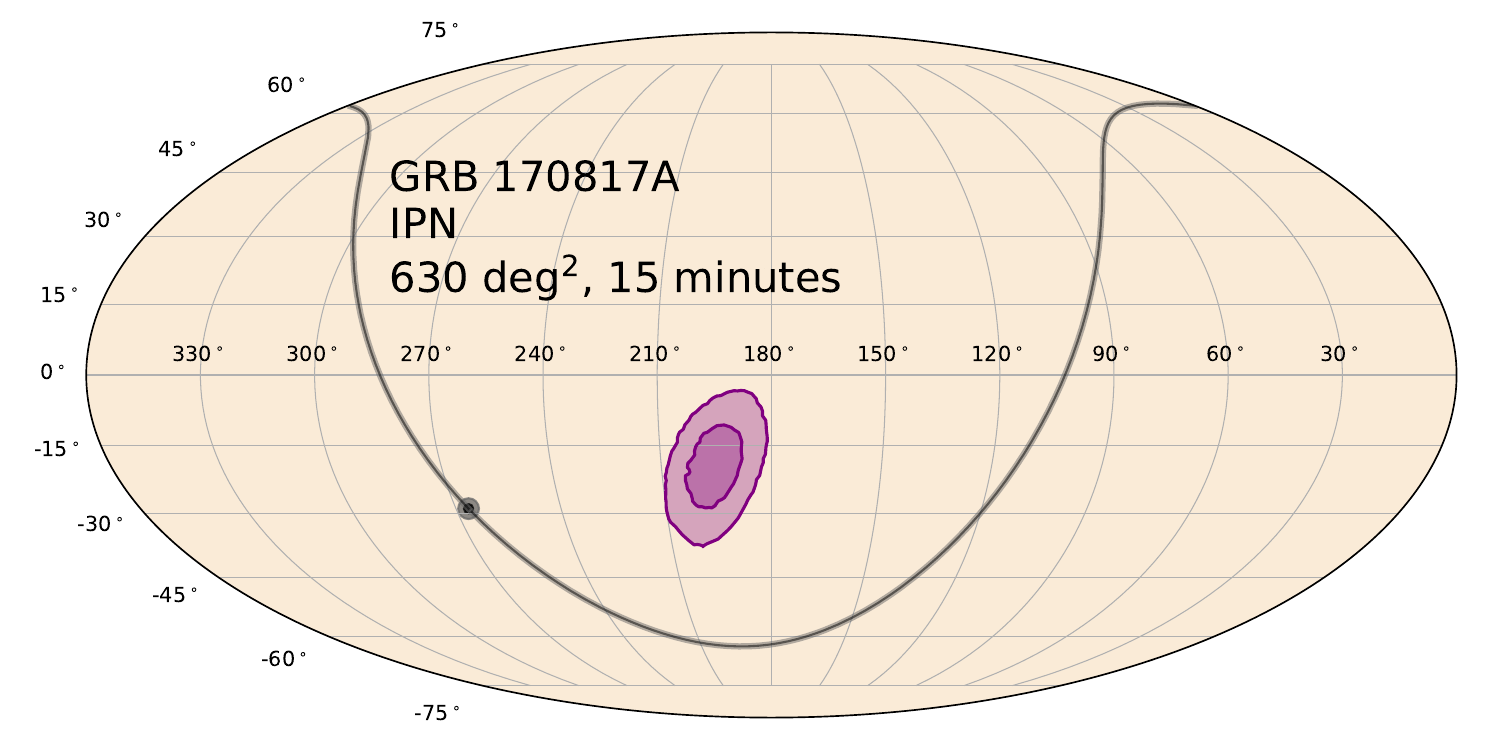}
\end{minipage}\par
\begin{minipage}{0.49\textwidth}
\includegraphics[width=\textwidth]{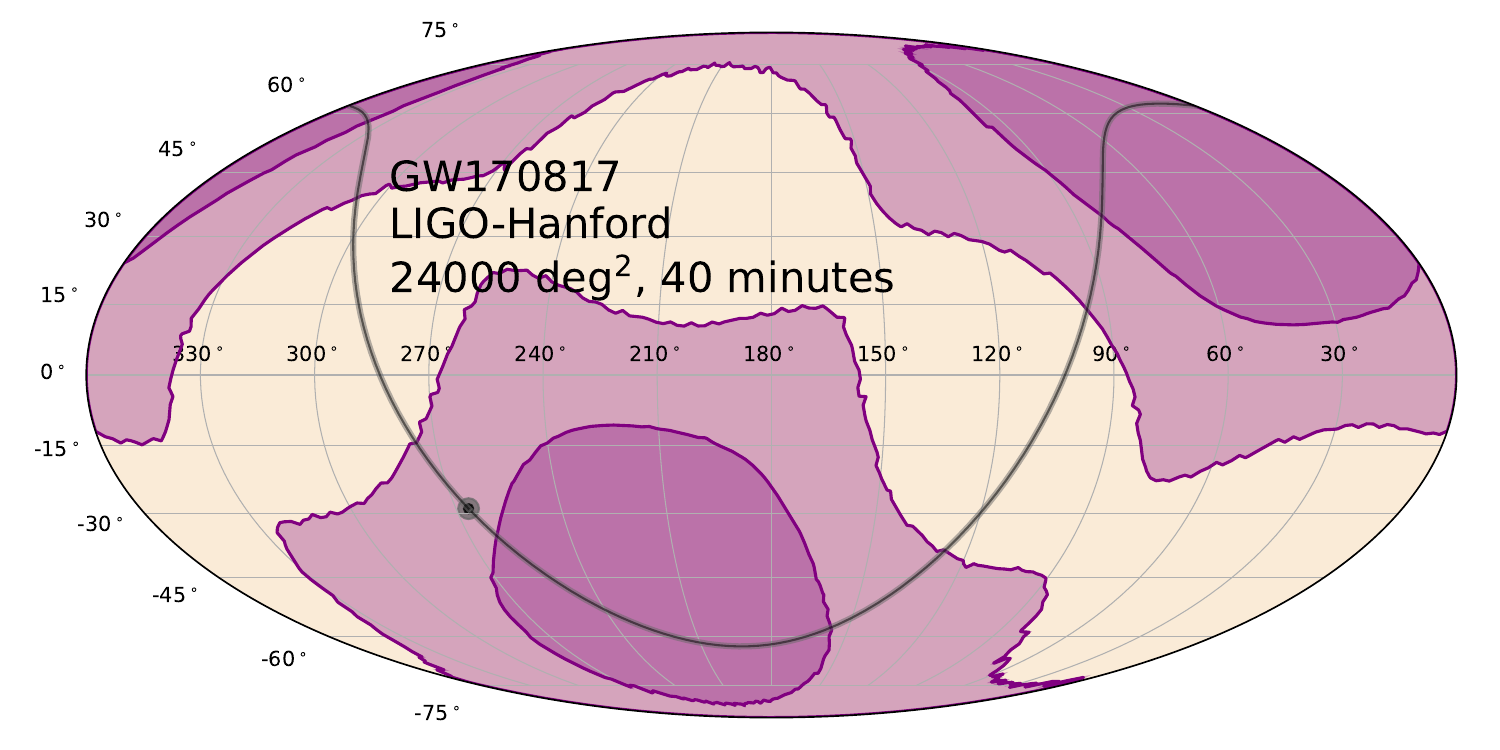}
\end{minipage}\hfill
\begin{minipage}{0.49\textwidth}
\includegraphics[width=\textwidth]{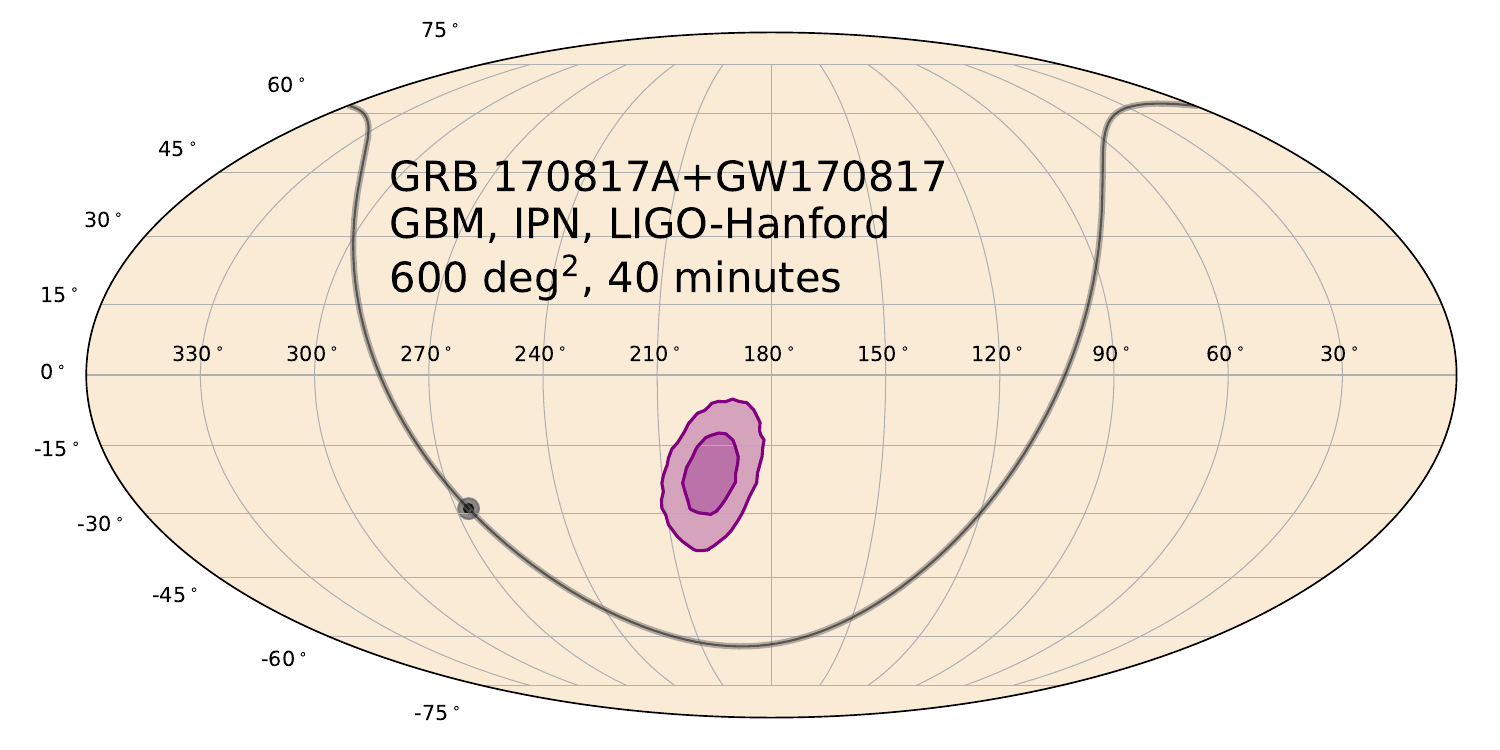}
\end{minipage}\par
\begin{minipage}{0.49\textwidth}
\includegraphics[width=\textwidth]{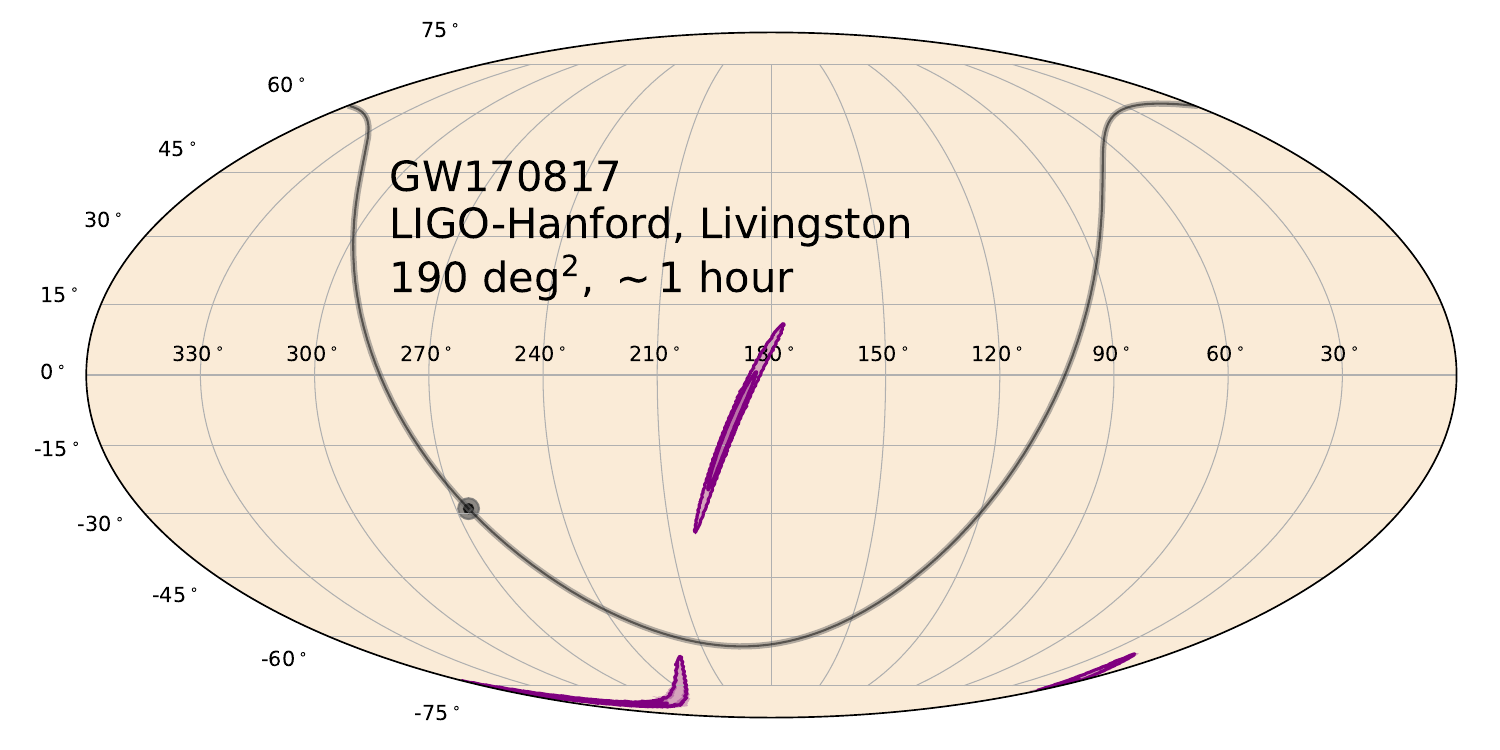}
\end{minipage}\hfill
\begin{minipage}{0.49\textwidth}
\includegraphics[width=\textwidth]{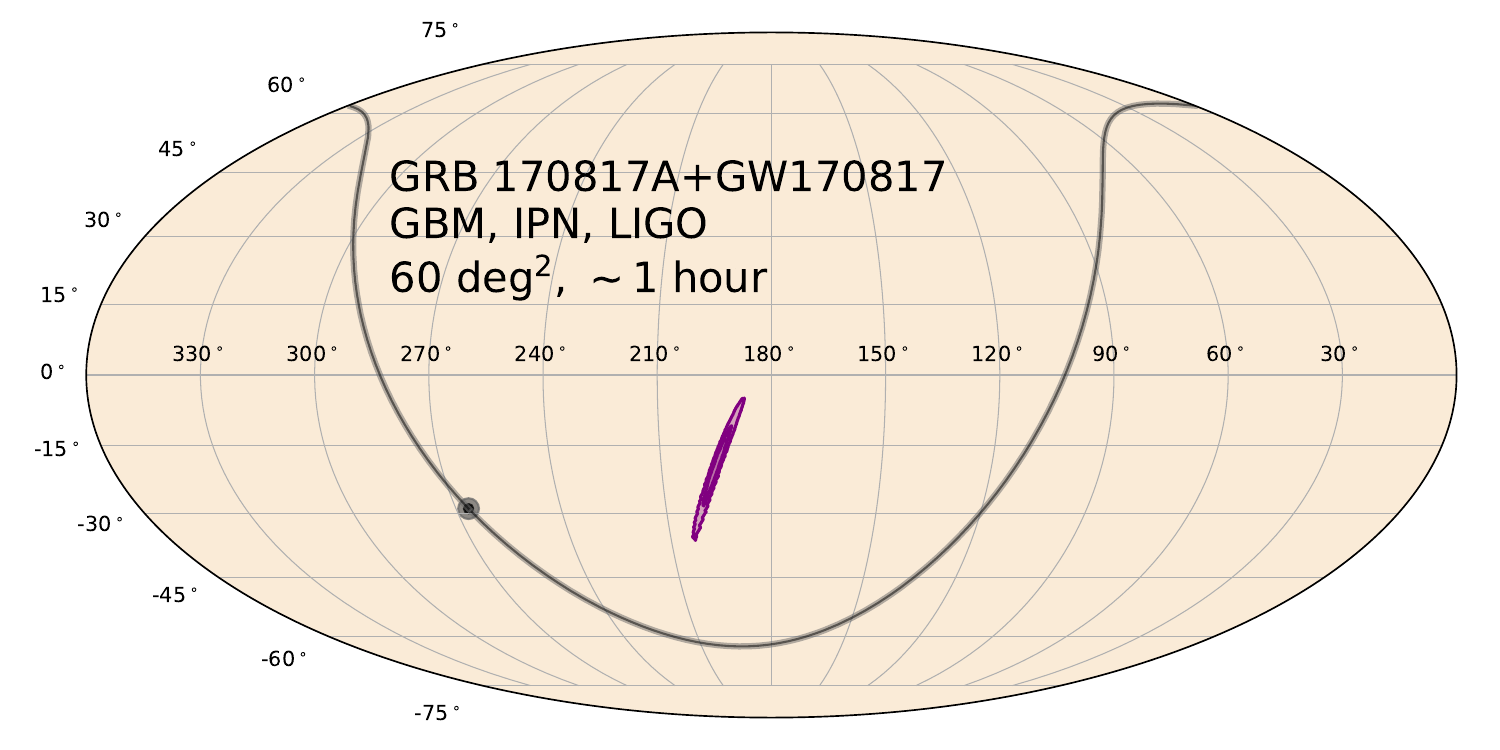}
\end{minipage}\par
\begin{minipage}{0.49\textwidth}
\includegraphics[width=\textwidth]{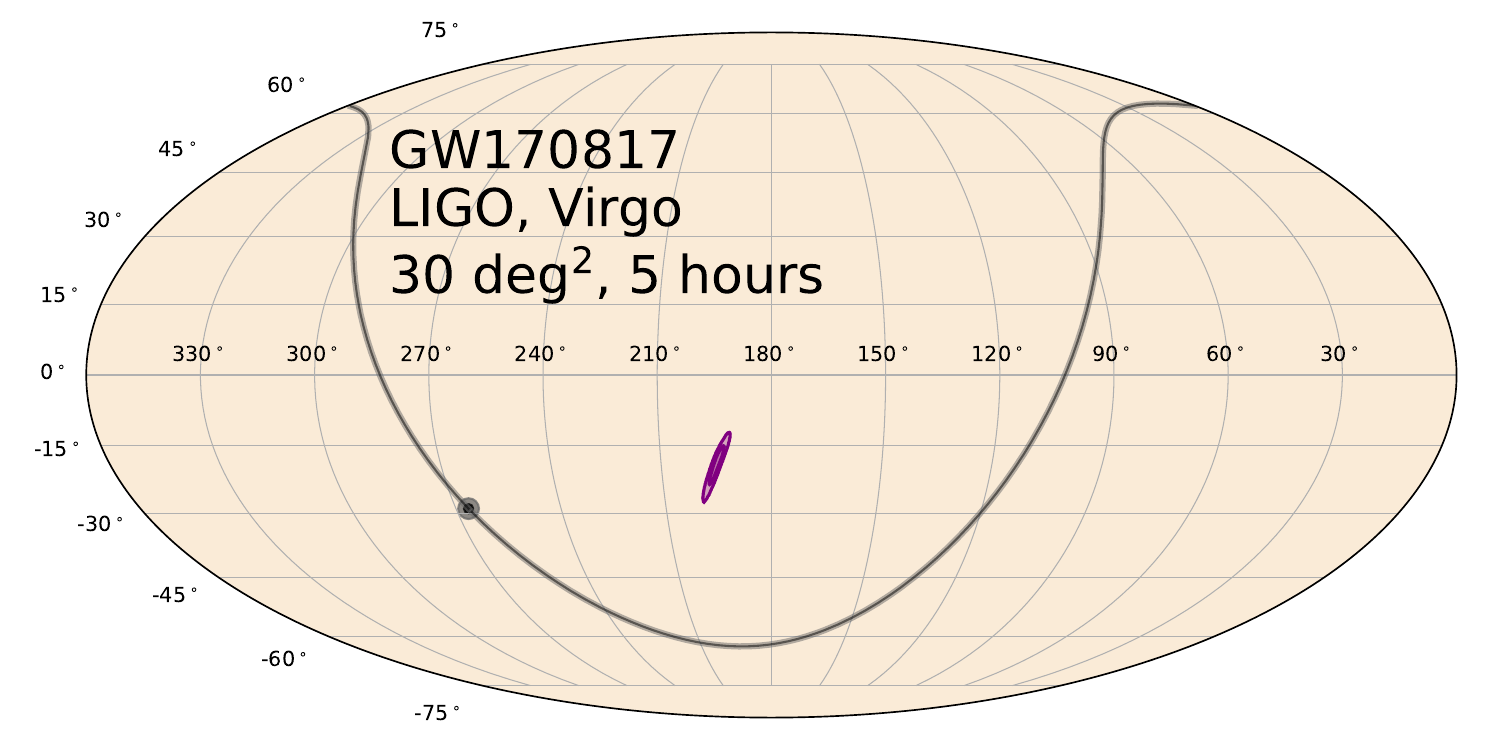}
\end{minipage}\hfill
\begin{minipage}{0.49\textwidth}
\includegraphics[width=\textwidth]{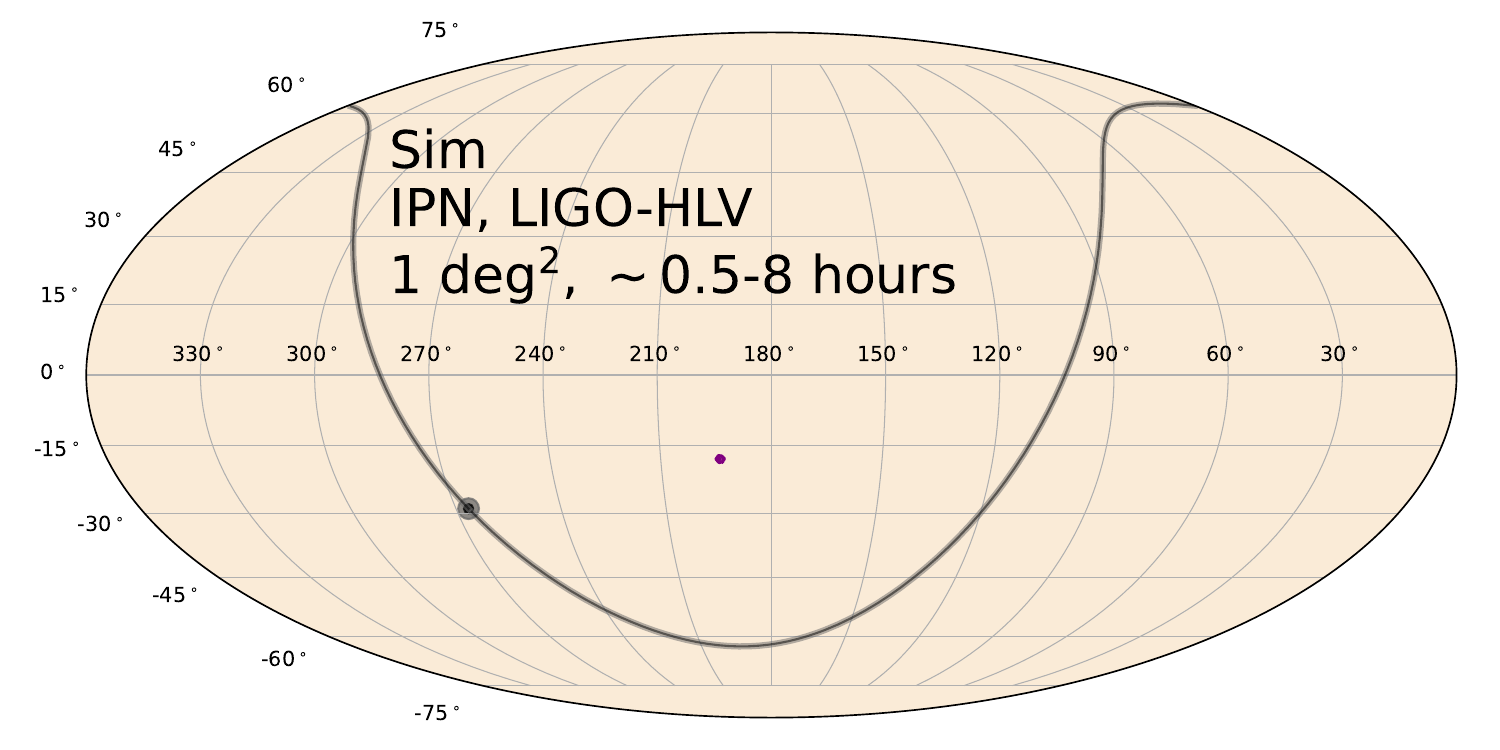}
\end{minipage}\par
\caption{The ideal joint localization steps for GW170817 and GRB\,170817A that could have occurred. The first 7 images are described in the text. The last image shows what would have occurred if the GRB was bright enough to be recovered in a distant IPN instrument.}
\label{fig:maps_170817a}
\end{figure}

The joint detection of a BNS merger as GW170817 and GRB\,170817A led to one of the greatest observing campaigns in history and major scientific advances in several fields of physics. The three-interferometer GW localization took hours to compute and release owing to a major glitch during the signal and data access issues. The initial GW localization was on the order of $\sim$24,000\,deg$^2$. This could have been automatically combined with the GBM localization of the GRB resulting a 90\% area of $\sim$1,100\,deg$^2$. Automation of the Fermi-INTEGRAL timing annulus could have reduced the area to $\sim$600\,deg$^2$. This could have been released approximately as quickly as LIGO and Virgo confirmed the GW event. About an hour later, the two-interferometer localization was produced, but not yet released. It could have been combined with the GBM localization or the IPN localization and achieved $\sim$60\,deg$^2$ where the host galaxy NGC\,4993 would have been one of the top-ranked candidates, allowing for recovery of the afterglow and kilonova $\sim$10\,hours earlier than. The final three-interferometer localization was 30\,deg$^2$. These are shown visually in Figure\,\ref{fig:maps_170817a}.

The scientific loss here is depressing. Recovery of the precise position within one hour instead of eleven would have likely directly answered if the prompt gamma-ray signature could have been cocoon, would have vastly constrained the viable contributions to the early UV emission, and would have also allowed recovery of the earliest afterglow signature before it faded below detectability. This would have fundamentally altered our understanding of these events and led to 6\,years of more informed advances in theory and simulations. Furthermore, the distance of the event suggests it is approximately a once per decade event, which, given the GW observing livetimes, may result in decades of waiting for another event as near, meaning a unique opportunity for full characterization was lost. Ultimately, the limitation was in handling of data from existing facilities, and it is critical that we must minimize the chance of this problem occurring again.

\subsubsection{GRB\,230307A}\label{sec:grb230307a}
\begin{figure}[htp]
\centering
\vskip\floatsep
\includegraphics[width=0.65\textwidth]{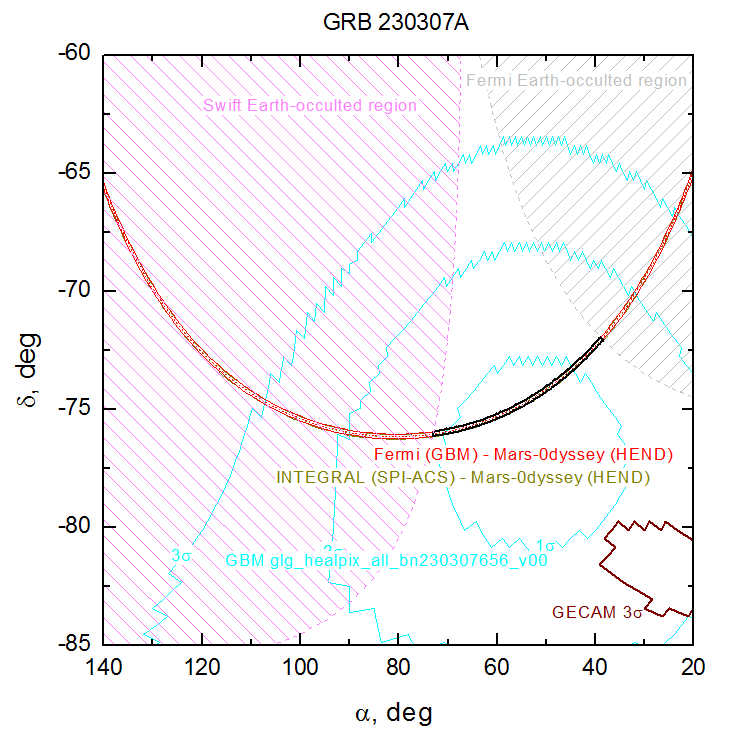}
\begin{minipage}{0.49\textwidth}
\includegraphics[width=\textwidth]{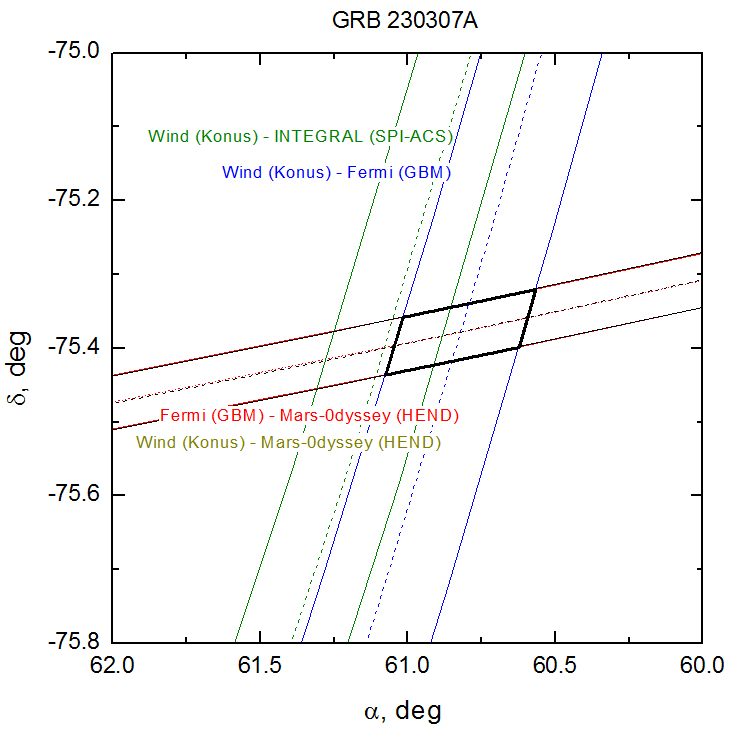}
\end{minipage}\hfill
\begin{minipage}{0.49\textwidth}
\includegraphics[width=\textwidth]{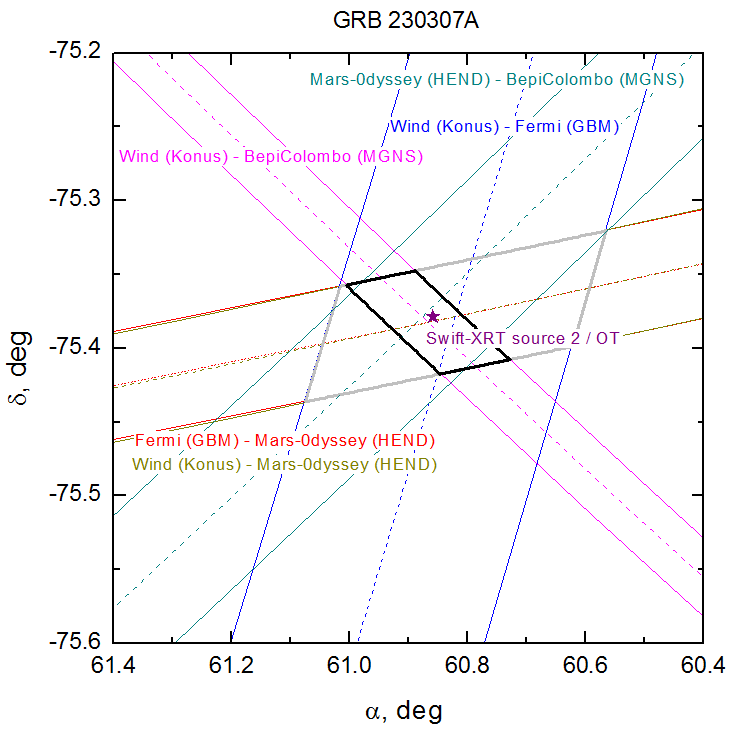}
\end{minipage}\par
\caption{The IPN triangulation of GRB\,230307A. Upper left is the initial skymap, with information available in under an hour and the upper right shows the Swift XRT tiling of this localization. The lower  left shows the refined skymap available within $\sim$1\,day and the bottom right the final localization available after 3\,days.}\label{fig:ipn_230307a}
\end{figure}

GRB\,230307A is one of the best demonstrations of the need to invest in the IPN in the current era and shows that the greater TDAMM community is open to following up interesting events without precise localizations. The timeline of the announcement, localization, and recovery of the afterglow is shown in Table\,\ref{tab:230307a}. The corresponding sky maps and tiling patterns are shown in Figure\,\ref{fig:ipn_230307a}.

\begin{table}[htp]
\begin{tabular}{|r|l|l|l|}
\hline
Report                                          & Reality & Actual Public Announcement                            & Ideal Ground System \\ \hline
Discovery Notice                                & 11\,s                             & 11\,s                                               & 11\,s                \\
Identification as very bright GRB               & $<$1\,hour                       & $\sim$1\,hour \citep{2023GCN.33406....1X}          & $<$1\,minute        \\
First IPN Map (narrow annulus)                  & 3.5\,hours                       & 8\,hours \citep{2023GCN.33413....1K}               & $<$1\,hour          \\
Solar Orbiter STIX annulus                                    &                                  & 7\,hours \citep{2023GCN.33410....1X}               & $<$7\,hour          \\
Identification as 2nd brightest GRB &                                  & 12\,hours \citep{burns2023grb,2023GCN.33427....1S} & $<$4\,hours         \\
Swift XRT, UVOT Follow-up                       &                                  & 1\,day \citep{2023GCN.33465....1B}                 & $\sim$1\,hour       \\
Second IPN Map                                  & 26\,hours                        & 28\,hours \citep{2023GCN.33425....1K}              & 26\,hours           \\
Ground based optical follow-up                  &                                  & 34\,hours \citep{2023GCN.33439....1L}              & $\sim$hours         \\
Third IPN Map                                   & 39\,hours                        & 45\,hours \citep{2023GCN.33461....1K}              & 39\,hours           \\ \hline
\end{tabular}
\caption{The summary timeline of early reports on GRB\,230307A. The first column contains the specific piece of information reported. The second column contains the geocentric time since the GRB was detected (at Earth) when this was reported, either publicly or through private methods (e.g. the Time Domain Astronomy Slack). The third column marks the first public announcement. The last column shows what could have occurred with the described IPN enhancements (i.e. improvements to the ground segment, but leaving the space-based communication timeline unaffected). The major improvements are in the earliest hours, enabling recovery of several key diagnostics that are otherwise missed.}
\label{tab:230307a}
\end{table}

The last column reports when these steps could have occurred with an ideal ground segment. The first IPN localization could have been completed in under an hour given the prompt downlink of HEND data from Mars, but the real-time data products had incorrect timing, causing hours of delay. Follow-up groups were interested in utilizing this information, but many did not have tiling generation code for narrow, elongated annuli. These factors prevented observations on the first night (including delays due to sleep schedules). Increasing the accuracy and robustness of real-time reporting of data from IPN instruments will vastly enhance the science return of existing assets. Regularly reporting these early localizations will encourage the follow-up community to build the necessary software pipelines to observe these events. 

This event is unique for several reasons. It has the second highest prompt fluence ever identified, out of more than 10,000 prompt GRB detections. By chance, the burst occurred within the field of view of TESS \citep{2023GCN.33453....1V} giving an evident detection of the prompt emission in optical as well as the onset of afterglow \citep{2023RNAAS...7...56F}. Also by luck, the burst was within the field of view of LEIA giving 0.5--4\,keV coverage \citep{sun2023magnetar}. The prompt gamma-ray emission is $\sim$200\,s in duration \citep{2023GCN.33427....1S}, but follow-up observations have found what appears to be a kilonova, strongly suggesting a compact merger origin for this long GRB \citep{2023GCN.33569....1L,2023GCN.33578....1B}. There were subsequent photometric and spectral observations by JWST at $\sim$30 and $\sim$60\,days \citep{2023GCN.33569....1L,2023GCN.33580....1L,2023GCN.33747....1L}. This is the first late-time near-infrared spectra of r-process creation, which would be the most direct measure of heavy element nucleosynthesis ever captured \citep[e.g.][]{metzger2020kilonovae}. It would similarly imply the TESS observation presents the first prompt optical detection of a compact merger GRB. 

The brightness of this GRB requires this to be an exceptionally rare event, as confirmed by the putative host galaxy at $\sim$300\,Mpc. The lack of early broadband observations is limiting the full characterization and understanding of this likely r-process event. This event may be the single best example of why investment in the ground segment of active missions is critical for the new TDAMM era. Broadband characterization could have been possible beginning in 1\,hour instead of 1\,day if the proposed products were available. The successful recovery of the afterglow from both X-ray and optical tiling of early IPN localizations demonstrates the broader community is ready and willing to observe these events, especially when provided prioritization guidance by the high energy monitor teams. The IPN enabled JWST to observe a unique event that will drive the theory and simulation of r-process creation for years, before the start of the O4 observing run, but investment in the IPN ground segment would have enabled key additional observations.

\subsubsection{GRB\,221009A}\label{sec:grb221009a}
GRB\,221009A is the brightest GRB ever observed by the measures of fluence, peak flux, and $E_{\rm iso}$ \citep{burns2023grb}. The identification of the burst as interesting took several hours longer than should have been necessary. The prompt burst was discovered by Fermi-GBM \citep{lesage2023fermi}, but telemetry issues due to TDRS prevented automated dissemination of alerts. About an hour later, Swift reported the discovery of a bright Galactic transient \citep{williams2023grb}. This classification was reported since the BAT trigger was significant only over long periods and showing little variation, the localization was near the Galactic Plane, and because of the particularly bright X-ray, UV, and optical signals seen in the follow-up instruments. The automated Fermi-LAT processing of this Swift transient uncovered a bright GeV signal.

\begin{figure*}[htp]
	\centering
	\includegraphics[width=\textwidth]{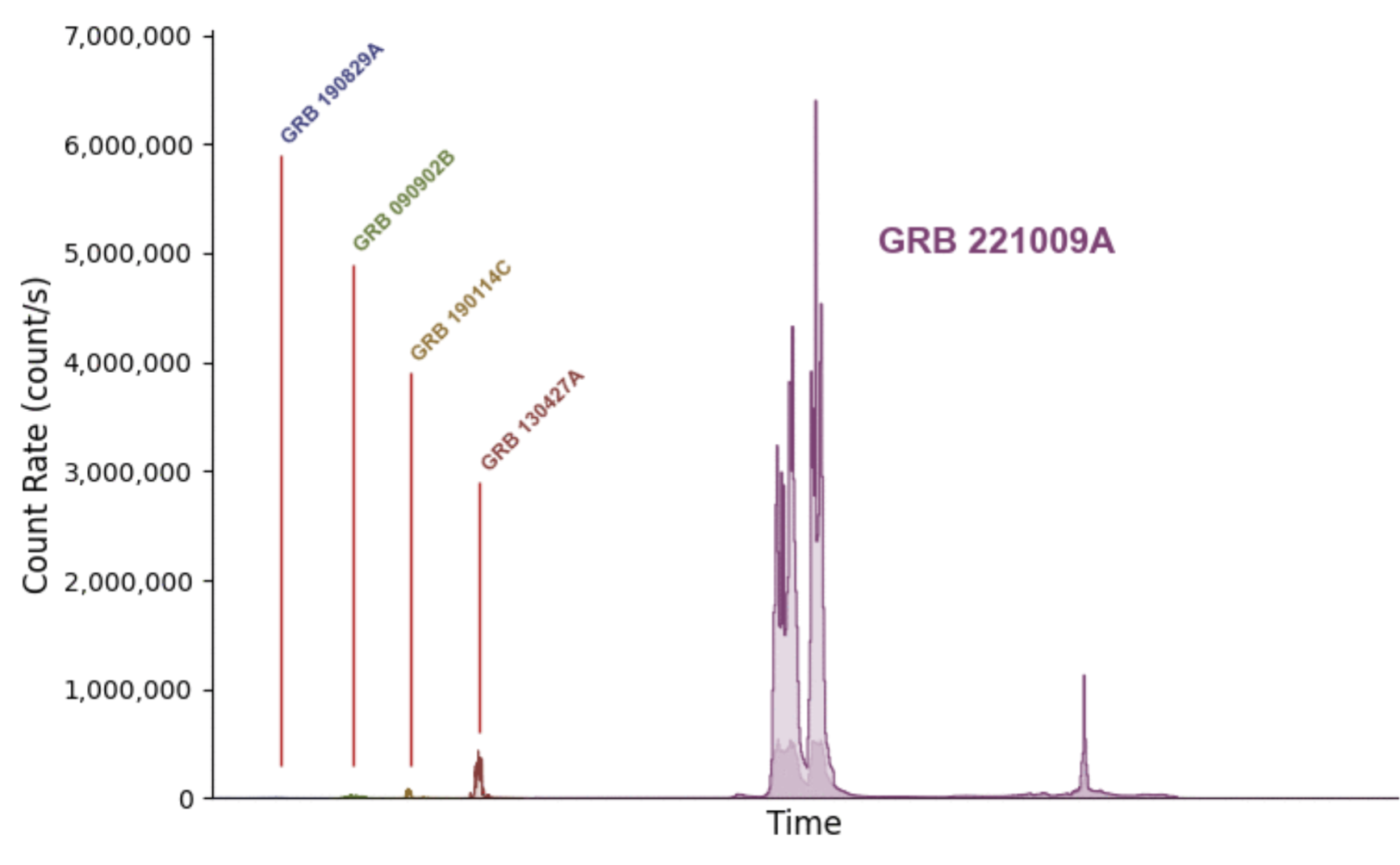}
    \caption{A comparison of the Fermi-GBM observations of GRB\,221009A against historic bursts. Credit: Adam Goldstein (USRA).}
    \label{fig:boat}
\end{figure*}

The reporting of the LAT analysis caused the LAT Burst Advocate to notice the incredibly bright Fermi-GBM signal and alert others to determine if the GBM and Swift signals may be related. The initial comparisons of the GBM and Swift localizations were inconsistent, leading to belief that they may be unrelated, but this confusion was caused by saturation of the GBM detectors which affected the localization. Internal discussions between the GBM and LAT teams determined it was possible that they were related and that the LAT signal had a typical $-2$ power law observed in prior GRBs. Private discussions, again enabled through Slack, allowed the GBM and Swift teams to compare results. This led to the realization that the BAT trigger was an image trigger, which had not been mentioned in the GCN circular. The combined discussions between the teams resulted in the early conclusion that Swift-BAT had triggered on afterglow from the incredibly bright GRB, which had not previously occurred in 17\,years of operation. The joint discovery of the brightest GRB ever detected was then announced, roughly 8\,hours after GRB time \citep{2022GCN.32635....1K,2022GCN.32636....1V}. 

This GRB demonstrates that rare and unique events do happen. The amount of data is so large that analysis is severely limited by the fidelity of the available models and will lead to continued simulation and theory advances for several years. While real-time data issues are an inescapable problem faced by space missions, the majority of the delay to announcement could have been avoided if a shared database was operational. This could have flagged the events as potentially related as soon as the Swift trigger arrived, saving perhaps seven of the eight hours. Having a more formal method of discussion between GRB monitor teams would have facilitated earlier discussions. In this case, we were fortunate that the individuals who shared information between the Swift and GBM teams were both checking Slack at the same time, otherwise this single-point failure may have led to an additional eight-hour delay due to sleep schedules. More intentional technical and programmatic integration will result in more important scientific data sets for events of interest.

\subsubsection{GRB\,200612A and GRB\,220627A}\label{sec:grbs200612a_220627a}
GRB\,200612A was detected by Swift-BAT and was promptly reported \citep{2020GCN.27915....1S}. Fermi-GBM triggered on a burst ten minutes later \citep{2020GCN.27934....1M}, which was confirmed to be associated with the Swift signal by the full coverage from Konus-Wind \citep{2020GCN.27922....1S}. Separately, GRB\,220627A triggered Fermi-GBM twice, pointing to $\sim$1,000\,s duration \citep{2022GCN.32288....1R}. This was again confirmed by Konus who noted a duration up to 3,700\,s \citep{2022GCN.32295....1F}.

A collated GRB alert stream can check new triggers against recent triggers and automatically flag possible ultra-long GRBs within $\sim$15\,minutes of event time, providing key prioritization information for the worldwide follow-up community. This results in the capture of key early information of these unique events and recovery of key information such as redshift, host galaxy type, and potential offset from the host galaxy. The ultra-long classification can be performed in higher latency. 

\subsubsection{GRB\,200826A}\label{sec:grb200826a}
GRB\,200826A was discovered by Fermi-GBM, which identified the temporal behavior as consistent with a short GRB origin. The ZTF tiled $\sim$180\,deg$^2$ and found a single optical counterpart candidate that survived their filter criteria, representing the first recovery of an afterglow via wide-field tiling of a GBM short GRB localization. The prompt emission was seen by four additional instruments in the IPN, which resulted in a localization spanning a 90\% localization region of $\sim$100\,arcminute$^2$. This enabled a robust association of the optical signal as the confirmed counterpart and the direct claim of the shortest confirmed collapsar identified \citep{ahumada2021discovery}. The IPN localization also allowed Swift tiling of the region, recovering the X-ray afterglow as well. This example demonstrates the new capability to observe large regions of the sky in optical and the use of higher latency IPN alerts for robust association of possible counterparts.

\subsubsection{GRB\,170114A}\label{sec:grb170114a}
Observations of prompt polarization is done using Compton telescopes. Spectropolarimetric studies allow for greater insight into the underlying prompt emission mechanism of GRBs but requires large photon statistics to perform polarimetry. The greatest understanding of prompt emission will come from time-resolved spectropolarimetric studies, requiring even larger photon statistics. Providing external spectral information or additional counts through multi-mission analysis can greatly enhance the return from a given GRB polarimetric observation. The polarization observation of GRB\,170114A by POLAR is one example, where the time-resolved spectropolarimetric study benefited from the inclusion of Fermi-GBM data \citep{burgess2019time}.

These studies promise great scientific advancement but are tedious. They are further complicated by the current inter-calibration uncertainty, which would be reduced, or at least properly characterized, through a broad, coordinated effort. This work will directly increase the robustness of the polarization study results of NASA's forthcoming COSI mission.

\subsubsection{GRB\,200415A}\label{sec:grb200415a}
GRB\,200415A is the single most convincing identification of an extragalactic MGF \citep{svinkin2021bright}. With the final localization of 0.013\,deg$^2$ the chance alignment with a local, star forming galaxy is 5E-6 \citep{burns2021identification}. While the initial Fermi-GBM localization of 277\,deg$^2$ provides only marginal evidence in favor of an MGF origin, the first arrival of data from a distant spacecraft is capable of both strongly suggesting an MGF origin and correctly identifying the putative host galaxy. X-ray observations could begin before the final localization is available. Automatic identifying of extragalactic MGF candidates may enable new insight into these events.

Further, automatic identification of Galactic MGFs may be even more fruitful. This would require checking new events as being extremely bright and rapidly pointing X-ray telescopes at the currently active magnetar(s). This is yet to be succesfully achieved.

\subsection{Additional Benefits}
Implementation of these community products by a multi-mission consortium will produce a number of additional benefits. Building a tech stack designed for various missions will ease the required workload to bring new instruments to scientific readiness, and adding such missions to the consortium will create a more powerful network, enabling new science with the same instruments for less effort. This may be particularly helpful for smallsat gamma-ray missions, both from NASA and other countries. The collated GRB stream will ensure maximal use of their data and the collated GRB catalog will ensure their observational results are archived for future analysis. A sufficiently advanced shared database may reduce the number of total active Burst Advocates for real-time monitoring of transients. It may also remove the need for separate websites for each instrument team.

A critical aspect of this idea is ensured credit to early-career scientists. A sufficiently organized group can engage with the follow-up community in a method analogous to the organization of the Exoplanet community. The community organized observing campaigns around known transit times of exoplanets observed by TESS (and previously Kepler). For the resulting published literature the community groups reach out to the TESS team for instrument experts to contribute to their manuscript. Here, the numerous GRB follow-up groups publish multiple papers a year. Each could have a small number of prompt GRB experts, with preference towards early-career members as co-authors. This ensures visibility for the prompt GRB community and inclusion of gamma-ray expertise in multiwavelength published literature.

Intentional planning for proper credit is key to the success of TDAMM in the open-science era. If the prompt GRB community builds tools to identify interesting transients and reports results to enable follow-up, those scientists will lose the visibility of papers unless field dynamics change. An equally critical consideration is funding for this kind of support and infrastructure work, a mechanism that currently does not exist, which will enhance science return of operating and future missions. 

The IPN demonstrates a method of utilizing otherwise private data for public alerts and results. The agreements that exist were constructed to enable this while ensuring the contributions of the various instrument teams are visible to their respective funding agencies. As NASA implements its Transition to Open Science, the IPN is a model of how international partnerships with private data may work, but growth of the IPN responsibilities requires greater investment from NASA. 

Lastly, IGWN is now the obvious group to begin discussions of relevance to the ground-based GW community. The consortium enables organized advocacy and engagement with funding agencies in different countries to seek support for continued investment in the area. IGWN is greater than the sum of its parts, so individual funding agencies get a greater return for their investment. We may expect similar benefits for a broader gamma-ray transient network. These can include enhanced return on education and public outreach activity, coordinating press releases on GRB-related discoveries, providing the single access point for future requests of the prompt high-energy monitors, and acting as a general venue for continuously improving the community products from these instruments.

\section{Conclusions}\label{sec:conclusions}
High-energy monitors are a foundational pillar of the modern time-domain and multimessenger ecosystem. They are the preeminent discoverer of relativistic transients, and the signatures they detect carry unique diagnostic information necessary to understand these cosmic explosions. The scientific results promised in the coming years include identifying the progenitors and physical origin of a broad set of phenomenologically classified events, the sources of GWs and neutrinos, and several probes of fundamental physics. Their continued need for the foreseeable future is inescapable. 

This report details the broad sources of interest, the key observational signatures and corresponding instrument requirements, and details what kinds of missions are needed to accomplish these goals. We find that meeting the full community needs this decade requires the recommended NASA TDAMM ecosystem described in the Astro2020 Decadal. Furthermore, we find that the IPN presents a unique set of capabilities, including total coverage of the gamma-ray sky, that must be intentionally maintained this decade and in the decades that follow. Support from NASA is critical for long-term planning, maximal scientific return from existing and forthcoming instruments, and to ameliorate the problems inherent in programmatic stovepipes.

\newpage 

\bibliographystyle{aasjournal}
\bibliography{bibliography}

\end{document}

%% file: authors.tex
\author[0000-0002-2942-3379]{Eric~Burns} 
\affil{Department of Physics \& Astronomy, Louisiana State University, Baton Rouge, LA 70803, USA}

\author[0000-0002-8262-2924]{Michael~Coughlin}
\affil{School of Physics and Astronomy, University of Minnesota, Minneapolis, Minnesota 55455, USA}

\author[0000-0002-8648-0767]{Kendall~Ackley}
\affiliation{Department of Physics, University of Warwick, Gibbet Hill Road, Coventry CV4 7AL, UK}

\author[0000-0002-8977-1498]{Igor~Andreoni}
\affil{Joint Space-Science Institute, University of Maryland, College Park, MD 20742, USA}
\affil{Department of Astronomy, University of Maryland, College Park 
College Park, MD 20742, USA}
\affil{Astrophysics Science Division, NASA Goddard Space Flight Center, Greenbelt, MD 20771, USA}

\author[0000-0002-4618-1674]{Marie-Anne~Bizouard}
\affiliation{Artemis, Universit\'e C\^ote d’Azur, Observatoire de la C\^ote d’Azur, CNRS, Nice 06300, France}

\author[0000-0002-4421-4962]{Floor~Broekgaarden}
  \affiliation{Center for Astrophysics $|$ Harvard $\&$ Smithsonian, 60 Garden St, Cambridge, MA 02138}

\author[0000-0002-6870-4202]{Nelson~L.~Christensen}
\affiliation{Artemis, Universit\'e C\^ote d’Azur, Observatoire de la C\^ote d’Azur, CNRS, Nice 06300, France}

\author[0000-0001-7618-7527]{Filippo~D'Ammando}
\affiliation{Italian National Institute for Astrophysics, Institute for Radioastronomy, Bologna, 40129, Italy}

\author[0000-0001-5229-1995]{James~DeLaunay}
\affiliation{Department of Physics \& Astronomy, University of Alabama, Tuscaloosa, AL 35487, USA}

\author[0000-0002-0794-8780]{Henrike~Fleischhack}
\affil{Catholic University of America, Washington, DC, USA}
\affil{Astrophysics Science Division, NASA Goddard Space Flight Center, Greenbelt, MD 20771, USA}
\affil{Center for Research and Exploration in Space Science and Technology, NASA/GSFC, Greenbelt, MD 20771, USA}
\affil{Physikalisch-Technische Bundesanstalt (National Metrology Institute), Braunschweig, Germany}

\author[0000-0003-0341-2636]{Raymond~Frey}
\affil{Department of Physics, University of Oregon, Eugene, OR 97403, USA}

\author[0000-0003-2624-0056]{Chris~L.~Fryer}
\affiliation{Center for Non Linear Studies, Los Alamos National Laboratory, Los Alamos, NM, 87545, USA}

\author[0000-0002-0587-7042]{Adam~Goldstein}
\affiliation{Science and Technology Institute, Universities Space and Research Association, 320 Sparkman Drive, Huntsville, AL 35805, USA.}

\author[0000-0003-4179-7063]{Bruce~Grossan}
\affil{Space Sciences Laboratory, University of California, Berkeley, Berkeley, CA 94720, USA}

\author[0000-0003-0761-6388]{R.~Hamburg}
\affiliation{Universit\'e Paris-Saclay, CNRS/IN2P3, IJCLab, 91405 Orsay, France}

\author[0000-0002-8028-0991]{Dieter~H.~Hartmann}
\affiliation{Clemson University, Department of Physics \& Astronomy, Clemson, SC 29634-0978, USA}

\author[0000-0002-9017-3567]{Anna~Y.~Q.~Ho}
\affiliation{Department of Astronomy, Cornell University, Ithaca, NY 14853, USA}

\author[0000-0001-7891-2817]{Eric~J.~Howell}\affiliation{OzGrav-UWA, School of Physics \& Astrophysics, University of Western Australia, Crawley WA 6009, Australia}

\author[0000-0002-0468-6025]{C.~Michelle~Hui}\affiliation{ST12 Astrophysics Branch, NASA Marshall Space Flight Center, Huntsville, AL 35812, USA}

\author[0000-0002-9212-2034]{Leah~Jenks} 
\affiliation{Kavli Institute for Cosmological Physics, University of Chicago, Chicago, IL 60637, USA}

\author[0000-0001-5783-8590]{Alyson~Joens}
\affil{Space Sciences Laboratory, University of California, Berkeley, Berkeley, CA 94720, USA}

\author[0000-0001-8058-9684]{Stephen~Lesage}
\affiliation{Department of Space Science, University of Alabama in Huntsville, 320 Sparkman Drive, Huntsville, AL 35899, USA}
\affiliation{Center for Space Plasma and Aeronomic Research, University of Alabama in Huntsville, Huntsville, AL 35899, USA}

\author{Andrew~J.~Levan}
\affiliation{Department of Astrophysics/IMAPP, Radboud University, P.O. Box 9010, NL-6500 GL Nĳmegen, The Netherlands}

\author[0000-0002-7851-9756]{Amy~Lien}
\affiliation{University of Tampa, Department of Chemistry, Biochemistry, and Physics, 401 W. Kennedy Blvd, Tampa, FL 33606, USA}

\author[0000-0002-3064-5307]{Athina~Meli}
\affil{College of Science and Technology, Department of Physics, North Carolina A\&T State University, Greensboro, NC 27411, USA}
\affil{STAR Institute, Institut d'Astrophysique et de Geophysique, University of Liege, Liege 4000, BE}

\author[0000-0002-6548-5622]{Michela~Negro}
\affiliation{University of Maryland, Baltimore County, Baltimore, MD 21250, USA}
\affiliation{Astrophysics Science Division, NASA Goddard Space Flight Center, Greenbelt, MD 20771, USA}
\affiliation{Center for Research and Exploration in Space Science and Technology, NASA/GSFC, Greenbelt, MD 20771, USA}

\author[0000-0002-4299-2517]{Tyler~Parsotan}
\affiliation{Astrophysics Science Division, NASA Goddard Space Flight Center, Greenbelt, MD 20771, USA}

\author[0000-0002-7150-9061]{Oliver~J.~Roberts}
\affiliation{Science and Technology Institute, Universities Space and Research Association, 320 Sparkman Drive, Huntsville, AL 35805, USA.}

\author[0000-0001-7297-8217]{Marcos~Santander}
\affiliation{Department of Physics \& Astronomy, University of Alabama, Tuscaloosa, AL 35487, USA}

\author[0000-0002-3594-6133]{Jacob~R.~Smith}
\affiliation{George Mason University, Department of Physics \& Astronomy, 4400 University Drive, Fairfax, VA 22030, USA}
\affiliation{Naval Research Laboratory, 4555 Overlook Avenue SW, Washington, DC 20375, USA}

\author[0000-0002-2810-8764]{Aaron~Tohuvavohu}
\affil{Department of Astronomy \& Astrophysics, University of Toronto, Toronto, ON M5S 3H4}

\author[0000-0001-5506-9855]{John~A.~Tomsick}
\affil{Space Sciences Laboratory, University of California, Berkeley, Berkeley, CA 94720, USA}

\author[0000-0002-9249-0515]{Zorawar~Wadiasingh}
\affiliation{Department of Astronomy, University of Maryland, College Park 
College Park, MD 20742, USA}
\affiliation{Astrophysics Science Division, NASA Goddard Space Flight Center, Greenbelt, MD 20771, USA}
\affiliation{Center for Research and Exploration in Space Science and Technology, NASA/GSFC, Greenbelt, MD 20771, USA}

\author[0000-0002-2149-9846]{P\'eter~Veres}
\affiliation{Department of Space Science, University of Alabama in Huntsville, 320 Sparkman Drive, Huntsville, AL 35899, USA}
\affiliation{Center for Space Plasma and Aeronomic Research, University of Alabama in Huntsville, Huntsville, AL 35899, USA}

\author[0000-0002-5814-4061]{Ashley~V.~Villar}
\affil{Department of Astronomy and Astrophysics, Pennsylvania State University, 525 Davey Laboratory, University Park, PA 16802, USA}
\affil{Institute for Computational \& Data Sciences, The Pennsylvania State University, University Park, PA, USA}
\affil{Institute for Gravitation and the Cosmos, The Pennsylvania State University, University Park, PA 16802, USA}

\author[0000-0001-9826-1759]{Haocheng~Zhang}
\affiliation{Astrophysics Science Division, NASA Goddard Space Flight Center, Greenbelt, MD 20771,
USA}

\author[0000-0002-6468-8292]{Sylvia~J.~Zhu}
\affiliation{Deutsches Elektronen-Synchrotron (DESY), D-15738 Zeuthen, Germany}